\documentclass[aps,pra,reprint,superscriptaddress,nofootinbib,longbibliography]{revtex4-2}

\usepackage{amsmath,amssymb,amsthm}
\usepackage{bm,bbm}
\usepackage{physics,braket}
\usepackage{xcolor,graphicx}
\usepackage{booktabs}
\usepackage{enumitem}
\usepackage[colorlinks=true,linkcolor=blue!55!black,citecolor=blue!55!black,
            urlcolor=blue!55!black]{hyperref}
\usepackage{orcidlink} 

\newcommand{\sys}{\mathrm{sys}}
\newcommand{\bath}{\mathrm{bath}}

\newcommand{\fwd}{\mathrm{fwd}}
\newcommand{\bwd}{\mathrm{bwd}}
\newcommand{\dF}{\Delta F}
\newcommand{\lNH}{\lambda_{\mathrm{NH}}}
\newcommand{\lNESS}{\lambda_{\mathrm{NESS}}}
\newcommand{\lOTOC}{\lambda_{\mathrm{OTOC}}}
\newcommand{\lcone}{\lambda_{\mathrm{cone}}}
\newcommand{\lleak}{\lambda_{\mathrm{leak}}}
\newcommand{\cdag}{c^{\dagger}}
\newcommand{\Heff}{H_{\mathrm{eff}}}
\newcommand{\Vf}{\mathcal{V}_{\mathrm{fwd}}}
\newcommand{\Vb}{\mathcal{V}_{\mathrm{bwd}}}
\newcommand{\Lf}{\mathcal{L}_{\mathrm{fwd}}}
\newcommand{\Lb}{\mathcal{L}_{\mathrm{bwd}}}
\newcommand{\teta}{\tilde\eta}
\newcommand{\one}{\mathbbm{1}}
\newcommand{\dEF}{\Delta_{\eta}F}

\begin{document}

\title{Quantum Information Scrambling in Liouvillian Skin Effects}

\author{Mahn-Soo Choi\,\orcidlink{0000-0001-6901-5233}}
\email{choims@korea.ac.kr}
\affiliation{Department of Physics, Korea University, Seoul 02841, Republic of Korea}

\date{\today}

\begin{abstract}
The non-Hermitian skin effect belongs to the evolution of an open
quantum system conditioned on the absence of quantum jumps, and its
observation requires postselection at an exponential cost. Its
counterpart in the full Lindblad dynamics, the Liouvillian skin effect,
requires no postselection but calls for an observable that resolves its
nonreciprocal dynamics. We show that the dissipative out-of-time-ordered
correlator (OTOC), the interferometric echo used to measure quantum
information scrambling, reveals the core dynamical features of this
effect, and we construct measure-and-reset quantum circuits that realize
it on current quantum processors. For the Hatano--Nelson Lindbladian of
free fermions, solved exactly, the effect of the butterfly operator
separates into a part carrying the memory of the reference state and a
chiral part generated by the dissipation alone, and two reset strategies
of the bath give the steady-state skin profile. For a local XX spin
chain directly simulable on qubit-based quantum computers, an exact
symmetry selection rule determines when the chirality is visible in the
OTOC, and a Monte Carlo simulation of the circuit resolves it with about
a thousand trajectories per time slice.
\end{abstract}

\maketitle

\section{Introduction}
\label{sec:intro}

Non-Hermitian Hamiltonians arise as effective descriptions of open
quantum systems and display phenomena that have no Hermitian
counterpart~\cite{Ashida20a,Kawabata19a,Bergholtz21a}. A prominent
example is the non-Hermitian skin
effect~\cite{Hatano96a,Yao18a,Gong18a,Okuma20a}. On a chain with
nonreciprocal hopping, of which the Hatano--Nelson
model~\cite{Hatano96a} is the paradigm, an extensive number of
eigenstates are exponentially localized at one boundary under open
boundary conditions, the spectrum depends strongly on the boundary
conditions, and the conventional bulk--boundary correspondence has to be
reformulated on a generalized Brillouin zone~\cite{Yao18a}. The effect
has a topological origin in the point-gap winding of the complex
spectrum~\cite{Gong18a,Okuma20a}. In a quantum system with Markovian
dissipation, however, a non-Hermitian Hamiltonian generates only the
evolution conditioned on the absence of quantum jumps, that is, the
postselected part of the dynamics described by a Lindblad master
equation. The probability of such jump-free trajectories decays
exponentially with time and with the system size, so that the
localization length of the non-Hermitian eigenmodes is not by itself a
practically observable quantity.

The counterpart of the non-Hermitian skin effect in the full Lindblad
dynamics is the Liouvillian skin
effect~\cite{Song19a,Haga21a,Yang22a,Zhou22b,Feng24a,Lee23a}, which
manifests itself in chiral damping and the resulting nonreciprocal
dynamics, in steady states with a boundary accumulation of particles,
and in a relaxation time under open boundary conditions that is not set
by the Liouvillian gap. Its signatures have so far been sought mainly in
one-point observables such as populations, currents and relaxation
rates, a notable exception being the chiral damping of the two-point
correlation function~\cite{Song19a}. These studies leave open how a
local perturbation spreads under the nonreciprocal dissipation and how
to separate what the dynamics retains of the initial state from what the
dissipation itself generates; moreover, the lengths that govern the
one-point observables need not coincide with the localization length of
the non-Hermitian Hamiltonian~\cite{Haga21a,Lee23a}. Two questions therefore remain open:
which observable resolves the dynamical features of the Liouvillian skin
effect, and how it can be measured in an experiment without
postselection.

The purpose of this work is to answer these two questions in the
setting of quantum simulation. Specifically, we pursue two objectives.
The first is to establish the dissipative out-of-time-ordered correlator
(OTOC) as an observable that reveals the core dynamical features of the
Liouvillian skin effect: the nonreciprocal spreading of a local
perturbation, the separation of the memory of the initial reference
state from the contribution generated by the chiral dissipation, and
the skin profile and the relaxation modes of the full Lindblad dynamics.
The second is to construct concrete quantum circuits, executable on
currently available quantum processors, that realize the required
nonreciprocal dissipation and measure the dissipative OTOC.

The OTOC is measured on quantum processors by an interferometric
echo~\cite{Swingle16a,Mi21a,Abanin25a}: the system is evolved forward in
time, perturbed by a local \emph{butterfly} operator, evolved backward,
and read out through a local \emph{probe} operator that is conditioned
on a control qubit. In an open system the backward evolution can invert
the Hamiltonian of the system but not its coupling to the environment,
and the protocol then measures the dissipative OTOC introduced by Zhang,
Huang, and Chen (ZHC)~\cite{Zhang19a}. Two properties make this quantity
suitable for our purpose. First, as a four-point interferometric
correlator it contains more information about the dissipation than
one-point observables. We show that it obeys exact symmetry selection
rules, that it has a chiral component independent of the reference
state, and that it gives the Liouvillian skin profile without
postselection. Second, the circuit that measures it is itself a
well-defined model of nonreciprocal dissipation. It consists of the
system register, bath registers that play the role of the reservoirs and
are measured and reset after every time step, and one control qubit. The
measurement record of the bath registers unravels the Lindblad dynamics
into quantum trajectories, and the non-Hermitian evolution is recovered
only if all records except the jump-free one are discarded.

We study two models. The first is the Hatano--Nelson
Lindbladian~\cite{Kawabata23a}, a chain of spinless fermions with
Hermitian hopping and, on every bond, a loss channel
$A_l\propto c_l+\eta e^{i\phi}c_{l+1}$ and a gain channel
$B_l\propto\cdag_l+\eta e^{i\phi}\cdag_{l+1}$ of the \emph{same}
chirality. Its dynamics is Gaussian and number conserving, so that the
dissipative OTOC can be evaluated exactly at a cost polynomial in the
chain length $L$. This allows analytical statements about the structure
of the OTOC and numerical results for chains of about $40$ sites, far
beyond what the $4^L$-dimensional Liouville space permits in general.
The steady state of this model carries a chiral current and a skin
profile at the boundaries. The second model is a local XX spin chain
coupled to a thermal bath through the bond operators
$S^-_l+\eta e^{i\phi}S^-_{l+1}$ and their adjoints. After a
Jordan--Wigner transformation its effective non-Hermitian Hamiltonian is
the same Hatano--Nelson chain, but its jump operators are \emph{local}
spin operators, which can be implemented directly with nearest-neighbor
gates and measure-and-reset bath qubits on a quantum processor. Its
steady state, however, is the uniform thermal product state for every
$\eta$, so that the chirality of this model appears only in transient
dynamics or at finite bath temperature. The comparison of the two models
shows how strongly the full Lindblad dynamics can differ for one and the
same non-Hermitian Hamiltonian.

Our main results are as follows. For the fermionic model, we derive the
Lindbladian from a Hermitian system--bath model in the wideband
(singular-coupling) limit, in which the nonreciprocity is a flux of a
quarter flux quantum through the plaquettes formed by the chain and its
reservoirs. We evaluate the dissipative OTOC by three methods: an exact
solution for the single-particle correlation matrix at a cost of
$O(L^3)$ per time point; a Stinespring dilation in which the two
branches of the interferometer share their bath registers (a dilation
with separate registers for each branch gives the postselected OTOC
instead); and an unbiased quantum-trajectory estimator within the
fermionic Gaussian formalism, which corresponds to the hardware-efficient
measure-and-reset circuit. We then analyze the butterfly-induced
deviation $\dF$, defined as the OTOC minus the leak factor, that is, the
same echo measured without the butterfly. It separates exactly into a
reflection-symmetric part that carries the memory of the reference state
and an antisymmetric part that is generated by the chiral dissipation
and is independent of the reference state up to a known sign. The
nonreciprocity thus appears in the light cone as a sign reversal between
its two arms, which propagate at the same velocity and decay at the same
rate. In the steady state, the difference between two reset strategies
of the bath in the backward evolution reproduces the Liouvillian skin
profile. Its depth $\lOTOC$ equals the depth $\lNESS$ of the steady
state, for which we obtain a closed form, and differs from the
non-Hermitian depth $\lNH$, which requires postselection, in both its
magnitude and its dependence on the nonreciprocity.
For the spin model, we establish an exact symmetry selection rule that
determines for which reference states the chirality is visible in the
OTOC and identify a reference state for which it is isolated exactly. A
Monte Carlo simulation of the measure-and-reset circuit with thermal
reset of the bath qubits reproduces the light cone of an $11$-site chain
and resolves the chirality at more than $10\sigma$ with about $10^3$ trajectories
per time slice. At finite bath temperature the OTOC of the achiral
thermal steady state becomes chiral, and at late times the leak factor
displays the profile of the skin-localized relaxation modes of the
Liouvillian. Throughout, we use the difference $\dF$ rather than the
ratio of the OTOC to the leak factor considered by ZHC, because it
separates the effect of the butterfly from the decay of the echo in a way
that preserves the exact symmetries.

The paper is organized as follows. Section~\ref{sec:Model} introduces
the Hatano--Nelson Lindbladian and its microscopic system--bath model.
Section~\ref{sec:OTOC} defines the dissipative OTOC and the quantum
circuit that measures it, the observables derived from it, their
symmetries, and the reset strategies of the bath.
Section~\ref{sec:Methods} presents the three methods of evaluation: the
exact Gaussian solution, the Stinespring dilation and the quantum
trajectories. Section~\ref{sec:Results} contains the results for the
fermionic model, namely the anatomy of $\dF$, the light cones and the
Liouvillian skin effect in the steady state. Section~\ref{sec:LocalSpinModel} is devoted to the local
spin model. Section~\ref{sec:Summary} summarizes the results and lists
open questions. Technical derivations are given in the Appendices.

\section{Model and Observables}
\label{sec:Model}

\subsection{Hatano--Nelson Hamiltonian and Lindbladian}
\label{sec:HatanoNelson}

The Hatano--Nelson model~\cite{Hatano96a} is the paradigmatic
non-Hermitian tight-binding chain. For spinless fermions on an open chain
of $L$ sites, $l=1,\dots,L$, it reads
\begin{equation}
  H_{\rm HN}=-\sum_{l=1}^{L-1}\Big[J_L\,\cdag_lc_{l+1}+J_R\,\cdag_{l+1}c_l\Big],
  \qquad J_L\neq J_R,
  \label{eq:HN_form}
\end{equation}
where $c_l$ and $\cdag_l$ annihilate and create a fermion on site $l$ and
the left and right hopping amplitudes $J_L$ and $J_R$ differ. On the
finite open chain the imaginary-gauge similarity transformation
$S^{-1}H_{\rm HN}S$ with $S=\mathrm{diag}[(J_R/J_L)^{l/2}]$ is Hermitian,
so every eigenmode is exponentially localized at the boundary toward
which the stronger hopping points, at the common rate
\begin{equation}
  \frac{1}{\lNH}=\frac12\,\Big|\ln\frac{J_R}{J_L}\Big|.
  \label{eq:kappa_general}
\end{equation}
This is the non-Hermitian skin
effect~\cite{Hatano96a,Yao18a,Gong18a,Okuma20a} (for reviews, see
Refs.~\cite{Ashida20a,Kawabata19a,Bergholtz21a}). The
localization of the eigenmodes is the generalized-Brillouin-zone skin
effect~\cite{Yao18a}, and $\lNH$ is the Kawabata--Ryu nonunitary
localization length of the single-particle generator~\cite{Kawabata21a}.

A non-Hermitian Hamiltonian of this form arises as the effective
Hamiltonian of a Lindblad equation with nonreciprocal jump
operators~\cite{Kawabata23a,Kawabata22a}: it generates the evolution
conditioned on the absence of quantum jumps, which we call the no-jump
evolution (also known as the postselected or, in experiments that
monitor the jumps~\cite{Liu25b}, the no-click evolution). Consider the same chain with the
Hermitian tight-binding Hamiltonian
\begin{equation}
  H_{\sys}=-\frac{J}{2}\sum_{l=1}^{L-1}\big(\cdag_l c_{l+1}+\mathrm{h.c.}\big)
  \label{eq:Hsys}
\end{equation}
(we set $J=1$ as the unit of energy unless stated otherwise), and on every
bond $(l,l{+}1)$ a loss and a gain channel with the same
complex relative amplitude $\eta e^{i\phi}$ between the two end sites,
\begin{align}
  A_l&=\frac{c_l+\eta e^{i\phi}\,c_{l+1}}{\sqrt{1+\eta^2}}\qquad(\text{loss}),
  \label{eq:jumpA}\\
  B_l&=\frac{\cdag_l+\eta e^{i\phi}\,\cdag_{l+1}}{\sqrt{1+\eta^2}}\qquad(\text{gain}),
  \label{eq:jumpB}
\end{align}
$l=1,\dots,L-1$, normalized so that $\{A_l,A_l^\dagger\}=\{B_l,B_l^\dagger\}=\one$.
The master equation is
\begin{equation}
  \dot\rho=-i[H_{\sys},\rho]
  +\gamma\sum_{l=1}^{L-1}\Big(\mathcal D[A_l]+\mathcal D[B_l]\Big)\rho,
  \label{eq:LindbladHN}
\end{equation}
with $\mathcal{D}[X]\rho=X\rho X^\dagger-\tfrac12\{X^\dagger X,\rho\}$.
This is the Hatano--Nelson Lindbladian, whose consequences range from
non-Bloch band theory to entanglement
transitions~\cite{Kawabata23a,Kawabata22a}. The
normalization in Eqs.~\eqref{eq:jumpA} and~\eqref{eq:jumpB} fixes the bulk
on-site damping at $\gamma$ per channel for every $\eta$ (with the
unnormalized choice it would grow as $\gamma(1+\eta^2)$), so that $\gamma$
is the on-site damping rate itself and \emph{all} of the $\eta$
dependence enters through the bounded combination
\begin{equation}
  \teta := \frac{2\eta}{1+\eta^2},
  \label{eq:teta}
\end{equation}
maximal at the self-dual point $|\eta|=1$. 

Two remarks on the model are in order. First, many works, including
Refs.~\cite{Kawabata23a,Kawabata22a}, are concerned with the skin effect
or with the topology of the effective non-Hermitian Hamiltonian and
accordingly keep only the loss channels $A_l$ of Eq.~\eqref{eq:jumpA};
keeping only the gain channels $B_l$ of Eq.~\eqref{eq:jumpB} instead
gives equivalent results in that context. Here both channels are
retained, so that the dissipation does not simply drain the chain: with
equal loss and gain rates the fermion number is preserved on average, and
the full Lindblad dynamics has a nontrivial steady state at half filling,
which is where the Liouvillian skin effect arises. It is moreover the
interplay of the two channels---whose chiral cross terms add in $\Heff$,
Eq.~\eqref{eq:AA_BB} below, but cancel in the damping of the
single-particle correlations (Sec.~\ref{sec:GaussianSolution})---that
will separate the Liouvillian skin depth from the non-Hermitian one in
Sec.~\ref{sec:NESS}. Second, we keep two parameters, the amplitude ratio
$\eta$ and the relative phase $\phi$, where the literature typically
fixes the maximally nonreciprocal point $\eta e^{i\phi}=\pm i$. Both are
physically realizable, as the microscopic model of
Sec.~\ref{sec:SystemBath} shows, and both serve as \emph{in situ} control
knobs in the quantum-circuit realizations of
Secs.~\ref{sec:QuantumTrajectoryMethod} and~\ref{sec:LocalSpinModel}:
$\eta$ is a ratio of couplings, and $\phi$ can be set in software as a
virtual phase rotation of a bath mode, so that the chirality can be
tuned, and reversed, without any change of the hardware.

The no-jump evolution of Eq.~\eqref{eq:LindbladHN} is generated by the
effective non-Hermitian Hamiltonian
\begin{equation}
  \Heff=H_{\sys}-\frac{i\gamma}{2}\sum_l\big(A_l^\dagger A_l+B_l^\dagger B_l\big).
  \label{eq:Heff_raw}
\end{equation}
By the anticommutation relations the number-operator pieces cancel
between loss and gain,
\begin{equation}
  A_l^\dagger A_l+B_l^\dagger B_l=\one+i\teta\sin\phi\,\big(\cdag_lc_{l+1}-\cdag_{l+1}c_l\big),
  \label{eq:AA_BB}
\end{equation}
leaving a c-number and a current-like Hermitian term whose anti-Hermitian
contribution to $\Heff$ shifts the hopping asymmetrically:
$\Heff=H_{\rm HN}-\tfrac{i\gamma(L-1)}{2}\one$ with
\begin{equation}
  J_{L/R}=\frac{J\mp\gamma\teta\sin\phi}{2}.
  \label{eq:JLR}
\end{equation}
The asymmetry $J_R-J_L=\gamma\teta\sin\phi$ is maximal at $\eta=1$,
$\phi=\pm\pi/2$, with the sign of $\phi$ deciding the favored direction,
and the uniform imaginary shift gives the no-jump probability
$e^{-\gamma(L-1)t}$. We focus mainly on the maximally nonreciprocal case
$\phi=\pm\pi/2$, where $\eta e^{i\phi}=\pm i\eta$, and from
Sec.~\ref{sec:OTOC} onward we write the relative amplitude simply as
$i\eta$ with $\eta\in\mathbb R$ of either sign,
$\mathrm{sgn}\,\eta=\mathrm{sgn}\,\phi$, as the figures do. For $\phi=\pi/2$
the skin depth
\eqref{eq:kappa_general} reads
\begin{equation}
  \frac{1}{\lNH}=\frac12\ln\frac{J(1+\eta^2)+2\gamma\eta}{J(1+\eta^2)-2\gamma\eta},
  \label{eq:kappa}
\end{equation}
which is exactly self-dual under $\eta\leftrightarrow1/\eta$, and finite
for \emph{all} $\eta$ when $\gamma<J$, since the fully chiral point
$J_L=0$ requires $\gamma=\tfrac J2(\eta+1/\eta)\ge J$; $\eta=1$ is
therefore a usable, maximally localized operating point rather than a
singular one.

We stress that the length $\lNH$ is a property of the
\emph{postselected} no-jump evolution: isolating it requires the
jump-free trajectory, whose probability decays exponentially in time and
system size, so $\lNH$ is not by itself a practically observable length.
Physically, the primary question is which length the \emph{full} Lindblad
dynamics imprints on observables. This is the
Liouvillian skin effect~\cite{Song19a,Haga21a,Yang22a,Lee23a}: the steady
state of Eq.~\eqref{eq:LindbladHN} is a current-carrying state with a
boundary accumulation of occupation, and the slowest relaxation modes of
the Liouvillian are themselves skin-localized, which separates the
Liouvillian gap from the relaxation time under open boundaries. We will
show in Sec.~\ref{sec:NESS} that the steady-state profile decays on a
length $\lNESS$ that is set by the uniform damping rather than by the
spectral asymmetry of $\Heff$, and that the dissipative OTOC measures
exactly this length.

\subsection{Microscopic System--Bath Model}
\label{sec:SystemBath}

\begin{figure}
\centering
\includegraphics[width=8cm]{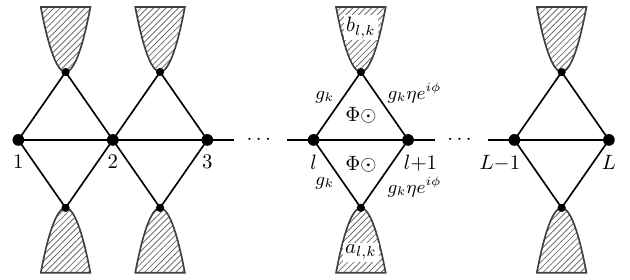}
\caption{Geometry of the microscopic system--bath model. Every bond $(l,l{+}1)$
of the chain is the base of two triangular plaquettes, closed by the
gain reservoir $b_{l,k}$ above and the loss reservoir $a_{l,k}$ below
(hatched lobes: the continua of reservoir modes, filled for $b$, empty
for $a$). Each reservoir couples to both end sites of its bond, with
amplitudes $g_k$ and $g_k\eta e^{i\phi}$ for each reservoir mode $k$
($g_k=g^{(a)}_k$, $g^{(b)}_k$ of Eq.~\eqref{eq:Vab}). The relative
phase $e^{i\phi}$ is the Peierls phase of a uniform magnetic flux
$\Phi=\phi\,\Phi_0/2\pi$ threading every triangle, with $\Phi_0=h/e$ the
electron flux quantum: $\Phi=\pm\Phi_0/4$ for $\phi=\pm\pi/2$. $\Phi>0$
means a field pointing out of the page ($\odot$, as drawn), $\Phi<0$ into
it. The chain bonds bound no plaquette, so the system hopping stays real.}
\label{fig:flux}
\end{figure}

The master equation~\eqref{eq:LindbladHN} can be derived as the reduced
dynamics of a fully Hermitian microscopic model. The derivation shows
where the nonreciprocity comes from and decides \emph{which} Markovian
limit is the physically correct one, a point that matters for the skin
effect. More importantly for our purposes, the microscopic bath model is
the physical template for the quantum-circuit realization constructed in
Sec.~\ref{sec:QuantumTrajectoryMethod}, in which the reservoirs become
bath registers that are measured and reset, and which is directly
simulable on a quantum computer. We summarize the derivation here and
give the details in Appendix~\ref{app:HNbath}.

We attach to each bond two independent free-fermion reservoirs with a
continuum of modes, $H_{\bath}^{(a,l)}=\sum_k\omega^{(a)}_ka^\dagger_{l,k}a_{l,k}$
and $H_{\bath}^{(b,l)}=\sum_k\omega^{(b)}_kb^\dagger_{l,k}b_{l,k}$, the
$a$-reservoirs in their vacuum [$f_a(\omega)=0$ for all $\omega$] and the
$b$-reservoirs fully inverted [$f_b(\omega)=1$ for all $\omega$];
physically one may
think of $b$ as a Fermi sea well above the system band and $a$ as one well
below. The asymmetric choice is what makes one channel a loss and the
other a gain. Note that the occupation functions $f_a$ and $f_b$ are
constant in energy rather than Fermi functions of a common temperature:
each reservoir feeds exactly one channel, and no Kubo--Martin--Schwinger
relation ties the gain rate to the loss rate, in contrast to a single
thermal reservoir, for which the two would be locked by detailed balance
(the situation of the spin model, Sec.~\ref{sec:spin:SystemBath}). Each
reservoir
couples to a \emph{coherent two-site superposition} of system operators
with a definite relative phase,
\begin{align}
  V_{a,l}&=\sum_k g^{(a)}_k\Big[a^\dagger_{l,k}\,A_l+\mathrm{h.c.}\Big],
  \label{eq:Vab}\\
  V_{b,l}&=\sum_k g^{(b)}_k\Big[b_{l,k}\,B_l+\mathrm{h.c.}\Big],
  \label{eq:Vb}
\end{align}
with $A_l$, $B_l$ exactly the operators of Eqs.~\eqref{eq:jumpA}
and~\eqref{eq:jumpB}. Both couplings are bilinear and number conserving in
the joint Fock space; no anomalous terms appear.

The relative phase $e^{i\phi}$ is the entire source of the
nonreciprocity: without it the bath would couple to a real symmetric or
antisymmetric superposition and there would be no preferred direction,
whereas the phase makes the bath sensitive to a chiral current rather than
to a charge density. In the geometry of Fig.~\ref{fig:flux}, where the
reservoir modes of bond $l$ sit above and below the bond and couple to
both of its end sites with real $g_k$, the phase is the Peierls phase of a
magnetic flux through the triangular plaquettes. Circulating
counterclockwise around either triangle accumulates the same phase
$-\phi$, so the configuration is a uniform field threading all triangles
of the ladder while the chain bonds themselves bound no plaquette and
$H_{\sys}$ stays real. For an electron, $\sum_{\rm loop}\theta=-2\pi\Phi/\Phi_0$,
hence
\begin{equation}
  \frac{\Phi}{\Phi_0}=\frac{\phi}{2\pi}=\pm\frac14\qquad(\phi=\pm\pi/2),
  \label{eq:quarterflux}
\end{equation}
a quarter of a flux quantum per triangle, pointing out of the page for
$\phi=+\pi/2$. The sign has a transparent semiclassical reading: for
$\mathbf B$ out of the page the Lorentz force deflects a right-mover
upward, toward the gain reservoir, and a left-mover downward, toward the
loss reservoir---right-movers are weakly lost and strongly fed, which is
the chiral damping of Eq.~\eqref{eq:chiral_rates}. Reversing the field
reverses the skin direction: by Eq.~\eqref{eq:AA_BB},
\begin{equation}
  J_R-J_L=\gamma\,\teta\,\sin\phi=\gamma\,\teta\,\sin\!\big(2\pi\Phi/\Phi_0\big),
  \label{eq:JRJL_flux}
\end{equation}
odd in $\Phi$ and maximal at a quarter flux quantum; at $\Phi=0$ or
$\Phi_0/2$ the coupling acts as an achiral energy filter and the
nonreciprocity vanishes. Time-reversal symmetry must be broken by the flux
for the skin effect to appear. The dimensionless $\eta$ is independent of
the flux and simply measures the ratio of the two coupling magnitudes.
Other realizations of the phase---a directional coupling to a chiral edge
mode in photonic implementations, or Raman coupling of reservoir modes to
a pair of sites with a controlled spatial phase---lead to the same
Lindbladian.

The Born--Markov elimination of the reservoirs
(Appendix~\ref{app:HNbath}) differs from the textbook route in one step:
the final approximation is a \emph{wideband} (flat spectral density)
limit, $J_\alpha(\omega)=\gamma_\alpha/2\pi$ over a bandwidth
$\Lambda\gg J,\gamma$, \emph{not} a secular approximation. The bath
correlation functions then collapse to $\gamma_\alpha\delta(\tau)$, every
two-time product in the second-order equation becomes an equal-time one,
and one obtains exactly Eq.~\eqref{eq:LindbladHN} with the \emph{local}
jump operators $A_l$, $B_l$ and $\gamma_a=\gamma_b=:\gamma$. This is an
instance of the singular-coupling limit of
Refs.~\cite{Gorini76a,Palmer77a,Breuer02a}, in which the reduced dynamics
converges rigorously to the Gorini--Kossakowski--Sudarshan--Lindblad
(GKSL) semigroup~\cite{Gorini76a,Lindblad76a} whose jump operators are the
bare coupling operators. The derivation requires precisely one hierarchy,
\begin{equation}
  \tau_B\ll\min(1/J,\,1/\gamma),
  \label{eq:validity}
\end{equation}
with $\tau_B$ the bath memory time, i.e., the decay time of the bath
correlation functions $C_{a,b}(\tau)$, which is the inverse of the width
of the energy window over which the spectral densities $J_{a,b}(\omega)$
and the occupations $f_{a,b}(\omega)$ are simultaneously flat. In words:
the bath correlations are $\delta$-like (``white'') on the system's
internal time scales $1/J$ and $1/\gamma$, which keeps the jump operators
local, and the Born--Markov truncation holds. Two absences are as
important as the
conditions. \emph{No relation between $\gamma$ and $J$} is required, so
the strongly nonreciprocal regime $\gamma\teta\sim J$ lies fully within
validity; and \emph{no condition involves $L$}, so the limit commutes
with the thermodynamic limit, which matters because the Liouvillian skin
effect is intrinsically a large-$L$ phenomenon. A secular (Davies)
treatment would instead need $\gamma\ll2\pi J/L$ and, more importantly,
would replace the local jump operators by energy-filtered, delocalized
ones and erase precisely the physics of interest
(Appendix~\ref{app:secular}). The per-step dilations and measure-and-reset
circuits of Sec.~\ref{sec:Methods} realize the same hierarchy with
$\tau_B=\Delta t$, and converge, under the Trotter conditions
$J\Delta t,\gamma\Delta t\ll1$, to the nonsecular, local-jump
Lindbladian~\eqref{eq:LindbladHN}.

\section{The Dissipative OTOC}
\label{sec:OTOC}

\begin{figure*}
\centering
\includegraphics[width=\textwidth]{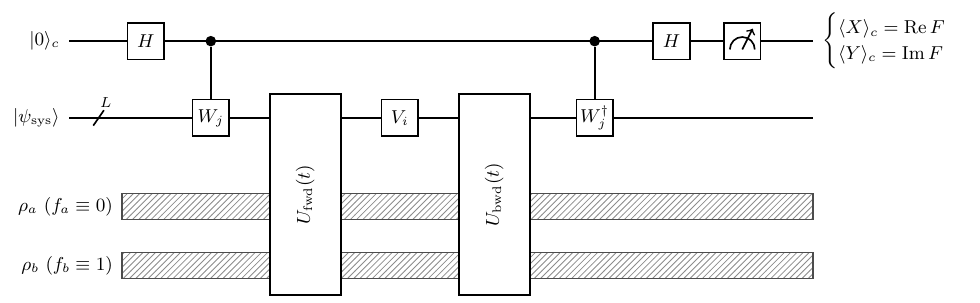}
\caption{Concept quantum circuit for the dissipative OTOC of the fermionic
model, with the butterfly $V_i$ at site $i=l_V$ and the probe $W_j$ at
site $j=l_W$. The control qubit $c$, prepared in $|+\rangle$ by a
Hadamard, selects the two branches---with and without the probe pair
$W_j$, $W_j^\dagger$ around the evolution---through the controlled $W_j$
and $W_j^\dagger$; after the final Hadamard, a measurement of the
control qubit in the $X$ or the $Y$ basis gives
$\langle X\rangle_c=\Re F$ or $\langle Y\rangle_c=\Im F$. The shaded
bands are the reservoirs themselves, continua of fermion modes, and the blocks
are the true evolutions of system plus reservoirs,
$U_\fwd(t)=e^{-it(H_{\sys}+H_{\bath}+V)}$ and $U_\bwd(t)$ with
$H_{\sys}\to-H_{\sys}$; the Lindbladians are what they reduce to in the
wideband limit. The two reservoirs are in the states of
Sec.~\ref{sec:SystemBath}, the $a$-reservoirs in their vacuum, $\rho_a$
with $f_a(\omega)=0$, and the $b$-reservoirs fully inverted, $\rho_b$
with $f_b(\omega)=1$. In the Trotterized, measure-and-reset realization
of this circuit (Fig.~\ref{fig:trajectory_fermi} and
Sec.~\ref{sec:QuantumTrajectoryMethod}) each band becomes a register of
$O(L)$ bath modes that is reset every $\Delta t$. The counterpart for
the spin model is Fig.~\ref{fig:circuit_spin}.}
\label{fig:circuit}
\end{figure*}

Out-of-time-ordered correlators~\cite{Larkin69a,Maldacena16a} quantify
the scrambling of quantum information. In this section we recall the
dissipative OTOC of ZHC and the quantum circuit that measures it
(Sec.~\ref{sec:Circuit}), define the observables derived from it that we
analyze in the rest of the paper (Sec.~\ref{sec:Observables}), establish
their exact symmetry properties (Sec.~\ref{sec:Symmetries}), and
introduce the bath-reset strategies that distinguish the forward and
backward blocks of the circuit (Sec.~\ref{sec:BathReset}). One purpose of
this work is to propose a concrete quantum circuit, simulable on current
quantum computers, that reveals the Liouvillian skin effect. The circuit
of this section, with its reservoirs replaced by bath registers that
are measured and reset after every step (Sec.~\ref{sec:QuantumTrajectoryMethod}),
is that proposal, and the observables and symmetries introduced below
are what one reads off from it.

\subsection{Quantum Circuit Model}
\label{sec:Circuit}

For two local unitaries
$V$ and $W$, the decay of the four-point correlator
$\langle V^\dagger(t)W^\dagger V(t) W\rangle$,\footnote{Some authors use the convention $\langle W^\dagger V^\dagger(t)WV(t)\rangle$, which is the complex conjugate of ours.}
with $V(t)=U^\dagger(t) VU(t)$ and unitary time-evolution operator $U(t)$,
tracks the growth of the commutator $[V(t),W]$: the \emph{butterfly}
perturbation $V$, applied at site $l_V$, spreads under the dynamics, and
its arrival at site $l_W$ is registered by the fixed \emph{probe} $W$. In
systems with spatial locality this growth is confined to a ballistic
information light cone bounded by Lieb--Robinson
physics~\cite{Lieb72a,Roberts16a,Hosur16a,Nahum18a,Keyserlingk18a}.
Throughout, both operators are Hermitian unitaries, $V=\one-2n_{l_V}$ and
$W=\one-2n_{l_W}$ for the fermionic model and $V=\sigma^z_{l_V}$,
$W=\sigma^z_{l_W}$ for the spin model of Sec.~\ref{sec:LocalSpinModel}.

On quantum hardware the OTOC is measured
interferometrically~\cite{Swingle16a,Mi21a,Abanin25a}: the system is
entangled with a control qubit and evolved forward and then backward in time,
the protocol being a time-domain analogue of the spin echo
(Fig.~\ref{fig:circuit}). The two branches of the control qubit share the sequence
forward evolution--$V$--backward evolution, which composes to $V(t)$, and
differ by the probe: $W$ before this sequence and $W^\dagger$ after it in
one branch, and neither in the other (equivalently, $W$ before the
sequence in one branch and $W$ after it in the other). The control qubit thus
measures the overlap of the two branches,
$\langle V^\dagger(t)W^\dagger V(t)W\rangle$. Without
the butterfly the forward and backward evolutions cancel, the two
branches coincide, and the echo is perfect; the butterfly spoils the echo
to the extent that $V(t)$ fails to commute with $W$. The probe is the
readout, and that it is applied first in one branch does not make it the
cause.

Real devices, however, are open: the backward evolution can invert the
system Hamiltonian, but neither the system--bath coupling nor the bath
itself. Zhang, Huang, and Chen~\cite{Zhang19a} formalized what such
a protocol measures and defined the dissipative OTOC
\begin{equation}
  F(t;V,W)=\Tr\left\{\Vb^{\dagger}(t)[W]\;V\,\Vf(t)[W\rho_0]\,V^{\dagger}\right\},
  \label{eq:ZHC}
\end{equation}
where $\Vf$ is the Lindblad propagator generated by $+H_{\sys}$ together
with the dissipator, $\Vb$ the one generated by $-H_{\sys}$ with the
\emph{same} dissipator, and $\Vb^\dagger$ its Heisenberg-picture adjoint.
The question
addressed in this paper is what Eq.~\eqref{eq:ZHC} sees of
\emph{structured, nonreciprocal} dissipation.

Operationally, Eq.~\eqref{eq:ZHC} is measured by the circuit of
Fig.~\ref{fig:circuit}: a single control qubit
is put in $|+\rangle$ by a Hadamard; a controlled-$W$ applies the probe in
the $|1\rangle$ branch; the system, coupled all along to its reservoirs,
undergoes the forward evolution, the butterfly $V$, and the backward
evolution with $H_{\sys}\to-H_{\sys}$; a controlled-$W^\dagger$ applies the
probe again in the same branch, while the $|0\rangle$ branch carries no
probe at all; and a final Hadamard maps the control's $\langle X\rangle$
($\langle Y\rangle$) to $\Re F$ ($\Im F$). Propagating the coherence block
of the control through the circuit and tracing out the reservoirs gives
exactly $\langle X\rangle_c=\Re\Tr\{W^\dagger\Vb[V\,\Vf[W\rho_0]\,V^\dagger]\}=\Re F$,
with $\Vf$ and $\Vb$ the reduced forward and backward channels. The
control/anti-control version, with $W$ before the evolution in one branch
and $W$ after it in the other, is equivalent~\cite{Zhang19a}. The
circuit makes explicit why the OTOC is dissipative: the backward block
inverts $H_{\sys}$ but not the coupling to the reservoirs.

A few remarks on Eq.~\eqref{eq:ZHC} and Fig.~\ref{fig:circuit} are in
order. First, in the closed-system limit Eq.~\eqref{eq:ZHC} reduces to
$\langle V^\dagger(t)W^\dagger V(t)W\rangle$; for $l_V=l_W$ it further
reduces to a simple expectation value and is analytically tractable for
free fermions~\cite{Lin18a}. Second, the \emph{leak factor}
$F(t;\one,W)$---the same quantity with the butterfly removed---measures
how much of the echo signal survives dissipation at all. It is a function
of the probe alone,
\begin{equation}
  \begin{split}
  F_I(t,l_W) &:= F(t;\one,W_{l_W})\\
  &=\Tr\left\{\Vb^\dagger(t)[W_{l_W}]\,\Vf(t)[W_{l_W}\rho_0]\right\},
  \end{split}
  \label{eq:leak}
\end{equation}
and it decays not because of any postselection but because
$\Vb^\dagger[W]$ shrinks under dissipative Heisenberg evolution
(dissipative channels are contractive on the operator norm); it usually
decays far more slowly than the no-jump probability
$e^{-\gamma(L-1)t}$.\footnote{At $L=21$, $\gamma=0.05$, for example, it
falls from $0.92$ at $t=0.4$ to $0.45$ at $t=4$.} Third, the inner $\Vf$
acts on the non-Hermitian operator $W\rho_0$, not on $W\rho_0W^\dagger$;
this asymmetry is essential. Finally, the two-argument function
$F(l_V,l_W):=F(t;V_{l_V},W_{l_W})$ is not symmetric under
$l_V\leftrightarrow l_W$, even at infinite temperature, so the presentation
is a genuine choice: unless stated otherwise, the maps below fix the
butterfly at the chain center and scan the probe, displaying where the
perturbation has arrived by time $t$.

\subsection{Observables}
\label{sec:Observables}

The OTOC $F(t;V,W)$ and the leak factor $F_I$ are the direct outputs
of the circuit of Fig.~\ref{fig:circuit}. The quantities we actually
analyze are three combinations of them, which isolate, respectively, the
effect of the butterfly, the reflection asymmetry about the butterfly,
and the dependence on the sign of the nonreciprocity. We define them
here and collect their general properties; their exact symmetries are
the subject of Sec.~\ref{sec:Symmetries}.

\paragraph*{Butterfly-induced deviation.} Since the leakage damps the
signal irrespective of the butterfly, the quantity of interest is
\begin{equation}
  \dF(t,l_W) :=  F(t;V,W)-F(t;\one,W).
  \label{eq:dF}
\end{equation}
The deviation has a useful operator form. Let
$G_{l_W}(t) := \Vb^\dagger(t)[W]$ denote the Heisenberg-evolved probe and
$\rho_\fwd(t) := \Vf(t)[\rho_0]$ the forward-evolved reference state.
For $V^2=\one$ one has the identity $V\rho V-\rho=-\tfrac12[V,[V,\rho]]$.
Consider further a reference state on which the probe acts as a sign,
$W\rho_0=s_{l_W}\rho_0$ with $s_{l_W}=\pm1$; this is the case for any $\rho_0$
diagonal in the occupation basis, in particular for a product state in
the occupation basis, which we call a \emph{canonical reference state}.
Then Eq.~\eqref{eq:dF} becomes
\begin{equation}
  \begin{split}
  \dF(t,l_W)
  &=-\frac12 s_{l_W}\Tr\left\{[V,[V,G_{l_W}(t)]]\,\rho_\fwd(t)\right\}\\
  &=-2s_{l_W}\Tr\left\{G_{l_W}(t)\,\rho_\fwd^{\rm off}(t)\right\},
  \end{split}
  \label{eq:dFcomm}
\end{equation}
where $\rho_\fwd^{\rm off}$ is the part of $\rho_\fwd$ off-diagonal
in the occupation of site $l_V$.

Two consequences follow. (i) $\dF$ is a
double-commutator correlator of the Heisenberg-evolved probe with the
butterfly: it vanishes identically as long as the operator support of
$G_{l_W}(t)$ has not reached $l_V$, so that, by the Lieb--Robinson
bound, it is confined to a cone with its apex at the butterfly, with the
leak \emph{subtracted} rather than divided out [$\dF(0)=0$ for every state]. (ii) When $G_{l_W}$ is
Hermitian and $\rho_0\propto\one$ (infinite temperature, equal rates),
\begin{equation}
  \dF(t,l_W)=-\frac{\|[V,G_{l_W}(t)]\|_F^2}{2^{L+1}}\le0,
  \label{eq:dFinf}
\end{equation}
where $\|\cdots\|_F$ denotes the Frobenius norm of an operator. That is,
$\dF$ is minus the squared commutator norm familiar from Lieb--Robinson
bounds in open systems~\cite{Poulin10a,Barthel12a}; ZHC's relation
$2(1-F/F_I)=\|[G,V]\|_F^2/\|G\|_F^2$ is the same statement normalized by
the leak. For a pure state $\dF$ is sign-indefinite.

ZHC~\cite{Zhang19a} mainly examined the normalized OTOC
$F(t;V,W)/F(t;\one,W)$. In this work we focus on the deviation
\eqref{eq:dF} instead, because it exposes the symmetry properties of the
dissipative OTOC (Sec.~\ref{sec:Symmetries}) more directly. It has two
further technical advantages. First, nothing is divided, so $\dF$ stays
meaningful where $F_I$ passes through zero, as happens, for example, for
pure reference states (Sec.~\ref{sec:spin:LightCones}). Second, being linear
in $F$ and $F_I$, it inherits every exact symmetry of the pair term by
term and splits every asymmetry of $F$ exactly into a leak part and a
commutator part,
$\mathrm{asym}_d[F]=\mathrm{asym}_d[F_I]+\mathrm{asym}_d[\dF]$ with
$\mathrm{asym}_d[X] := X(t,l_V{+}d)-X(t,l_V{-}d)$. The ratio conveys the
same information as long as $F_I$ is well behaved, but it amplifies the
late-time residue to $O(1)$, whereas $\dF$ simply lets it decay.

\paragraph*{Symmetric and antisymmetric parts.} For a butterfly at the
center, $l_V=(L+1)/2$, and writing $x=l_W-l_V$, we define the
reflection-even (symmetric) and reflection-odd (antisymmetric) parts of
$\dF$,
\begin{align}
  S(t,x)&=\frac12\big[\dF(t,x)+\dF(t,-x)\big],
  \label{eq:Sdef}\\
  A(t,x)&=\frac12\big[\dF(t,x)-\dF(t,-x)\big].
  \label{eq:Adef}
\end{align}
At $\gamma=0$ both models are reflection symmetric and $A$ vanishes
identically for a reflection-symmetric reference state; at $\gamma>0$ reflection reverses
$\eta$, so $A$ is a pure chirality signal.

\paragraph*{$\eta$-odd part.} The direct comparison of the two chiralities,
\begin{equation}
  \dEF(t,l_W) :=  F_{+\eta}(t; V, W_{l_W})-F_{-\eta}(t; V, W_{l_W}),
  \label{eq:dEF}
\end{equation}
vanishes in the closed limit and in the absence of nonreciprocal jump
operators and is a clean diagnostic of the nonreciprocal dissipation; it
requires two runs where $A$ requires one.

\subsection{Symmetries}
\label{sec:Symmetries}

The dissipative OTOC inherits exact symmetries from the Hamiltonian, the
jump set and the reference state, and these decide what of the chirality
of the dissipation is visible in it, and where. Three are used
throughout: a mirror covariance that relates the two chiralities to
spatial reflection, anti-chiral symmetries that can hide the chirality
altogether, and, for the fermionic model with occupation probes, an
eigenstate identity that reduces the echo to a one-point measurement.

\paragraph*{Mirror (parity) covariance.} Spatial reflection $P:l\to L+1-l$
preserves $H_{\sys}$ and maps the jump set to itself with $\eta\to-\eta$.
Hence for a reflection-symmetric reference state and a central butterfly,
\begin{equation}
  F_{+\eta}(t,l_W)=F_{-\eta}(t,L{+}1{-}l_W),
  \label{eq:mirror}
\end{equation}
and therefore
\begin{equation}
  A_{+\eta}(t,x)=\frac12\big[\dF_{+\eta}-\dF_{-\eta}\big](t,x),
  \label{eq:mirrorA}
\end{equation}
the same holding for $F_I$ per probe: the reflection-odd part of a single
scan \emph{is} the $\eta$-odd part at the same probe. This is verified to
$10^{-14}$ in all symmetric cases below, and it is the operational reason
why $A$ is a chirality observable needing neither an $\eta=0$ control nor
a $1/F_I$ correction. For a parity-breaking reference state (e.g., a state with a single
domain wall) $A$ is \emph{not} a chirality observable, and chirality must
be read from $\dEF$ or from scalars (Sec.~\ref{sec:spin:LightCones}).

\paragraph*{Anti-chiral symmetries.}
We call \emph{anti-chiral} any operation $\mathcal S$ that preserves the
Hamiltonian, maps the probe and butterfly to $\pm$ themselves (the signs
cancel in $F$) and maps the jump set onto itself with $\eta\to-\eta$. If
$\rho_0$ is invariant under an anti-chiral symmetry, $F_{\eta}=F_{-\eta}$
identically and the chirality is invisible in the OTOC. Whether such
symmetries exist depends on the jump algebra. If the jump set is closed
under Hermitian conjugation with equal rates, then
$\sum_\mu L_\mu^\dagger XL_\mu=\sum_\mu L_\mu XL_\mu^\dagger$ and
$\sum_\mu L^\dagger_\mu L_\mu=\sum_\mu L_\mu L_\mu^\dagger$, so
\begin{equation}
  \Vb^\dagger(t)=\Vf(t)
  \label{eq:identity}
\end{equation}
exactly, as superoperators, for any $\eta$---the analogue of the property
ZHC noted for phase-damping and depolarizing channels~\cite{Zhang19a}.
With Eq.~\eqref{eq:identity} the probe is propagated by the forward
Lindbladian and the dissipative OTOC needs only forward evolutions; at infinite
temperature it takes the two-$G$ form
$F=2^{-L}\Tr[G_{l_W}(t)VG_{l_W}(t)V]$ with $G_{l_W}(t)=\Vf(t)[W]$, for which
transposition is anti-chiral. This is the situation of the local spin
model at equal rates (Sec.~\ref{sec:spin:LocalSpin}), where the global spin
flip provides a second anti-chiral symmetry. For the fermionic model the
jump set $\{A_l,B_l\}$ is \emph{not} self-adjoint ($B_l\neq A_l^\dagger$),
Eq.~\eqref{eq:identity} fails, and no anti-chiral symmetry exists even at
infinite temperature: the chirality is always visible in the fermionic
OTOC, and its chiral component turns out to be universal
(Sec.~\ref{sec:Anatomy}).

\paragraph*{Fermionic occupation factorization.} For the fermionic
model with a canonical reference state, $W\rho_0=s_{l_W}\rho_0$ with the
sign factor $s_l:=1-2n_l^0$, where $n_l^0\in\{0,1\}$ is the occupation
of site $l$ in the reference state. The OTOC then reads
\begin{equation}
  F(t,l_W)=s_{l_W}\left(1-2\langle n_{l_W}\rangle_t\right),
  \label{eq:eigenstate_identity}
\end{equation}
where $\langle n_l\rangle_t$ is the occupation of the echo-evolved state
$\Vb[V\,\Vf[\rho_0]\,V]$. We refer to Eq.~\eqref{eq:eigenstate_identity}
as the eigenstate identity. It holds at \emph{every} $t$ (verified against the
exact dilation deep in the transient to $10^{-15}$): it merely collapses
the front probe to its eigenvalue, leaving all OTOC content in the
occupation of the echo-evolved state. It renders this class of dissipative OTOCs
\emph{interferometer-free}: prepare the product reference, run the echo
with the butterfly applied between the forward and backward blocks,
measure $n_{l_W}$, multiply by the
recorded $s_{l_W}$. The control qubit of Fig.~\ref{fig:circuit} is
required only when the front probe fails to commute with the
reference---superposition references, or probes not diagonal in the
measured basis.

\subsection{Bath Reset Strategies}
\label{sec:BathReset}

In this work we are interested in realizations in which the reservoirs
are controlled through macroscopic parameters, such as the occupations of
the bath modes, and in which these parameters can be set independently
for the forward and backward blocks $U_\fwd$ and $U_\bwd$ of
Fig.~\ref{fig:circuit}; in the measure-and-reset circuit of
Sec.~\ref{sec:QuantumTrajectoryMethod} they are simply the states to which
the bath registers are reset. In the context of the Liouvillian skin
effect two strategies are of particular interest: the \emph{same} reset,
in which the backward reservoirs are prepared exactly as the forward ones
(loss reservoirs empty, gain reservoirs filled), and the \emph{flipped}
reset, in which the occupations of the backward loss and gain reservoirs
are interchanged (loss filled, gain empty). The flipped reset is a
particle--hole transformation of the backward reservoirs, and at the level
of superoperators, to machine precision and for all $\gamma$ and
$\Delta t$, $\Vb^{\rm flip}(+\eta)=\Vb^{\rm same}(-\eta)$: the flip
reverses the effective chirality of the dissipation in the backward block
only. It is \emph{not} a global $\eta\to-\eta$:
$F^{\rm flip}(+\eta)\neq F^{\rm same}(-\eta)$ in general. Because the
dissipative nonreciprocity reverses, the flipped backward map is a
strictly better partial time reversal of the forward map,
$\|\Vb^{\rm flip}\Vf-\one\|<\|\Vb^{\rm same}\Vf-\one\|$ for all $\gamma>0$,
both differences vanishing at $\gamma=0$. Two facts, established in
Secs.~\ref{sec:LightCones} and~\ref{sec:NESS}, determine what the choice
affects. First, the butterfly-induced deviation $\dF$ is
\emph{independent} of the reset strategy, so the difference between the
two strategies resides entirely in the leak factor; it vanishes for a
probe at the center of the chain and is exponentially localized at the
boundaries. Second, in the steady state this difference measures the
Liouvillian skin profile. For a particle--hole-symmetric
reference (half-filled Fermi sea) the two strategies coincide exactly.
For the spin model with a single thermal bath there is no such choice;
the thermal reset is the only consistent one
(Sec.~\ref{sec:spin:Methods}).

\section{Methods}
\label{sec:Methods}

The Hatano--Nelson Lindbladian is quadratic and number conserving, so its
dynamics stays within the fermionic Gaussian (Wick)
family~\cite{Bravyi05a,Prosen08a,Cao19a,Feng24a,ChoiMS26a} and the dissipative OTOC
can be evaluated \emph{exactly}, polynomially in $L$. We use three
complementary routes: a channel-level propagation of the single-particle
correlation matrix, which is the primary tool for the light cones and the
steady state; an explicit-bath Stinespring dilation, which represents the
whole echo as a single Slater-determinant overlap, serves as the
benchmark, and motivates the third route; and a quantum-trajectory Monte
Carlo of the measure-and-reset circuit, which is the hardware-efficient
protocol, directly simulable on quantum computers. The brute-force
state-vector evaluation requires the full $2^L\times2^{2(L-1)}$ space and
becomes infeasible beyond $L\approx4$; ``exact'' below means the Gaussian
evaluation, which
agrees with brute force to $\sim10^{-15}$ wherever the latter is
feasible. Throughout, $N_t$ denotes the number of Trotter steps of a
discretized evolution to time $t$ and $\Delta t:=t/N_t$ the step.

Most of the numerical results below were obtained with a Wolfram Language
package built on the Q3 framework~\cite{Q3,Choi22a}; in particular,
its \texttt{Wick} module provides the free-fermion Gaussian
simulation~\cite{ChoiMS26a}.

\subsection{Exact Solution to the Lindblad Equation}
\label{sec:GaussianSolution}

Because the Lindbladian is quadratic and number conserving, the
correlation matrix 
\begin{math}
  C_{ij}[\rho] := \Tr[\rho\,\cdag_ic_j]
\end{math} 
obeys the closed affine
equation
\begin{equation}
  \dot C=DC+CD^\dagger+\gamma M,
  \label{eq:affineC}
\end{equation}
with the drift and source matrices
\begin{align}
  D&=ih^{T}-\gamma\Re M,
  \label{eq:Ddef}\\
  M&=\alpha^\dagger\alpha,
  \label{eq:Mdef}
\end{align}
where $h$ is the single-particle hopping matrix of Eq.~\eqref{eq:Hsys}
and the rows of $\alpha$ are the bond amplitudes of the jump operators
\eqref{eq:jumpA}. The form of the drift matrix $D$ deserves a comment.
The loss and gain channels contribute the dampings
$-\tfrac\gamma2(M^TC+CM^T)$ and $-\tfrac\gamma2(MC+CM)$, respectively.
In their sum the chiral imaginary part of $M$ cancels, which is why only
$\Re M$ appears in Eq.~\eqref{eq:Ddef}: the damping matrix is the
diagonal matrix
$\gamma\Re M=\gamma\,\mathrm{diag}\big(\tfrac{1}{1+\eta^2},1,\dots,1,
\tfrac{\eta^2}{1+\eta^2}\big)$, $\eta$-independent except on the two
terminated edge sites, and the chirality survives only in the source
term $\gamma M$. This is the
algebraic complement of Eq.~\eqref{eq:AA_BB}, where the anti-Hermitian
part of $\Heff$ is built from the \emph{difference} of the same dyads and
the chirality survives there instead. This single fact underlies the
exact anatomy of Sec.~\ref{sec:Anatomy} and the $\eta$-independent skin
depth of Sec.~\ref{sec:NESS}. The correlation
matrix of the nonequilibrium steady state (NESS) is the solution of the
Lyapunov equation 
\begin{equation}
  \label{eq:Lyapunov}
  DC+CD^\dagger+\gamma M=0.
\end{equation}

Equation~\eqref{eq:affineC} is linear in $C$ and is solved in closed form.
Writing 
\begin{equation}
  X_{\fwd,\bwd}(t)=e^{tD_{\fwd,\bwd}}
\end{equation}
for the propagators of the
forward and backward blocks ($D_\bwd$ has $h\to-h$) and
\begin{equation}
  Y_{\fwd,\bwd}(t)=\int_0^t{ds}\, 
  e^{sD_{\fwd,\bwd}}\,\gamma M\,e^{sD_{\fwd,\bwd}^\dagger}
  \label{eq:Yf}
\end{equation}
for the accumulated source, the correlation matrix after a forward
evolution of duration $t$ is
\begin{equation}
  C\big[\Vf(t)[\rho]\big]=X_\fwd(t)\,C[\rho]\,X_\fwd^\dagger(t)+Y_\fwd(t),
  \label{eq:Csol}
\end{equation}
and likewise for the backward block; both $X$ and $Y$ are obtained at
once from a single $2L\times2L$ matrix exponential of
$\big(\begin{smallmatrix}D&\gamma M\\0&-D^\dagger\end{smallmatrix}\big)t$,
whose upper blocks are $X$ and $Y(X^{\dagger})^{-1}$. The butterfly acts on
$C$ as the sign flip $S_V=\mathrm{diag}(1,\dots,-1,\dots,1)$ of row
and column $l_V$, so the echo-evolved reference state
$\rho_\mathrm{echo} := \Vb(t)\big[V\,\Vf(t)[\rho_0]\,V^\dagger\big]$ (forward
evolution, butterfly, backward evolution) has the correlation matrix
\begin{equation}
  C[\rho_\mathrm{echo}]=X_\bwd\big(S_V\,C\big[\Vf[\rho_0]\big]\,S_V\big)X_\bwd^\dagger+Y_\bwd.
  \label{eq:Cecho}
\end{equation}
For a canonical reference state the eigenstate identity
\eqref{eq:eigenstate_identity} then gives the whole probe profile from a
single evaluation, $F(l)=(1-2C_{ll}[\rho_0])(1-2C_{ll}[\rho_\mathrm{echo}])$
for the probe at site $l$.

Two remarks extend the scope of this solution. First,
Eq.~\eqref{eq:affineC} follows from the Lindbladian by linear identities
alone---the equations of motion of the bilinears $\cdag_ic_j$ and the
cyclicity of the trace---and never invokes Wick factorization of the
evolving object. It therefore propagates the two-point correlation data
$\Tr[\mathcal O\,\cdag_ic_j]$ of \emph{any even operator} $\mathcal O$
exactly, not only of states. This is precisely what Eq.~\eqref{eq:ZHC}
requires: its forward block carries the non-Hermitian operator $W\rho_0$.
Second, for a
general Gaussian reference---a Fermi sea, an $H_{\sys}$ ground state, or a
\emph{mixed} thermal state or steady state, none of which is an
eigenstate of the probe---linearity of $W_{l_W}\rho_0=\rho_0-2n_{l_W}\rho_0$
still applies: the correlation data of the operator $n_{l_W}\rho_0$ differ
from $C[\rho_0]$ by a rank-one Wick contraction, and because the solution
\eqref{eq:Csol} is affine that rank-one piece evolves through the
homogeneous part alone, carried by the single-particle echo propagator
$X_\mathrm{echo}=X_\bwd S_VX_\fwd$. The entire probe profile then follows
from a single evaluation,
\begin{multline}
  F(l)=\big(1-2C_{ll}[\rho_0]\big)\big(1-2C_{ll}[\rho_\mathrm{echo}]\big)\\{}
  +4\big(X_\mathrm{echo}C[\rho_0]\big)_{ll}\big((\one-C[\rho_0])X_\mathrm{echo}^{\dagger}\big)_{ll},
  \label{eq:collective}
\end{multline}
which involves no divisions (it is regular at $\langle n_l\rangle_0=0,1$);
for a canonical reference state the second term vanishes identically. The cost is
$O(L^3)$ per time point, independent of the chain's $2^L$-dimensional
Hilbert space. This direct solution of Eq.~\eqref{eq:affineC} agrees
with the exact $2^L$-dimensional Lindblad dynamics to $10^{-15}$ wherever
the latter is feasible, and it is the method behind the light cones of
Sec.~\ref{sec:LightCones} and the anatomy of Sec.~\ref{sec:Anatomy}.

\subsection{Stinespring Dilation Method}
\label{sec:DilationMethod}

The dilation is not a practical hardware protocol---its number of bath
modes grows with the circuit depth, i.e., with the number of Trotter
steps $N_t$---but it is an efficient classical benchmark and provides the insight on which the quantum-trajectory circuits of
Sec.~\ref{sec:QuantumTrajectoryMethod} are built.

\begin{figure*}
\centering
\includegraphics[width=\textwidth]{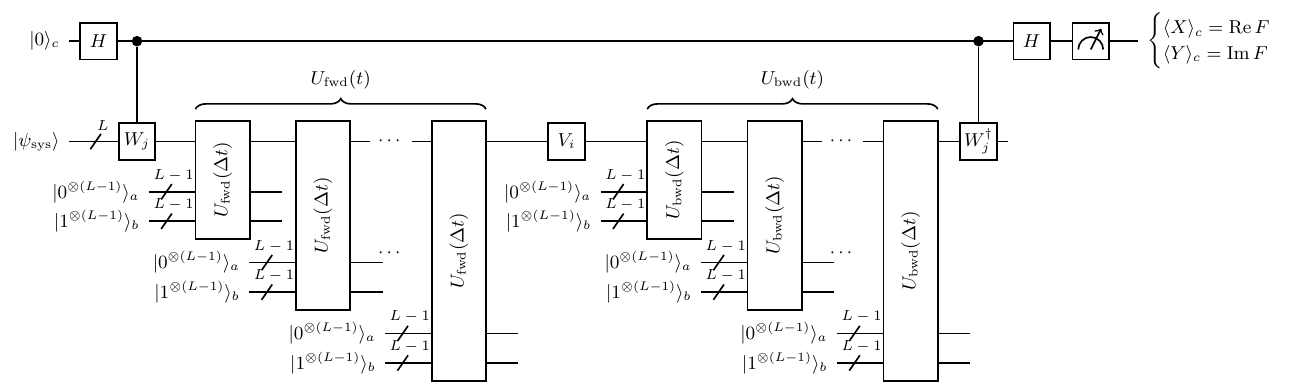}
\caption{The Stinespring dilation as a quantum
circuit, for the butterfly $V_i$ at site $i=l_V$ and the probe $W_j$ at
site $j=l_W$; the control qubit $c$ and the controlled $W_j$, $W_j^\dagger$
are those of Fig.~\ref{fig:circuit}. The system register of $L$ modes is
prepared in the reference state $|\psi_\sys\rangle$. Each of the $N_t$
forward Trotter steps
is the same unitary $U_\fwd(\Delta t)=\exp[-i\Delta t H^{(+)}_{\sys\bath}]$,
acting on the system and on a \emph{fresh} bath register of $2(L-1)$
modes---the $a$-modes in $|0^{\otimes(L-1)}\rangle$, the $b$-modes in
$|1^{\otimes(L-1)}\rangle$---so that the $N_t$ steps compose to
$U_\fwd(t)$ with $t=N_t\Delta t$; likewise each backward step
$U_\bwd(\Delta t)=\exp[-i\Delta t H^{(-)}_{\sys\bath}]$ acts on another
fresh register (prepared in the same or the flipped references of
Sec.~\ref{sec:BathReset}). The registers are never measured or reset and
stay coherent (their wires are truncated in the drawing once they are
idle), so the number of bath modes grows as $4(L-1)N_t$.}
\label{fig:dilation}
\end{figure*}

We begin with the dilation of a single Trotter step. The master equation
\eqref{eq:LindbladHN} admits a Stinespring dilation~\cite{Stinespring55a}
as a unitary on system$+$bath traced over the bath. For each bond and
each Trotter step we introduce two fresh bath modes: an $a$-mode (loss channel)
initialized in $|0\rangle$ and a $b$-mode (gain channel) initialized in
$|1\rangle$. The joint Hamiltonian for one step is
\begin{equation}
  H_{\sys\bath}^{(\pm)}=\pm H_{\sys}
  +\sqrt{\gamma/\Delta t}\sum_{l=1}^{L-1}\Big[A_la_l^\dagger+B_lb_l+\mathrm{h.c.}\Big],
  \label{eq:Hjoint}
\end{equation}
with the sign selecting forward or backward evolution. Both couplings are
hopping terms between system and bath, so the trajectories remain
representable in the number-conserving Wick framework; note that the gain
coupling is $B_lb_l+\mathrm{h.c.}$, not $B_lb_l^\dagger+\mathrm{h.c.}$ (which
would be anomalous): with the $b$-mode initialized in $|1\rangle$, the term
$(\cdag_l+i\eta\cdag_{l+1})b_l$ annihilates the bath fermion while creating
a system particle in the $B_l$ superposition. With
$U^{(\pm)}=\exp(-i\Delta t\,H^{(\pm)}_{\sys\bath})$ and the canonical bath
state $|0\rangle_{\bath}=|0\cdots0\rangle_a\otimes|1\cdots1\rangle_b$,
\begin{equation}
  \Vf^{\Delta t}[\rho]=\Tr_{\bath}\big[U^{(+)}(\rho\otimes|0\rangle\langle0|_{\bath})U^{(+)\dagger}\big],
  \label{eq:onestep}
\end{equation}
and the evolution to time $t$ is the composition of $N_t=t/\Delta t$
steps, each with its own fresh bath modes. The collision scaling
$\sqrt{\gamma/\Delta t}$ of the coupling is what fixes the dissipation
rate at $\gamma$: it is the discrete counterpart of the
$\delta$-correlated bath of Sec.~\ref{sec:SystemBath}.

Figure~\ref{fig:dilation} shows the dilation of the full echo as a
quantum circuit, in direct correspondence with the circuit of
Fig.~\ref{fig:circuit}: the two reservoirs of Fig.~\ref{fig:circuit}
have become the registers of fresh bath modes attached to each Trotter
step. In this form the circuit is not directly simulable by the
fermionic Gaussian method, because the controlled-$W$ and
controlled-$W^\dagger$ gates are not Gaussian, although every other
element is. This is overcome by evaluating the two branches of the control qubit
separately,
\begin{equation}
  |\Psi_{\rm ket}\rangle=W^\dagger U_\bwd\,V\,U_\fwd\,W\,|\Psi_0\rangle
  \label{eq:ket}
\end{equation}
for $\ket{1}_c$
and
\begin{equation}
  |\Psi_{\rm bra}\rangle=U_\bwd\,V\,U_\fwd\,|\Psi_0\rangle
  \label{eq:bra}
\end{equation}
for $\ket{0}_c$,
and taking their inner product, $F=\langle\Psi_{\rm bra}|\Psi_{\rm ket}\rangle$,
which is what the control qubit measures. Here
$\ket{\Psi_0}=\ket{\psi_\sys}\otimes\ket{0}_\mathrm{loss}\otimes\ket{1}_\mathrm{gain}$,
where $\ket{\psi_\sys}$ is the reference state of the system,
$\rho_0=\ket{\psi_\sys}\!\bra{\psi_\sys}$, and $\ket{0}_\mathrm{loss}$
and $\ket{1}_\mathrm{gain}$ are the canonical states of the bath
registers of every step, with all $a$-modes empty and all $b$-modes
filled, respectively. Written out,
\begin{equation}
  F(t;V,W)=\big\langle\Psi_0\big|\,
  U_\fwd^\dagger\,V\,U_\bwd^\dagger\,W\,U_\bwd\,V\,U_\fwd\,W\,
  \big|\Psi_0\big\rangle,
  \label{eq:2bath}
\end{equation}
which reduces to Eq.~\eqref{eq:ZHC} by the cyclicity of the trace
together with the identities
\begin{align}
  \Vf[\rho]&=\Tr_{\bath}\big[U_\fwd(\rho\otimes\pi_\bath)U_\fwd^\dagger\big],
  \label{eq:Vfwd_dilated}\\
  \Vb^\dagger[O]&=\Tr_{\bath}\big[U_\bwd^\dagger(O\otimes\one_\bath)U_\bwd\,
  (\one_\sys\otimes\pi_\bath)\big],
  \label{eq:Vbwd_dilated}
\end{align}
where $\pi_\bath=|0\rangle\langle0|_\bath$ projects onto the canonical
state of the bath registers of the respective block.
One point deserves emphasis.
The two branches share their bath registers: each register appears in a
$U\cdots U^\dagger$ pair sandwiching system operators, the bath being
entangled by $U$ and disentangled by $U^\dagger$, and what survives the
projection of each register onto its initial state is exactly the
bath-averaged dissipative OTOC.
Had each branch been given registers of its own,
every register would instead have contributed the no-jump matrix
element $\langle0|U|0\rangle_{\bath}$, a non-Hermitian operator generating
the no-jump evolution under $\Heff$, and the result would have been the
OTOC of the \emph{postselected} dynamics.
The postselected-versus-averaged distinction is the same one
emphasized for the skin effect itself in Refs.~\cite{Haga21a,Yang22a},
where the full Liouvillian and the postselected $\Heff$ are shown to
differ in their topology and in the relaxation they govern; OTOC studies
in non-Hermitian Hamiltonians~\cite{Zhai20a,Zhao22a} compute the
postselected quantity.

The two branches are evaluated efficiently in the Gaussian
representation. The initial state is a Slater determinant, all step
unitaries are quadratic, and $V$, $W$ are single-mode Gaussian unitaries
(column sign flips), so each branch is represented throughout by an
$N_\mathrm{tot}\times P$ matrix $\mathcal M$ whose columns are the
occupied orbitals. A step unitary $U^{(\pm)}$ acts as
$\mathcal M\to u\,\mathcal M$, where
$u=e^{-i\Delta t\,h^{(\pm)}_{\sys\bath}}$ is the $(3L-2)$-mode
single-particle unitary matrix of the quadratic step Hamiltonian
\eqref{eq:Hjoint},
$H^{(\pm)}_{\sys\bath}=\sum_{mn}\big(h^{(\pm)}_{\sys\bath}\big)_{mn}
d_m^\dagger d_n$ with $d_n$ running over the $L$ system modes and the
$2(L-1)$ fresh bath modes of the step, so that
$U^{(\pm)}d_n^\dagger U^{(\pm)\dagger}=\sum_md_m^\dagger u_{mn}$ (on the
modes of earlier steps $u$ acts as the identity). The OTOC is the
determinant of the overlap of the two branches,
\begin{equation}
  F=\det\!\big(\mathcal M_{\rm bra}^\dagger\,\mathcal M_{\rm ket}\big).
  \label{eq:slater}
\end{equation}
Here $N_{\rm tot}=L+4(L-1)N_t$ and $P=L/2+2(L-1)N_t$ at half filling; the
per-step cost is $O(P(3L-2)^2)$ and the terminal determinant costs
$O((N_tL)^3)$, so that at $L=21$, $N_t=10$ a full sweep over $l_V$ and
both $\eta=\pm1$ runs in seconds.

The dilation also has a correlation-matrix form, which is the
Trotterized counterpart of Sec.~\ref{sec:GaussianSolution}.
Since each step unitary is quadratic, one step acts on the system correlation matrix
as
\begin{equation}
  C\longmapsto\big[u^{*}\,(C\oplus R)\,u^{T}\big]_{\sys}
  =T^{*} C T^{T}+G_R,
  \label{eq:affine}
\end{equation}
where $u$ is the single-particle unitary matrix of the step introduced
above ($u^{*}$ its complex conjugate, $u^{T}$ its transpose),
$T=u_{\sys,\sys}$ is its $L\times L$ system block, and $R$ is the
correlation matrix of the fresh bath modes (loss modes empty, gain modes
filled; interchanged in the backward block for the flipped strategy).
Composing $N_t$ steps reproduces the affine structure of
Eq.~\eqref{eq:Csol} with $X_{\fwd,\bwd}\to(T_{\fwd,\bwd}^{*})^{N_t}$, and
Eq.~\eqref{eq:collective} holds with the echo propagator
$X_\mathrm{echo}\to(T_\bwd^{*})^{N_t}S_V(T_\fwd^{*})^{N_t}$. The cost is
$O(N_tL^3)$ per evaluation, versus $O((N_tL)^3)$ for the determinant
\eqref{eq:slater}, so this form is the method of choice at long times and
is what generated the Trotterized steady-state profiles of
Sec.~\ref{sec:NESS}: vectorized, the one-step map is an $L^2\times L^2$
superoperator that factorizes as $T\otimes T^{*}$ plus an affine shift,
whose diagonalization yields evolution by matrix powers, the Trotterized
relaxation spectrum $\{\lambda_i\lambda_j^{*}\}$, and the Trotterized
steady state as the Stein fixed point $C=T^{*}CT^{T}+G_R$.

In either form, Eq.~\eqref{eq:slater} or Eq.~\eqref{eq:affine}, the
dilation converges to the direct integration of
Eq.~\eqref{eq:LindbladHN} or \eqref{eq:affineC} with an $O(\Delta t)$
residual; with the combined-exponential step it agrees with brute-force
Lindblad evolution to $\sim10^{-15}$ for $L=2,3,4$, and it respects the
mirror covariance \eqref{eq:mirror} to machine precision.

\subsection{Quantum Trajectory Method}
\label{sec:QuantumTrajectoryMethod}

\begin{figure*}
\centering
\includegraphics[width=\textwidth]{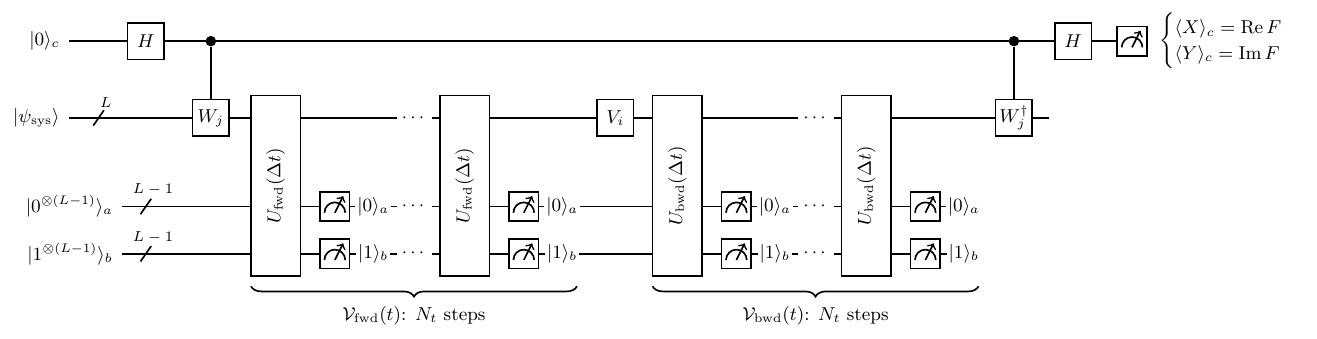}
\caption{Quantum-trajectory (measure-and-reset) realization of the
concept quantum circuit for the dissipative OTOC in Fig.~\ref{fig:circuit}. The two
reservoirs are replaced by two registers of $L-1$ bath modes, prepared
in $|0^{\otimes(L-1)}\rangle_a$ and $|1^{\otimes(L-1)}\rangle_b$, respectively; every
Trotter step applies the joint unitary $U_{\fwd,\bwd}(\Delta t)=
\exp[-i\Delta tH^{(\pm)}_{\sys\bath}]$ of Eq.~\eqref{eq:Hjoint}, with coupling
$\sqrt{\gamma/\Delta t}$, then measures both registers and resets them
to their reference states ($|0\rangle_a$, $|1\rangle_b$ on every mode; in
the backward block either the same or the flipped reference states,
Sec.~\ref{sec:BathReset}). The $N_t$ steps of a block compose, as
$\Delta t\to0$, to the channel $\mathcal V_\fwd(t)$ or
$\mathcal V_\bwd(t)$, and the measurement record of the registers
unravels each channel into quantum trajectories. The control qubit is
read out as in Fig.~\ref{fig:circuit}. Each register is reused in every
step, so the circuit width
stays $L+O(L)$ independently of $N_t$, in contrast to the dilation of
Fig.~\ref{fig:dilation}.}
\label{fig:trajectory_fermi}
\end{figure*}

Neither of the circuits in Figs.~\ref{fig:circuit} and \ref{fig:dilation}
is practical on a (fermionic) quantum computer: the reservoirs of
Fig.~\ref{fig:circuit} are infinite, and the number of bath modes in
Fig.~\ref{fig:dilation} grows with the number of Trotter steps. Guided by
the dilation circuit, we therefore propose the hardware-efficient circuit
of Fig.~\ref{fig:trajectory_fermi}, which runs on a quantum computer of
fixed size: the reservoirs of Fig.~\ref{fig:circuit} are replaced by
registers of bath modes, one per bond and channel, which are coupled
to the system for a time $\Delta t$ with coupling $\sqrt{\gamma/\Delta t}$
exactly as in Fig.~\ref{fig:dilation}, but are measured and reset to
their reference state before the next step, so that the same register
serves all $N_t$ steps of each block. The repeated
prepare--couple--measure sequence is a collision
model~\cite{Ciccarello22a}; measurement of the bath in any basis,
ensemble averaged, equals the partial trace, so the measurement record
simply unravels the channel into quantum
trajectories~\cite{Dalibard92a,Daley14a}. This is the hardware-efficient
route: the bath register ($O(L)$ modes) is \emph{reused} in every
step, so the circuit width stays fixed regardless of $N_t$, whereas the
deterministic dilation of Sec.~\ref{sec:DilationMethod} keeps every bath
of every step coherent and its number of bath modes grows as $O(N_tL)$.

The classical simulation of this circuit, which we use to validate the
protocol, meets an obstruction at the controlled gates, which are not
Gaussian. Immediately after the controlled-$W$ gate the joint state of
control and fermions is
\begin{equation}
  |\Phi\rangle=\frac{1}{\sqrt2}\big(|0\rangle_c|\psi_0\rangle+|1\rangle_c|\psi_1\rangle\big),
  \label{eq:branches}
\end{equation}
where $|\psi_{0,1}\rangle$ are the two branch states. Each branch is a
Slater determinant, but their control-entangled superposition is not: a
controlled-$W$ is not a quadratic operation on control$+$fermions. A naive
Gaussian simulator propagating one covariance matrix cannot carry the
circuit through the controlled gates---the concrete obstruction that
seems to make an exponentially costly brute-force simulation necessary.

The obstruction is removed by keeping the two branches as
\emph{separate} Slater determinants $\mathcal M_0$, $\mathcal M_1$ that
share one measurement record, never forming their superposition: the
control measurement only ever extracts the cross-overlap $\Re\langle\psi_0|\psi_1\rangle=\Re\det
(\mathcal M_0^\dagger \mathcal M_1)$. The one place the branches genuinely couple is the
mid-circuit bath measurement: there is a \emph{single} physical bath
register, measured once in the joint state \eqref{eq:branches}, so its
outcome $m$ is \emph{shared}---it projects both branches onto the same
$m$. This is not optional: if the branches were measured independently
with outcomes $m^{(0)}\neq m^{(1)}$, the bath factors would be orthogonal
and the cross-overlap---the entire OTOC---would collapse to zero. Because
the bath projector $\Pi_m$ annihilates the probe cross terms, the Born
probability of the shared outcome is the \emph{averaged} rule
\begin{equation}
  q(m)=\langle\Phi|\Pi_m|\Phi\rangle=\frac12\big[p_0(m)+p_1(m)\big],
  \label{eq:qm}
\end{equation}
with $p_s(m)=\|\Pi_m|\psi_s\rangle\|^2$, sampled without leaving the
Gaussian family: flip a fair coin to pick a branch $s$, draw the full
bath pattern $m$ from that branch's Green's function with the sequential
(Schur-complement) Wick sampler, and project \emph{both} branches onto
$m$; the coin is only a device for sampling the mixture, since summed
over its two outcomes the distribution of $m$ is exactly $q(m)$. A
subtlety that is easily overlooked is the reweighting. Each branch is held as a \emph{normalized} Slater
determinant, so projecting and reorthonormalizing renormalizes each
branch separately, discarding the relative amplitude the true joint state
carries. Restoring it as a per-step weight,
\begin{equation}
  w_k(m)
  := \frac{\sqrt{p_0(m)p_1(m)}}{q(m)}
  = \frac{2\sqrt{p_0(m)p_1(m)}}{p_0(m)+p_1(m)}\le1,
  \label{eq:weight}
\end{equation}
with $m$ the shared outcome of Trotter step $k$ (the branch probabilities $p_s$
depend on the step through the current branch states), makes the
average over $m\sim q$ telescope correctly, so that over the whole
circuit
\begin{equation}
  F=\Re\,\mathbb E_\mathrm{traj}
  \left[\left(\prod_{k=1}^{2N_t}w_k(m_k)\right)
  \det\left(\mathcal M_0^\dagger \mathcal M_1\right)\right]
  \label{eq:estimator}
\end{equation}
is unbiased, the index $k$ running over the $N_t$ steps of the forward
block and the $N_t$ steps of the backward block. Dropping the weights inserts a spurious factor
$(p_0+p_1)/2\sqrt{p_0p_1}\ge1$ once per step---a multiplicative bias that
does not vanish as $\Delta t\to0$. Per step the recipe is: (i) apply
$U_\fwd$ (or $U_\bwd$) identically to both branch orbital
matrices; (ii) coin-flip a branch and sample the shared bath pattern from
it; (iii) project both branches onto $m$ and reorthonormalize; (iv)
multiply the weight by \eqref{eq:weight}; (v) reset both baths to their
references.

In the classical simulation described above, the weight $w_k$ is a
bookkeeping artifact: it is \emph{computed}, not measured. On quantum
hardware there is one physical bath
register; measuring or resetting it once, in the joint state, forces a
single outcome on both branches with the true Born probability $q$, and
the post-measurement state retains the \emph{unnormalized} branch
amplitudes $\sqrt{p_0}|0\rangle_c|\psi_0'\rangle+\sqrt{p_1}|1\rangle_c|\psi_1'\rangle$,
because the device never renormalizes. The protocol on quantum hardware
therefore carries no weights at all: reset the bath registers at each step, run to the end,
measure the control in $X$ ($Y$), and average \emph{all} shots with equal
weight. The explicit $w_k$ is precisely the importance-sampling factor
that undoes the simulator's two classical choices---renormalizing each
branch, and drawing $m$ from one branch---both of which the Born rule
performs for free. Consequently the weight decay $\prod_kw_k\le1$
reappears on hardware as the shrinking control-qubit signal at long
times, that is, as a statistical cost paid in the number of shots. For unitary $V$, $W$ at infinite
temperature the weights are unity and the effective sample size is full.

We have verified the estimator \eqref{eq:estimator} against the exact
evaluation by the dilation method (itself checked against the
$2^L$-dimensional Lindblad dynamics) on short chains, for the N\'eel
reference state at $L=4$ and for the infinite-temperature reference at
$L=3$, the latter averaged over the $2^L$ basis states, with
$J=\gamma=1$, $\eta=1$, $\Delta t=0.1$ and $N_t=4$. The comparison
covers both reset strategies and both the OTOC and the leak factor. With
about $10^4$ trajectories the statistical errors are $0.004$--$0.009$,
and all estimates agree with the exact values within these errors, the
deviations ranging from $0.0\sigma$ to $1.3\sigma$. The weights did not
reduce the effective sample size in any of these cases.

\section{Results}
\label{sec:Results}

\subsection{Exact Anatomy of the Butterfly-Induced Deviation}
\label{sec:Anatomy}

With the propagators $X_{\fwd,\bwd}$ and accumulated sources $Y_{\fwd,\bwd}$
of Sec.~\ref{sec:GaussianSolution}, the forward-evolved correlation matrix
is $C_\fwd(t)=X_\fwd C_0X_\fwd^\dagger+Y_\fwd$ with $C_0=C[\rho_0]$,
and the butterfly acts as the sign flip $S_V$ of row and
column $l_V$. For a canonical reference state the eigenstate identity
\eqref{eq:eigenstate_identity} together with Eq.~\eqref{eq:Cecho} gives
\begin{equation}
  \dF(t,l)=-2s_l\Big[X_\bwd\big(S_V\,C_\fwd\,S_V-C_\fwd\big)X_\bwd^\dagger\Big]_{ll},
  \label{eq:dF_closed}
\end{equation}
the backward source $Y_\bwd$ having canceled in the difference. In
particular $\dF$ does not depend on how the reservoirs are reset in the
backward block---the structural fact announced in
Sec.~\ref{sec:BathReset}: the reset-strategy difference lives entirely in
the leak factor $F_I$. Splitting $C_\fwd$ into the \emph{memory term}
$X_\fwd C_0X_\fwd^\dagger$ and the \emph{source term} $Y_\fwd$ gives
$\dF=\dF_{\rm mem}+\dF_{\rm src}$ with the following properties.
\begin{enumerate}[label=(\roman*),leftmargin=*]
\item \emph{$\dF_{\rm mem}$ is reflection symmetric.} $D_{\fwd,\bwd}$
are achiral (both $h$ and $\Re M$ are inversion symmetric), and so are
$C_0$ (for a symmetric reference) and $S_V$.
\item \emph{$\dF_{\rm src}$ is reflection antisymmetric in the bulk.} On a
uniformly damped chain $e^{sD_\fwd}=e^{-\gamma s}e^{ish}$, so
$\Re Y_\fwd\propto\one$ and $S_VY_\fwd S_V-Y_\fwd$ involves
only $i\Im Y_\fwd$, the chiral, current-carrying coherences that build
up the nonequilibrium steady state [cf.\ the momentum-space occupation
\eqref{eq:nk_ss}].\footnote{The small symmetric residue ($10^{-9}$--$10^{-6}$) is
seeded by the damping deficit $\tfrac12$ at the two edges and arrives
from the boundaries.}
\item \emph{Decay rates.} $\dF_{\rm mem}$ carries four damped propagators
and decays as $e^{-4\gamma t}$; $\dF_{\rm src}$ carries two (the backward
block) acting on a source that saturates on the scale $1/2\gamma$, and
decays as $e^{-2\gamma t}$.
\item \emph{Universality.} $Y_\fwd$ does not involve $C_0$, so
$\dF_{\rm src}/s_l$ is the \emph{same function} of $t$ and $l$ for every
canonical reference state, and away from the boundaries it depends on the
butterfly position only through $x=l-l_V$; all state dependence resides
in $\dF_{\rm mem}$.
\item \emph{Sign structure.} The symmetric part of Eq.~\eqref{eq:Sdef}
satisfies $S(t,x)\le0$ (the memory term only reduces the OTOC, as
$\dF=F-1\le0$ at $\gamma=0$), while the antisymmetric part of
Eq.~\eqref{eq:Adef}, multiplied by the sign factor of the reference at the
probe site $l_W=l_V+x$, satisfies $s_{l_W}A(t,x)\le0$ for $x>0$ and
$\ge0$ for $x<0$ (for $\eta=+1$, until the front reaches the chain ends): any
checkerboard in the maps is entirely the sign factor $s_l$, and the
chirality itself is sign definite on each arm.
\end{enumerate}
Hence, for a reflection-symmetric reference state and a central butterfly,
$S=\dF_{\rm mem}$ and $A=\dF_{\rm src}$ in the bulk: \emph{the
antisymmetric part of the OTOC is literally the butterfly's imprint on the
chiral steady-state coherences}, which is why it vanishes identically at
$\gamma=0$, saturates on the scale $1/2\gamma$, decays at half the rate of
the symmetric part and dominates $\dF$ for $t\gtrsim1/2\gamma$, and is
independent of the reference state up to $s_l$. Two remarks are in
order. First, the
chirality appears as a \emph{sign reversal} between the two arms of the
cone, not as a rate asymmetry: both arms travel at the free velocity $J$
and decay at the same rate, because the damping matrix $\gamma\Re M$ is
achiral; a rate-level chiral cone (one-sided propagation) would require
moving the chiral part of $M$ into the damping, i.e., unequal loss and
gain rates, at the price of a trivial steady state---precisely what
happens in the spin model at finite temperature
(Sec.~\ref{sec:spin:LongTime}). Second, universality (iv) identifies a
reference-state-free observable: $A/s_l$, or directly $\dF_{\rm src}$,
measures the chiral steady-state coherences $i\Im Y_\fwd(t)$ alone, a
natural candidate for extracting $\teta$ and $\gamma$ from OTOC data
without knowledge of the reference state.

\subsection{Light Cones}
\label{sec:LightCones}

\begin{figure*}[tbp]
\centering
\includegraphics[width=0.7\textwidth]{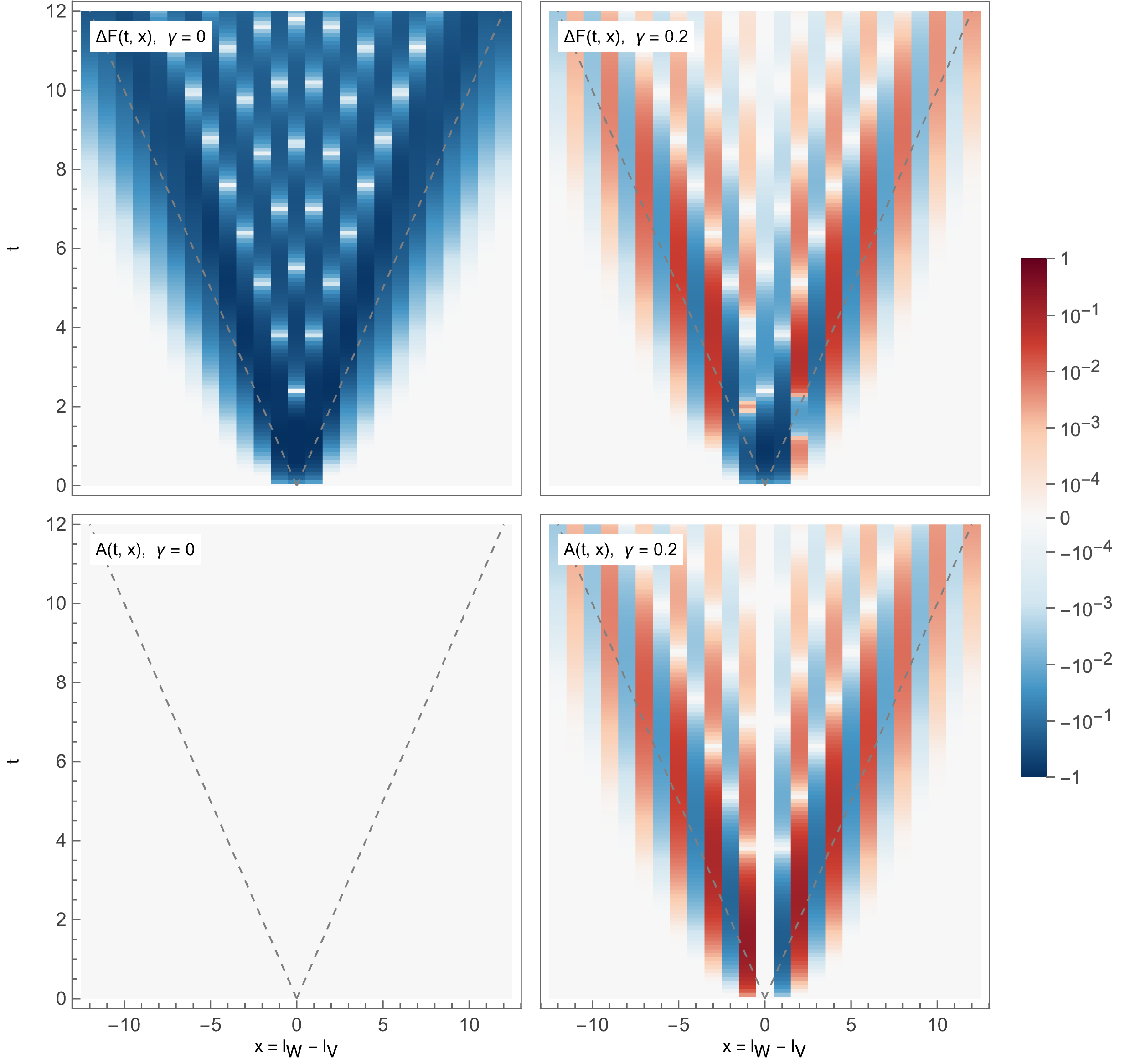}
\caption{Light cones of the Hatano--Nelson model for the N\'eel
reference state, with the butterfly at the center ($l_V=13$) of the
chain ($L=25$). Panels show the butterfly-induced deviation $\dF(t,x)$
(top row) and its antisymmetric part $A(t,x)$ (bottom row) for
$\gamma=0$ (left column) and $\gamma=0.2$ (right column), on a
signed-logarithmic color scale (linear below $10^{-4}$). Dashed lines
mark the free cone $|x|=Jt$.}
\label{fig:fermi_lc25}
\end{figure*}

\begin{figure*}[tbp]
\centering
\includegraphics[width=0.7\textwidth]{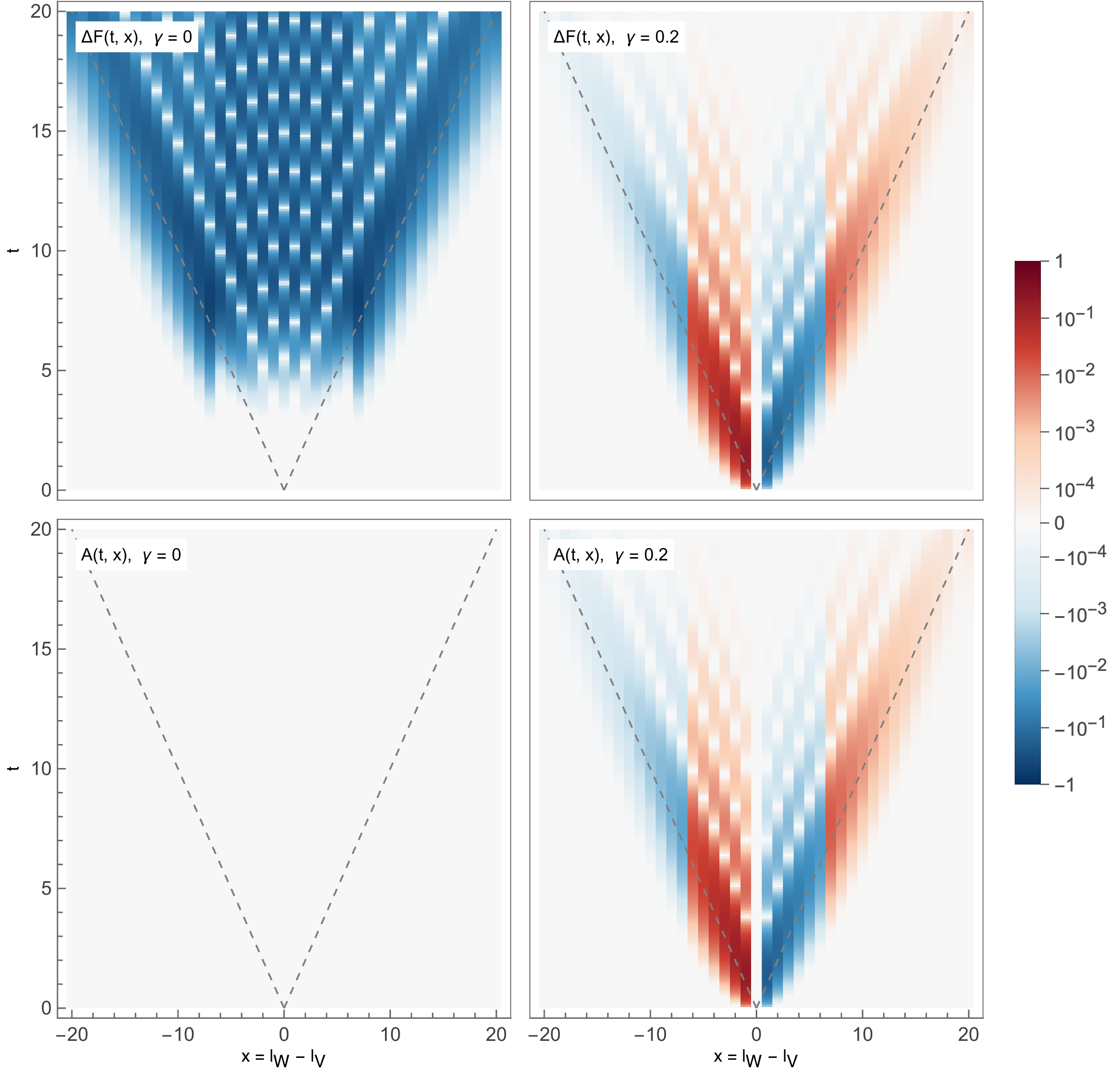}
\caption{Light cones of the Hatano--Nelson model for the two-wall
reference state $|1^{\otimes 14}0^{\otimes 13}1^{\otimes 14}\rangle$,
with the butterfly at the center ($l_V=21$) of the chain ($L=41$), in
the middle of the empty region. Rows and columns as in
Fig.~\ref{fig:fermi_lc25}.}
\label{fig:fermi_twowall}
\end{figure*}

\begin{figure*}[tbp]
\centering
\includegraphics[width=0.78\textwidth]{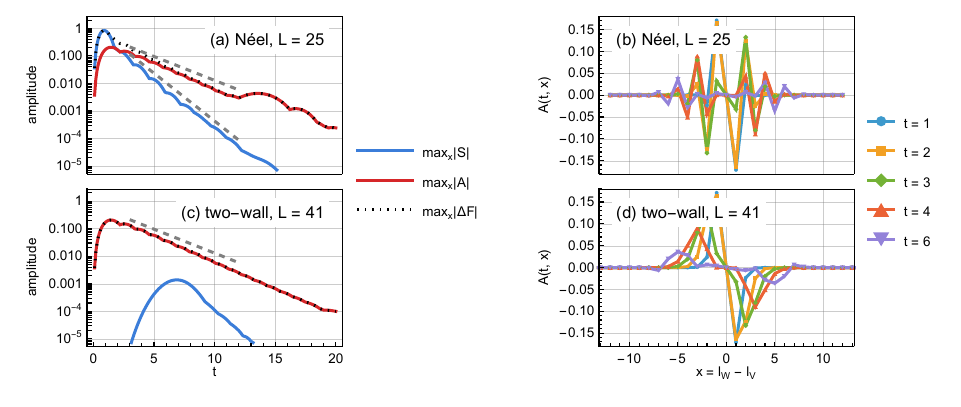}
\caption{Anatomy of $\dF$ at $\gamma=0.2$ for the N\'eel reference state of
Fig.~\ref{fig:fermi_lc25} ($L=25$, top row) and the two-wall reference state of
Fig.~\ref{fig:fermi_twowall} ($L=41$, bottom row). (a),~(c) $\max_x|S|$,
$\max_x|A|$ and $\max_x|\dF|$ versus time, with the $e^{-4\gamma t}$ and
$e^{-2\gamma t}$ guides of Sec.~\ref{sec:Anatomy}. 
(b),~(d) Cuts of $A(t,x)$ at $t=1,\dots,6$.}
\label{fig:fermi_anatomy}
\end{figure*}

\begin{figure*}[tbp]
\centering
\includegraphics[width=\textwidth]{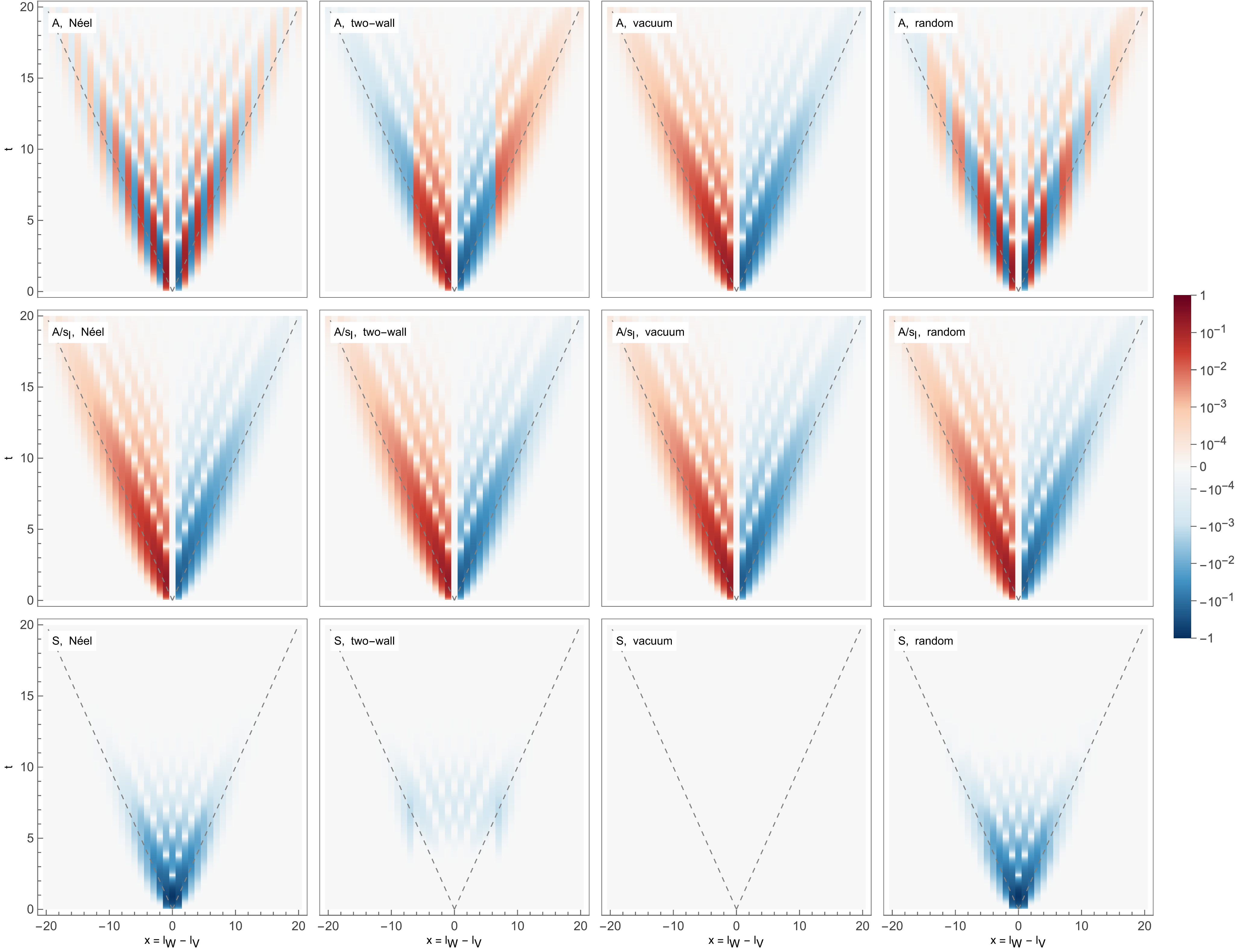}
\caption{Reference-state independence of the antisymmetric part ($L=41$,
$l_V=21$, $\gamma=0.2$) for four reflection-symmetric basis states: N\'eel,
two-wall, vacuum, and a random symmetric occupation pattern. Top:
antisymmetric part $A(t,x)$. Middle: $A/s_l$, the antisymmetric part
divided by the sign factor of the reference state at the probe site.
Bottom: symmetric part $S(t,x)$, with $\max|S|=0.838$, $0.001$, $0$ and
$0.838$ for the four states.}
\label{fig:fermi_universality}
\end{figure*}

Figures~\ref{fig:fermi_lc25} and~\ref{fig:fermi_twowall} show the
light cones obtained from the exact solution of the Gaussian Lindblad
equation~\eqref{eq:LindbladHN} described in
Sec.~\ref{sec:GaussianSolution}. The parameters are $J=1$ and $\eta=+1$,
with open boundaries; the butterfly sits at the center of the chain and
the probe site is scanned. The maps use a signed-logarithmic color scale
(linear below $10^{-4}$) that exposes the whole dynamic range, and the
dashed lines mark the free cone $|x|=Jt$. We have chosen a longer chain
for the two-wall reference state ($L=41$) than for the N\'eel reference
state ($L=25$) to exhibit more clearly the features discussed below.
For the N\'eel reference state (Fig.~\ref{fig:fermi_lc25}), the isolated chain
($\gamma=0$, left panels) shows a symmetric ballistic cone: the
excitations of the free-fermion band $\varepsilon(k)=-J\cos k$ of the
hopping Hamiltonian~\eqref{eq:Hsys} travel at most at the group velocity
$\max_k|\partial_k\varepsilon|=J$. Inside the cone $\dF\le0$ and, the chain
being integrable, oscillates rather than settling on a scrambled plateau.
At $\gamma=0.2$ both $\dF$ and $A$ still fill the free cone
$|x|\le Jt$: dissipation damps the front but neither slows nor smears it,
and $A$, which vanishes to round-off at $\gamma=0$, rises to $0.206$. 
For the two-wall reference state (Fig.~\ref{fig:fermi_twowall}) the
butterfly sits in the middle of the empty region, seven sites away from
either domain wall. At $\gamma=0$ the butterfly operator
$V=\one-2n_{l_V}$ acts trivially on the empty site, so $\dF$ vanishes
until the particles that spread ballistically from the two melting domain
walls into the empty region arrive at the butterfly site, at
$t\approx7/J$; only then does a cone emanate from the butterfly. At
$\gamma=0.2$, by contrast, the chiral response is present from $t=0$ and
$\dF$ is essentially its antisymmetric part; since $s_l=+1$ throughout
the empty region, $A$ is uniform in sign inside $|x|<7$ and flips once,
at the walls.
As a side remark, the dilation method of Sec.~\ref{sec:DilationMethod}
reproduces the light cones of Figs.~\ref{fig:fermi_lc25}
and~\ref{fig:fermi_twowall} within its $O(\Delta t)$ Trotter error.

Figures~\ref{fig:fermi_anatomy} and~\ref{fig:fermi_universality} analyze
these cones in the light of Sec.~\ref{sec:Anatomy}.
Recall that for a reflection-symmetric reference state the symmetric part
$S$ is the memory term $\dF_{\rm mem}$ and the antisymmetric part $A$ is
the source term $\dF_{\rm src}$. Figure~\ref{fig:fermi_anatomy} tracks
the two parts in time for both reference states: $\max_x|S|$ decays at a
rate of approximately $4\gamma$ and $\max_x|A|$ at approximately
$2\gamma$, so the cone becomes purely antisymmetric at late times. After
the front reaches the chain ends, $A$ is dominated by the boundary layer
of the current-carrying steady state and decays more slowly still. For
the two-wall reference state the memory term, and hence $S$, remains zero
until the particles spreading from the domain walls reach the butterfly
site ($t\approx7$), by which time it is damped by
$\approx e^{-4\gamma\cdot7}$; $S$ therefore never exceeds $0.0014$ and
$\dF$ is the chiral source term alone. 
The cuts of $A$ for the two reference states
[Figs.~\ref{fig:fermi_anatomy}(b) and~\ref{fig:fermi_anatomy}(d)] differ only by the sign
pattern $s_l$.
Figure~\ref{fig:fermi_universality} illustrates universality (iv) at
$L=41$ for four reflection-symmetric basis states: $A/s_l$ is the same
for all four, with $\max|A|=0.2056$ at $\gamma=0.2$ independently of $L$
and of the reference state. In contrast, $\max|S|$ ranges from $0$ for
the vacuum, for which the memory term vanishes identically and $\dF$ is
exactly the antisymmetric source term, to $0.84$ for the N\'eel and the
random states.
A scan over $\gamma$ with the same exact solution
shows that the chiral amplitude $\max_{t,x}|A|$ is nonmonotonic: it
grows with $\gamma$ at weak dissipation, the source term being
proportional to $\gamma$, peaks at $\approx0.21$ for
$\gamma\approx0.2$--$0.3\,J$, and decreases beyond (to $0.11$ at
$\gamma=J$), as the damping $e^{-2\gamma t}$ suppresses the response
before it can develop.

The chirality can also be exposed by comparing the two signs of $\eta$
directly. With the butterfly fixed at the center and a
reflection-symmetric reference state, the mirror covariance
\eqref{eq:mirror} gives $F_{-\eta}(t,x)=F_{+\eta}(t,-x)$, so the
$\eta$-odd part $\dEF(t,x)$ of Eq.~\eqref{eq:dEF} is twice the
reflection-odd part of $F$ itself. The disturbance created at the
butterfly site thus reaches the probes on either side at the same speed
$J$ but with unequal amplitudes, and reversing $\eta$ exchanges the two
arms of the cone. In closed systems a directional asymmetry of
operator spreading requires engineered
Hamiltonians~\cite{Stahl18a,Zhang20a} or anyonic
statistics~\cite{Liu18a}. Here the system Hamiltonian is Hermitian and
reflection symmetric, the front velocity is the same in both directions,
and the asymmetry, which resides in the amplitude, is carried entirely by
the dissipator.

Finally, we compare the two reset strategies of Sec.~\ref{sec:BathReset}.
The butterfly-induced deviation $\dF$, including its chiral part, is the
same for both, wherever the butterfly sits (the difference is of order
$10^{-16}$ at $L=11$, $J=\gamma=1$, $\eta=1$), in accordance with
Eq.~\eqref{eq:dF_closed}: the flip reverses $\eta$ in the backward block,
on which $X_\bwd$ does not depend. The two strategies therefore differ
only in the leak factor $F_I(t,l_W)$, which involves no butterfly, and
the strategy difference
\begin{equation}
  D(l_W) :=  F^{\rm flip}(l_W)-F^{\rm same}(l_W)
  \label{eq:D}
\end{equation}
is a function of the probe position alone. It vanishes identically when
the probe sits at the center of the chain, for any reference state, and
decays exponentially with the distance of the probe from the boundaries;
at the self-dual point $\eta=1$ it is in addition reflection
antisymmetric, $D(l_W)=-D(L{+}1{-}l_W)$. The flipped reset therefore does
not modify the light cone emanating from the butterfly: its effect is
confined to probes within about one site of the boundaries. As shown in
Sec.~\ref{sec:NESS}, in the long-time limit $D(l_W)$ gives direct access
to the steady-state skin profile.

\subsection{Liouvillian Skin Effects in the Nonequilibrium Steady State}
\label{sec:NESS}

Three length scales, defined by three different objects, must be kept
apart. The non-Hermitian skin depth $\lNH$ of Eq.~\eqref{eq:kappa} is the
localization length of the eigenmodes of $\Heff$, which is observable
only under postselection on the absence of quantum jumps. The
Liouvillian (NESS) skin depth $\lNESS$~\cite{Haga21a} is the decay length
of the boundary occupation deviation
$\delta(l)=\langle n_l\rangle_{\rm NESS}-\tfrac12$ in the steady state of the
full Lindblad equation. The OTOC skin depth $\lOTOC$ is the decay length
extracted from the reset-strategy difference $D(l_W)$ of the dissipative
OTOC. Since the no-jump probability is exponentially small, $\lNESS$ is the
physically meaningful depth in practice, and the OTOC provides an
interferometric way to detect it.

\begin{figure}[tbp]
\centering
\includegraphics[width=\columnwidth,height=0.85\textheight,keepaspectratio]{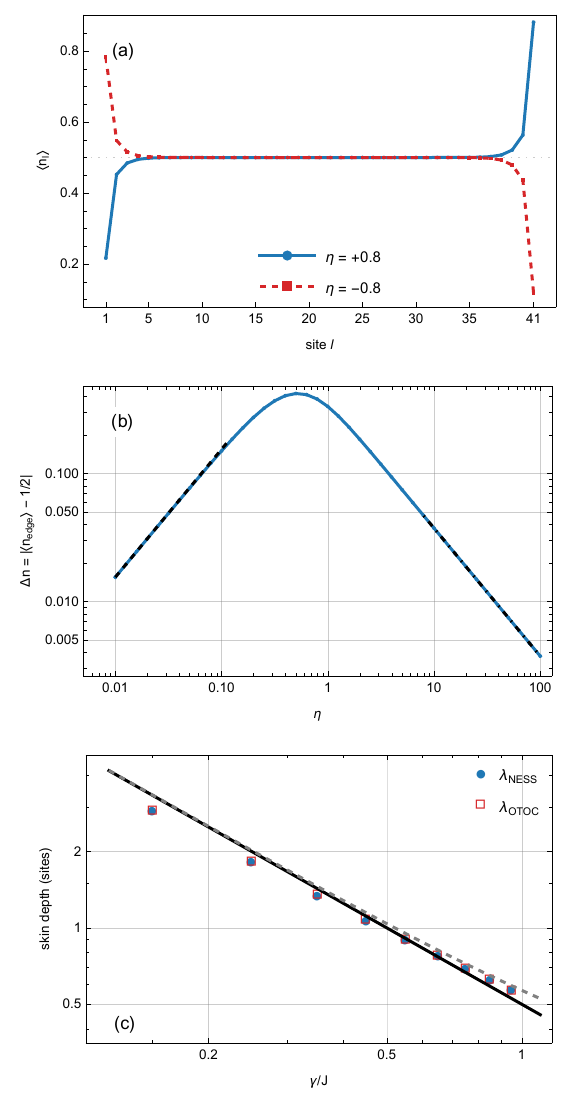}
\caption{Anatomy of the Liouvillian (NESS) skin effect.
(a) Steady-state occupation profiles at $\eta=\pm0.8$ ($L=41$,
$\gamma/J=0.5$). (b) Boundary accumulation height $\Delta n$ versus
$\eta$ for the same $L$ and $\gamma$. (c) $\lNESS$ (from the Lyapunov
steady state) and $\lOTOC$ (from the steady state of the Trotterized
echo, $\Delta t=0.1$) versus $\gamma/J$ at $\eta=0.8$, with the leading
behavior $J/2\gamma$ (solid) and the exact
$1/[2\,\mathrm{arcsinh}(\gamma/J)]$ (dashed).}
\label{fig:fermi_height}
\end{figure}

We first examine the Liouvillian skin effect itself. The exact
steady-state correlation matrix shows that the $\eta$ dependence of the
occupation profile separates into three attributes: direction
[Fig.~\ref{fig:fermi_height}(a)], height
[Fig.~\ref{fig:fermi_height}(b)] and depth
[Fig.~\ref{fig:fermi_height}(c)]. In \emph{direction}, the
accumulation edge is set by $\mathrm{sgn}\,\eta$, and reversing $\eta$
mirrors the profile exactly.
In \emph{height}, the boundary accumulation
$\Delta n=|\langle n_{\rm edge}\rangle-\tfrac12|$ vanishes in both
reciprocal limits, $\propto\eta$ as $\eta\to0$ and $\propto1/\eta$ as
$\eta\to\infty$, and is maximal at $\eta=O(1)$; the skin effect thus
disappears through its \emph{amplitude}, at fixed depth. 
In \emph{depth}, the decay rate of $\delta(l)$ is essentially independent of $\eta$ and is
set by the uniform damping,
\begin{equation}
  \frac{1}{\lNESS}=2\,\mathrm{arcsinh}\frac{\gamma}{J}
  =\frac{2\gamma}{J}-\frac13\Big(\frac{\gamma}{J}\Big)^3+\cdots,
  \label{eq:lness}
\end{equation}
that is, $\lNESS\simeq J/2\gamma$. This result is confirmed numerically
for $0.15\le\gamma/J\le0.95$ and derived in Appendix~\ref{app:kernel}.

The derivation rests on three observations. First, the damping matrix
of the full channel is \emph{exactly diagonal}: the Hatano--Nelson cross
terms cancel between loss and gain (Sec.~\ref{sec:GaussianSolution}),
leaving the $\eta$-independent bulk damping rate $2\gamma$, whereas
$\Heff$ carries their difference. Second, all $\eta$ dependence resides
in a current-like source, whose sign sets the direction and whose
amplitude $\gamma\eta/(1+\eta^2)$ sets the height. Third, the bulk part
of this source generates no density and can be subtracted exactly; what
remains is an edge-localized effective source, which penetrates
ballistically ($v_{\max}=J$) against the uniform damping, so that
$\lNESS\simeq v_{\max}/2\gamma$. The evanescent zeros of the bulk
resolvent give the exact rate, and a Jacobi--Anger kernel representation
supplies the closed form together with an $x^{-1/2}$ prefactor, which
biases log-linear fits over a finite window slightly \emph{above} the
asymptote.

The height, notably, is \emph{not} self-dual:
$\Delta n(\eta{=}1/2)=0.42$ versus $\Delta n(\eta{=}2)=0.19$ at
$\gamma/J=0.5$, even though $\lNH$ is exactly
$\eta\leftrightarrow1/\eta$ symmetric. The self-duality is a property of
the no-jump spectrum, not of the full steady state: $\eta\to1/\eta$ maps
the jump operators to different ones (the terminated edge bonds have
damping $2\gamma/(1+\eta^2)$ and $2\gamma\eta^2/(1+\eta^2)$, which are
interchanged by the duality), and the steady state reflects this
difference. The exact symmetry of the steady state is the combination of
reflection and particle--hole exchange, which maps $\eta$ to $1/\eta$,
\begin{equation}
  \delta_\eta(l)=-\delta_{1/\eta}(L{+}1{-}l).
  \label{eq:deltadual}
\end{equation}
The occupation profile is therefore antisymmetric about the center of
the chain only at the self-dual point $|\eta|=1$; at $\eta=0.8$ the
accumulation at one end ($0.38$) visibly exceeds the depletion at the
other ($0.28$) [Fig.~\ref{fig:fermi_height}(a)]. In this respect, too, $\lNESS$ carries more information than
$\lNH$.

\begin{figure}[tbp]
\centering
\includegraphics[width=\columnwidth,height=0.72\textheight,keepaspectratio]{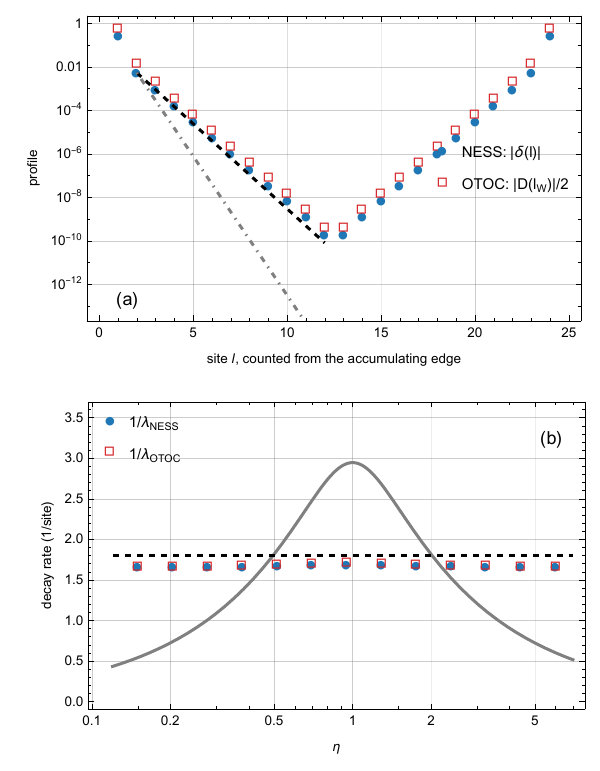}
\caption{Comparison of the three skin depths in the steady state
($J=1$, $\gamma=0.9$, $\Delta t=0.1$). (a) Profiles at the self-dual
point $\eta=1$ ($L=24$, $t=17$) of the NESS occupation deviation
$|\delta(l)|$ (Lyapunov solution of the backward channel) and of the
strategy difference $|D(l_W)|/2$ of the OTOC (correlation-matrix echo),
versus the site index counted from the accumulating edge. The dashed and
dash-dotted lines have slopes $2\gamma$ and $2/\lNH=2.94$, respectively.
(b) Fitted decay rates $1/\lNESS$ and $1/\lOTOC$ versus $\eta$ ($L=20$),
compared with $2/\lNH$ (solid) and $2\gamma$ (dashed).}
\label{fig:fermi_fit}
\end{figure}

The central result of this section is that $\lOTOC$ measures $\lNESS$.
As $t\to\infty$ the strategy difference reduces to a difference of
steady-state occupations of the backward channel~\cite{McDonald22a}. The
flipped reset is the exact particle--hole image of the same reset, so
that $\langle n_l\rangle_{\rm flip}=1-\langle n_l\rangle_{\rm same}$
(verified in the Trotterized echo to machine precision). By the
eigenstate identity \eqref{eq:eigenstate_identity} the two strategies
then yield steady-state OTOCs of equal magnitude and opposite sign,
$F^{\rm flip}=-F^{\rm same}$, and hence
\begin{equation}
  D(l_W)=F^{\rm flip}-F^{\rm same}=-2F^{\rm same}=4\,s_{l_W}\,\delta(l_W).
  \label{eq:Did}
\end{equation}
The strategy difference is thus the deviation profile of the NESS,
multiplied by the known sign factor $s_{l_W}$ of the reference state; of
the factor $4$, one factor of $2$ comes from the particle--hole flip and
one from the probe operator $\one-2n_{l_W}$. A polarized reference state
($s_l=1$ for all $l$) makes $D=4\delta$ a faithful signed image of the
profile, whereas the N\'eel reference state staggers it. At infinite
temperature $D$ vanishes in the steady state, so the steady-state
protocol requires a definite product reference state.

Figure~\ref{fig:fermi_fit} makes the three-way comparison quantitative
by independent calculations: $\lNESS$ from the Lyapunov steady state,
$\lOTOC$ from the strategy difference of the Trotterized echo (validated
against the dilation to $10^{-15}$), and $\lNH$ from
Eq.~\eqref{eq:kappa}. The result is
\begin{equation}
  \lOTOC=\lNESS\simeq\frac{J}{2\gamma}\qquad(t\to\infty),
  \label{eq:lotoc}
\end{equation}
with the fitted rates agreeing to within $0.7\%$ at every $\eta$.
Both are $\eta$-independent and do \emph{not} follow the peaked,
self-dual curve $2/\lNH$. This is the precise sense in which the OTOC
provides experimental access to the Liouvillian skin depth: it is an
interferometric observable that requires neither density microscopy of
the steady state nor postselection. The result is in line with the
observation that the steady-state and relaxation lengths of a
dissipative chain need not coincide with the bare skin localization
length~\cite{Haga21a,Lee23a}.

The transient behavior must be kept separate from this steady-state
statement: it is a wavefront, not a skin effect. At finite time, before the light cone has passed and the steady state is
established, the strategy-difference profile is a \emph{ballistic OTOC
wavefront}: super-exponential in space, with an apparent slope that
drifts with time and is far steeper than any of the three depths---the
open-system chiral wavefront of Ref.~\cite{Song19a}, where short-time
dynamics follows $\Heff$ while the long-time behavior requires the full
Lindbladian. A naive exponential fit at finite $t$ returns neither $\lNH$
nor $\lNESS$; the skin exponential emerges only in the steady state. A
length extracted from a steady-state population image is
$\lNESS\simeq J/2\gamma$ and must not be identified with $\lNH$ without
checking.

Two remarks on the dilation and trajectory methods close this
section. The steady-state results above were generated by the exact solution of
Sec.~\ref{sec:GaussianSolution}, and the two other methods provide
independent numerical verification. The dilation reproduces the same echo as a single
Slater-determinant overlap and agrees with the correlation-matrix echo to
$10^{-15}$ at every probe site and for both reset strategies, which
certifies that Eq.~\eqref{eq:Did} is a property of the echo generated
by the full Lindblad channels $\Vf$ and $\Vb$, that is, of the average
over all bath measurement records, and not of any postselected branch (a
dilation in which the two interferometer branches are given separate
bath registers would, by contrast, return the physics of $\Heff$ and the
length $\lNH$). The
trajectory estimator, in turn, reproduces the dilation for both reset
strategies within the statistical errors, with full effective sample
size, and we have shown in
Sec.~\ref{sec:QuantumTrajectoryMethod} that the hardware protocol
carries no weights at all. The reset-strategy difference is therefore
accessible on a quantum processor by running the measure-and-reset
interferometer twice, with the backward bath registers reprepared in
the same or in the flipped references, and reading the control qubit in
both runs; by the eigenstate identity it can even be read from the
occupations of the echo-evolved state without the control qubit. At long
times the price is the shrinking control-qubit signal, that is, a larger
number of shots, but no exponentially rare postselection is involved, in contrast to any
protocol that targets $\lNH$.

\section{Local Spin Model}
\label{sec:LocalSpinModel}

The fermionic model discussed so far is Gaussian and admits exact
analyses, which give transparent physical insight into the Liouvillian
skin effect as seen by the dissipative OTOC. Its hardware-efficient
circuit of Fig.~\ref{fig:trajectory_fermi}, however, cannot be run
directly on a quantum computer built from qubits: the fermion operators
must first be mapped to qubit operators, for example by the
Jordan--Wigner transformation
\begin{equation}
  c_l=\Big(\prod_{j<l}\sigma^z_j\Big)\sigma^-_l .
  \label{eq:JW}
\end{equation}
Under this transformation the Hamiltonian \eqref{eq:Hsys} becomes the XX
chain $H_0=-\tfrac J2\sum_l(\sigma^+_l\sigma^-_{l+1}+\mathrm{h.c.})$, and
the effective non-Hermitian Hamiltonian $\Heff$ becomes the same XX chain
with an imaginary Dzyaloshinskii--Moriya term, that is, a Hatano--Nelson
chain of spin excitations with the skin depth \eqref{eq:kappa}. The jump
operators, however, do not remain local:
\begin{equation}
  c_l+\eta e^{i\phi}c_{l+1}
  =\Big(\prod_{j<l}\sigma^z_j\Big)\big(\sigma^-_l+\eta e^{i\phi}\sigma^z_l\sigma^-_{l+1}\big)
  \label{eq:JWjump}
\end{equation}
carries the parity string of all sites to the left of the bond, and the
string survives in the recycling term $A_l\rho A_l^\dagger$ of the
dissipator. The same applies to the microscopic system--bath model of
Sec.~\ref{sec:SystemBath}, whose reservoirs are themselves fermionic.

In this section we therefore construct a \emph{local} spin model that
can be realized on a quantum processor with nearest-neighbor gates and
bath qubits coupled to pairs of neighboring sites: an XX chain with
bond dissipators $\propto S^-_l+\eta e^{i\phi}S^-_{l+1}$ and their
adjoints. It has the same bond structure as the fermionic model, but it
is generated by a single thermal bath coupled through two-site operators
rather than by two engineered reservoirs. Its effective non-Hermitian
Hamiltonian is, after the Jordan--Wigner transformation, the same
Hatano--Nelson chain, with the same postselected depth $\lNH$. The jump
operators, however, commute rather than anticommute on different sites,
and this changes the full Lindblad dynamics completely. Because the
model is directly simulable on currently available quantum computers, we
study it alongside the fermionic model.

\subsection{Local-Spin Lindbladian}
\label{sec:spin:LocalSpin}

The model is an open-boundary XX chain of $L$ spins-$\tfrac12$,
\begin{equation}
\begin{split}
  H_0 & = -J\sum_l(S^x_lS^x_{l+1}+S^y_lS^y_{l+1}) \\
  &=-\frac{J}{2}\sum_{l=1}^{L-1}\big(\sigma^+_l\sigma^-_{l+1}+\mathrm{h.c.}\big),
\end{split}
\label{eq:H0}
\end{equation}
with the bond operator ($\eta>0$, $\phi\in(-\pi,\pi]$)
\begin{equation}
  A^-_l=\big(A^+_l\big)^\dagger:=\frac{S^-_l+\eta e^{i\phi}\,S^-_{l+1}}{\sqrt{1+\eta^2}},
  \label{eq:ak}
\end{equation}
which carries the same complex relative amplitude as the fermionic jump
operators \eqref{eq:jumpA}. The normalization fixes the bulk on-site
damping at $\Gamma$ per channel for every $\eta$. Unlike in the fermionic
model, where $\{A_l,A_l^\dagger\}=\one$, it does not make
$A^+_lA^-_l+A^-_lA^+_l$ the identity: spin operators on different sites
commute rather than anticommute, so the cross terms add instead of
canceling,
\begin{equation}
  A^+_lA^-_l+A^-_lA^+_l=\one+\teta\big(e^{i\phi}S^+_lS^-_{l+1}+\mathrm{h.c.}\big),
  \label{eq:AAspin}
\end{equation}
and the surviving cross term is the origin of the imaginary
Dzyaloshinskii--Moriya coupling derived below. The chain is coupled, on every
bond, to a bath spin held at inverse temperature $\beta$
(Sec.~\ref{sec:spin:SystemBath}); Born--Markov elimination of the bath
spins in the singular-coupling limit (Appendix~\ref{app:spinbath})
yields~\cite{Breuer02a,Lindblad76a}
\begin{equation}
  \dot\rho=-i[H_0,\rho]
  +\Gamma_-\sum_l\mathcal{D}[A^-_l]\rho
  +\Gamma_+\sum_l\mathcal{D}[A^+_l]\rho,
  \label{eq:lindblad_spin}
\end{equation}
with the loss (gain) rate $\Gamma_-$ ($\Gamma_+$) proportional to the
thermal occupation $p_\downarrow$ ($p_\uparrow$) of a bath spin. Because
both channels are fed by the same thermal spin, the two rates are tied by
detailed balance at the bath temperature, which defines the inverse
temperature $\beta$ of the model,
\begin{equation}
  e^{\beta\omega}:=\frac{\Gamma_-}{\Gamma_+}=\frac{p_\downarrow}{p_\uparrow},
  \label{eq:beta_spin}
\end{equation}
with $\omega$ the bath-spin splitting. We write
$\Gamma := \tfrac12(\Gamma_-+\Gamma_+)$ and
$\delta := \tfrac12(\Gamma_--\Gamma_+)$. At infinite bath temperature,
$\Gamma_+=\Gamma_-=:\gamma$ ($\delta=0$), the jump set
$\{\sqrt\gamma A^-_l,\sqrt\gamma A^+_l\}$ is self-adjoint and the
backward--forward identity \eqref{eq:identity} holds.

The essential difference from the fermionic model is that the gain
operator is the \emph{adjoint} of the loss operator,
$A^+_l\propto S^+_l+\eta e^{-i\phi}S^+_{l+1}$. It therefore carries the
\emph{opposite} chirality ($\phi\to-\phi$), whereas the fermionic gain
operator $B_l$ has the same chirality as the loss operator. Both choices
are natural: the spin model is what a single thermal bath produces, and
the fermionic model is what two engineered reservoirs produce. As shown
below, the difference changes the steady state qualitatively.

We begin with the effective non-Hermitian Hamiltonian that governs the
no-jump evolution. Using Eq.~\eqref{eq:AAspin} and
$A^+_lA^-_l-A^-_lA^+_l=\frac{2}{1+\eta^2}(S^z_l+\eta^2S^z_{l+1})$, the effective
Hamiltonian at $\phi=\pm\pi/2$ is
\begin{multline}
  \Heff=H_0+\frac{\Gamma\teta\sin\phi}{2}\sum_{l=1}^{L-1}\big(S^+_lS^-_{l+1}-\mathrm{h.c.}\big)\\{}
  -i\delta\sum_{l=1}^{L}w_lS^z_l
  -\frac{i\Gamma}{2}(L-1)\one,
  \label{eq:heff_spin}
\end{multline}
with the site weights of the imaginary Zeeman term
\begin{equation}
  w_1=\frac{1}{1+\eta^2},\quad 
  w_l=1\ \ (1<l<L),\quad
  w_L=\frac{\eta^2}{1+\eta^2}.
  \label{eq:wk}
\end{equation}
(For general $\phi$ the hopping acquires in addition the common
imaginary part $-\tfrac i2\Gamma\teta\cos\phi$, an achiral,
momentum-dependent damping with no counterpart in the fermionic model.)
Three terms are added to $H_0$. The first is an \emph{imaginary}
Dzyaloshinskii--Moriya (DM) coupling~\cite{Dzyaloshinsky58a,Moriya60a},
$-i\Gamma\teta\sin\phi\sum_l(\mathbf S_l\times\mathbf S_{l+1})^z$,
proportional to the \emph{total} rate and independent of the bath
temperature; a Hermitian DM term would merely attach a Peierls phase to
the hopping, and it is precisely the imaginary coefficient that produces
a \emph{real} hopping asymmetry. The second is an \emph{imaginary
Zeeman} field, present only at finite bath temperature. It is uniform in
the bulk, where it merely shifts the decay rate within each sector of
$S^z_{\rm tot}$, and is reduced at the two ends to the chiral edge
weights $w_1\neq w_L$ (unless $\eta=1$); in particular, site $L$ is
undamped at $\eta=0$, which is the origin of the boundary effects
discussed in Sec.~\ref{sec:spin:LightCones}. The third is a uniform
decay, which is exact for the open chain. After
a Jordan--Wigner transformation the $\delta=0$ part of $\Heff$ is a
Hatano--Nelson chain with real asymmetric hoppings
\begin{equation}
  t_{R,L}=-\frac{J\mp\Gamma\teta\sin\phi}{2},
  \label{eq:hn_spin}
\end{equation}
the same asymmetry as Eq.~\eqref{eq:JLR} with $\gamma\to\Gamma$; under
open boundaries its spectrum is purely real by the imaginary gauge
transformation (valid for $\Gamma\teta<J$, verified to $10^{-15}$ at
$L=11$) and its eigenvectors are skin-localized with the same $\lNH$ as
Eq.~\eqref{eq:kappa}: $4.9$ sites at $\Gamma=0.2$, $\eta=1$.

In the full Lindblad dynamics, the physically meaningful chirality
resides in the channel-resolved damping rates: a plane wave at
momentum $q$ couples to the loss and gain channels with
\begin{equation}
  \kappa_{l,g}(q)=\Gamma_{-,+}\big[1+\teta\cos(q\pm\phi)\big],
  \label{eq:spectrum_spin}
\end{equation}
which at $\phi=\pm\pi/2$ become
$\kappa_{l,g}(q)=\Gamma_{-,+}(1\mp\teta\sin\phi\,\sin q)$ and obey the sum
rule
\begin{equation}
  \kappa_l(q)+\kappa_g(q)=2\Gamma-2\delta\,\teta\sin\phi\,\sin q .
  \label{eq:sumrule_spin}
\end{equation}
At $\eta=1$ the mode $q=\mathrm{sgn}(\phi)\,\pi/2$, which has the largest
group velocity, is a dark state of every loss operator, while its mirror
partner is lost at rate $2\Gamma_-$. This is the chiral-damping structure
of Refs.~\cite{Yao18a,Song19a}. At infinite bath temperature, however,
the gain chirality is exactly opposite to the loss chirality, and the two
cancel in the relaxation of the single-particle \emph{density}: wave
packets at $q=\pm\pi/2$ propagate in opposite directions but lose their
excess density at the common rate $2\gamma$ (fitted rates $0.393$ and
$0.415$, against $0.4$ at $\Gamma=0.2$). At finite temperature the
cancellation is incomplete by $2\delta\teta\sin\phi\sin q$, which is the
origin of the chiral relaxation of Sec.~\ref{sec:spin:LongTime}. The
chirality that many-body observables do detect at infinite temperature
relies on the occupation structure of the reference state. In a polarized
region the gain channel is blocked, which locally exposes the chiral loss
spectrum $\kappa_l(q)$. The exposure is maximal inside domain-wall
blocks, partial on a N\'eel pattern, and absent for spin-flip--invariant
states, in one-to-one correspondence with the selection rule of
Sec.~\ref{sec:spin:LightCones}.

In contrast to the fermionic model, the steady state of the spin model
is achiral. For the bond operators of
Eq.~\eqref{eq:ak} one has the operator identity
\begin{equation}
  \mathcal{D}[A^-_l]\one=A^-_lA^+_l-A^+_lA^-_l
  =-\frac{2}{1+\eta^2}\big(S^z_l+\eta^2S^z_{l+1}\big).
  \label{eq:DA1}
\end{equation}
The cross terms $\propto\eta e^{\pm i\phi}(S^-_lS^+_{l+1}-S^+_{l+1}S^-_l)$
cancel because spin operators on different sites \emph{commute}. For the
fermionic operators $c_l+\eta e^{i\phi}c_{l+1}$ they anticommute, the
cross terms do not cancel, and they survive as a bond-current source;
this is the point at which the two models differ. Hence
$\mathcal{D}[A^-_l]\one+\mathcal{D}[A^+_l]\one=0$, and for equal rates
the maximally mixed state is the unique steady state (Liouvillian gap
$0.30$--$0.36$ at $\Gamma=0.2$, $L=7$--$9$), whatever the chirality
assignment of gain and loss; we have verified this also for a gain
operator of the same chirality,
$\propto S^+_l+\eta e^{i\phi}S^+_{l+1}$. For unequal rates, every $A^-_l$
lowers $S^z_{\rm tot}$ by exactly one, so the pair
$\{\sqrt{\Gamma_-}A^-_l,\sqrt{\Gamma_+}A^+_l\}$ satisfies quantum detailed
balance bond by bond with respect to the uniform thermal product state
\begin{equation}
  \rho_\beta=\frac{e^{-\beta\omega S^z_{\rm tot}}}{Z}
  =\bigotimes_{l=1}^{L}\begin{pmatrix}p_\uparrow&0\\0&p_\downarrow\end{pmatrix},
  \label{eq:rhobeta}
\end{equation}
with $p_\downarrow=1-p_\uparrow=\Gamma_-/(\Gamma_-+\Gamma_+)$ and the
uniform magnetization
\begin{equation}
  \langle\sigma^z_l\rangle_\beta=-\tanh\frac{\beta\omega}{2}\qquad\text{for every }l,
  \label{eq:mbeta}
\end{equation}
for any $\eta$. Indeed,
$A^-_l\rho_\beta=e^{\beta\omega}\rho_\beta A^-_l$, so that the gain term
cancels the loss term on $\rho_\beta$ exactly, and $[H_0,\rho_\beta]=0$
(the numerical residual is below $10^{-16}$). The steady state of the
local spin model is therefore partially magnetized but \emph{uniform}: it
has no skin-like profile, carries no current and does not depend on
$\eta$, although $\Heff$ contains the chiral edge weights
$w_1\neq w_L$. The jumps compensate the edge imbalance exactly, because
detailed balance holds bond by bond. All chirality of this model is
therefore transient: it appears in the approach to the steady state or,
at finite temperature, in the OTOC of the steady state
(Sec.~\ref{sec:spin:LongTime}). This is opposite to the fermionic model,
in which the gain of the same chirality as the loss sustains a steady
current and a skin profile.

\subsection{Microscopic System--Bath Model for Local Spins}
\label{sec:spin:SystemBath}

\begin{figure}
\centering
\includegraphics[width=8cm]{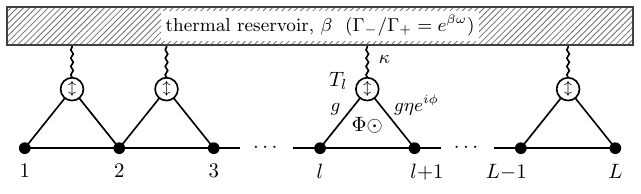}
\caption{Geometry of the microscopic system--bath model for the local
spin model. Every bond $(l,l{+}1)$ of the XX chain of $L$ spins
carries one bath spin $T_l$ (circle). The bath spin couples to both end
sites of its bond through the bond operator $A^-_l$ of
Eq.~\eqref{eq:ak}, with amplitudes $g$ and $g\eta e^{i\phi}$, and is
relaxed (wavy lines) at rate $\kappa\gg J,g,\omega$ toward the thermal
state of a common reservoir at inverse temperature $\beta$. Only the
bond $(l,l{+}1)$ is labeled; all bonds are identical. For the spin
excitations of the chain, the relative phase $e^{i\phi}$ is a synthetic
flux $\Phi=\phi\,\Phi_0/2\pi$ through the triangle $(l,l{+}1,T_l)$, drawn
for $\phi>0$ ($\odot$). The same bath spin provides both the loss and
the gain channel; since the gain operator is the \emph{adjoint} of the
loss operator, the two channels experience \emph{opposite} synthetic
flux, whereas the two triangles of a bond in the fermionic model of
Fig.~\ref{fig:flux} carry the same flux.}
\label{fig:spinbath}
\end{figure}

Every bond carries a bath spin-$\tfrac12$ $T_l$ with
$H_{\rm bath}=\omega\sum_lT^z_l$, coupled to the chain through
\begin{equation}
  H_{\rm int}=g\sum_{l=1}^{L-1}\big(A^+_lT^-_l+A^-_lT^+_l\big)
  \label{eq:Hint_spin}
\end{equation}
(Fig.~\ref{fig:spinbath}). All bath spins are in contact with a
\emph{common} thermal reservoir at inverse temperature $\beta$, which
relaxes each of them at a rate $\kappa\gg J,g,\omega$, so that the
stationary state of a bath spin is thermal with
$p_\downarrow/p_\uparrow=e^{\beta\omega}$.
The Born--Markov elimination of the bath spins
(Appendix~\ref{app:spinbath}) parallels that of the fermionic model, with
one structural difference that determines the steady state: there is a
\emph{single} bath spin per bond, so loss and gain are the two orderings
of the same correlation function and are related by the
Kubo--Martin--Schwinger condition. The memory kernels are one-sided
exponentials of width $\tau_B=2/\kappa$. For $\kappa\gg J,\omega$ they
are flat across the system band and reduce to
$\Gamma_{\mp}\delta(\tau)$, and Eq.~\eqref{eq:lindblad_spin} follows with
\begin{equation}
  \Gamma_{-,+}=\frac{4g^2}{\kappa}\,\frac{p_{\downarrow,\uparrow}}{1+(2\omega/\kappa)^2}
  \label{eq:rates_spin}
\end{equation}
and a Lamb shift of relative order $\omega/\kappa$. The hierarchy of time
scales is that of Eq.~\eqref{eq:validity} with $\tau_B=2/\kappa$, and
again no condition involves $L$. It should be noted that such a bath
thermalizes the magnetization, not the energy: its rates are frequency
independent, so it cannot resolve the Bohr frequencies of $H_0$, and the
steady state is $e^{-\beta\omega S^z_{\rm tot}}/Z$ rather than
$e^{-\beta H_0}$.

A spin carries no charge, so the phase $e^{i\phi}$ cannot be an
Aharonov--Bohm phase as in Sec.~\ref{sec:SystemBath}. It can nevertheless
be realized in three ways. The first is a synthetic flux for the spin
excitations. The full Hamiltonian conserves
$S^z_{\rm tot}+\sum_lT^z_l$, and the Matsubara--Matsuda
mapping~\cite{Matsubara56a} turns every spin into a hard-core-boson site,
$b^\dagger_l=S^+_l$. In this language $H_{\rm int}$ describes hopping
between the chain and the bath site with amplitudes $g$ and
$g\eta e^{-i\phi}$, and the only gauge-invariant content of $\phi$ is the
phase accumulated around the triangle $(l,l{+}1,T_l)$, that is, a
synthetic flux $\Phi/\Phi_0=\phi/2\pi$ per triangle, or a quarter flux
quantum at $\phi=\pm\pi/2$. In superconducting-qubit arrays a parametric
modulation of a tunable coupler with modulation phase $\varphi$ produces
the exchange $ge^{i\varphi}b^\dagger_ib_j+\mathrm{h.c.}$, and a
three-transmon loop with such a tunable synthetic flux was demonstrated
in Ref.~\cite{Roushan17a}; in cold atoms the same complex amplitudes
arise from laser-assisted tunneling~\cite{Goldman14a}. There is only
one triangle per bond, because the same bath site absorbs a particle
when empty and emits one when occupied. The opposite chirality of the
gain channel then follows automatically, since a particle traversing the
loop in the opposite sense experiences the opposite flux. The loss and
gain channels of a bond thus experience \emph{opposite} synthetic flux,
whereas the two triangles of a bond in the fermionic model
(Fig.~\ref{fig:flux}) carry the \emph{same} flux; this is the geometric
expression of the difference between the two models. The second
realization is a propagation phase in a waveguide. If
the bath of bond $l$ is a chiral photonic channel to which both emitters
couple, the collective jump operator carries the propagation phase
$\phi=k_0d$, as in chiral quantum optics~\cite{Pichler15a,Lodahl17a}.
The third realization is in the digital circuit itself. In the
measure-and-reset protocol of Sec.~\ref{sec:spin:Methods}, the step
$e^{-i\Delta tH_{{\rm int},l}}$ is Trotter-split into the two exchange
gates $T_l\leftrightarrow l$ and $T_l\leftrightarrow l{+}1$, and the
phase of the second gate is a virtual $Z$ rotation of the bath qubit
inserted between the two gates, which requires no additional hardware
operation. The chirality is thus set in software, and $\phi$ can be swept
freely, which is the counterpart of sweeping the flux in the fermionic
model.

\subsection{Methods for Local Spins}
\label{sec:spin:Methods}

\begin{figure*}
\centering
\includegraphics[width=\textwidth]{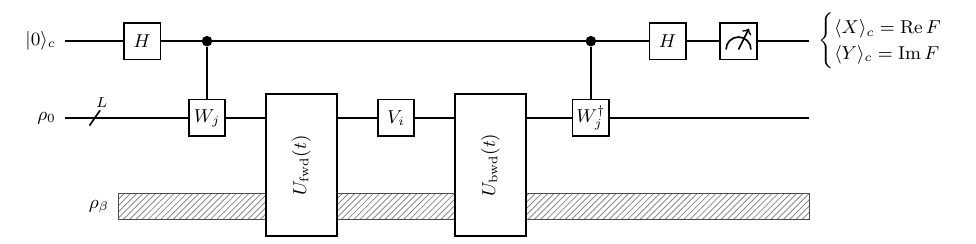}
\caption{Concept quantum circuit for the dissipative OTOC of the spin
model, the counterpart of Fig.~\ref{fig:circuit}, with the butterfly
$V_i$ at site $i=l_V$ and the probe $W_j$ at site $j=l_W$. The two
reservoirs of the fermionic model are replaced by a single thermal
reservoir $\rho_\beta$ at inverse temperature $\beta$
(Fig.~\ref{fig:spinbath}), and the reference state $\rho_0$ of the
system may be mixed, for example the thermal steady state itself
(Sec.~\ref{sec:spin:LongTime}). Since $V_i=Z_i$ and $W_j=Z_j$ are
single-qubit Pauli operators, the conditional probe is a controlled-$Z$
gate.}
\label{fig:circuit_spin}
\end{figure*}

\begin{figure*}
\centering
\includegraphics[width=\textwidth]{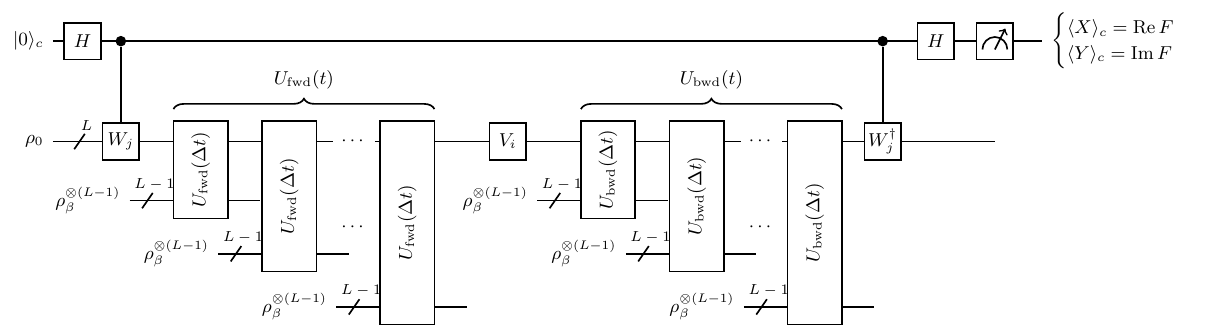}
\caption{The Stinespring dilation of the spin model as a quantum
circuit, the counterpart of Fig.~\ref{fig:dilation}; the control qubit
$c$ and the controlled $W_j$, $W_j^\dagger$ are those of
Fig.~\ref{fig:circuit_spin}. The system register of $L$ spins is
prepared in the reference state $\rho_0$. Each of the $N_t$ forward
Trotter steps is the same unitary
$U_\fwd(\Delta t)=e^{-i\Delta t(H_0+H_{\rm bath}+H_{\rm int})}$, acting
on the system and on a \emph{fresh} register of $L-1$ bath spins, one
per bond, each prepared in the single-spin thermal state with
$p_\downarrow/p_\uparrow=e^{\beta\omega}$; likewise each backward step
$U_\bwd(\Delta t)$, with $H_0\to-H_0$, acts on another fresh register
prepared in the same thermal state. The registers are never measured or
reset, so their number grows as $2N_t$.}
\label{fig:dilation_spin}
\end{figure*}

\begin{figure*}
\centering
\includegraphics[width=\textwidth]{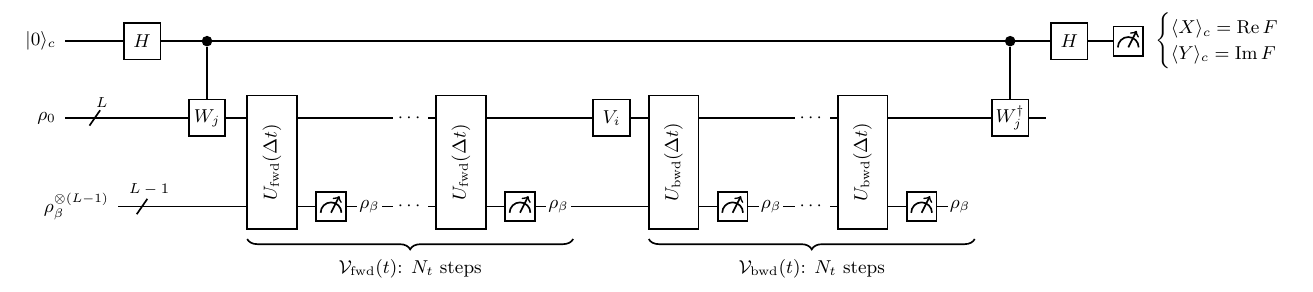}
\caption{Quantum-trajectory (measure-and-reset) realization of the
concept quantum circuit for the dissipative OTOC of the spin model in
Fig.~\ref{fig:circuit_spin}. The thermal reservoir is replaced by a
single register of $L-1$ bath spins, one per bond, each prepared in the
single-spin thermal state, that is, in $|\!\downarrow\rangle$ or
$|\!\uparrow\rangle$ with probabilities $p_\downarrow$ and $p_\uparrow$.
Every Trotter step applies the joint unitary $U_{\fwd,\bwd}(\Delta t)$
of Fig.~\ref{fig:dilation_spin}, then measures the register and resets
it \emph{thermally}, that is, prepares each spin again in the thermal
state. The control qubit is read out as in
Fig.~\ref{fig:circuit_spin}. The register is reused in every step, so
the circuit width stays $L+O(L)$ independently of $N_t$, in contrast to
the dilation of Fig.~\ref{fig:dilation_spin}.}
\label{fig:trajectory_spin}
\end{figure*}

The dissipative OTOC of the spin model has the same dilation structure
as the fermionic one, with a single thermal register per Trotter step in
place of the pair of empty and filled registers
(Fig.~\ref{fig:dilation_spin}); the Trotterized circuit is the one
sampled by the trajectory Monte Carlo described below. For the exact
reference values, however, we integrate the Lindblad equation directly
rather than through the dilation, exploiting a superselection structure.
Both Liouvillians conserve the difference $m$ between the magnetizations
of the bra and ket indices: $H_0$ conserves both magnetizations, the
terms $A^-_lXA^+_l$ ($A^+_lXA^-_l$) lower (raise) both by one, and the
anticommutators preserve both. Since the operators $G_{l_W}(t)$ lie in
the $m=0$ sector, only the $m=0$ component of the operators traced
against them is required. Its dimension is
$\sum_n\binom{L}{n}^2=\binom{2L}{L}$, that is, $705{,}432$ at $L=11$,
compared with $4^{11}\approx4.2\times10^6$ for the full superoperator.
We construct the sparse restricted superoperator ($1.96\times10^7$
nonzero elements at $L=11$) and integrate $\dot w=\Lf w$ with the
fixed-step fourth-order Runge--Kutta method ($\mathrm dt=0.05/J$,
$t_{\max}=12/J$). All $L$ probe operators $G_{l_W}(t)$ are propagated as
the columns of one matrix in a single pass, together with $\rho_\fwd$
for a $\sigma^z$-product state or with the $L$ operators
$X_{l_W}(t)=\Vf(t)[\rho_0W]$ for a general product state; $F$, $F_I$,
$\dF$ and the magnetization are obtained from the same run. At infinite
bath temperature the identity \eqref{eq:identity} allows every column to
evolve under the same forward generator, whereas for
$\Gamma_+\neq\Gamma_-$ the probe columns evolve under the restricted
$\Lb^\dagger$ and the state columns under $\Lf$. We have validated the
solver in three ways: the backward--forward identity holds at the level
of the generators to $9\times10^{-16}$; at $\Gamma=0$ the Runge--Kutta
evolution agrees with exact diagonalization to $1.4\times10^{-9}$; and
the sector solver agrees with a dense evolution of the full Lindbladian
to $3\times10^{-16}$.

The circuit Monte Carlo involves one choice that has no fermionic
counterpart: whether the bath qubit is only measured or is thermally
reset. In the Trotterized circuit (Fig.~\ref{fig:trajectory_spin}) each
layer couples the normalized bond operator to a freshly prepared bath
qubit for a time $\Delta t$, with coupling
$\lambda=\sqrt{(\Gamma_-+\Gamma_+)/\Delta t}$ and with the thermal
probability $p_\downarrow=\Gamma_-/(\Gamma_-+\Gamma_+)$ for the bath
qubit, so that $\Gamma_{-,+}=\lambda^2\Delta t\,p_{\downarrow,\uparrow}$.
A bath qubit in $|\!\downarrow\rangle$ can only absorb an excitation
(loss) and one in $|\!\uparrow\rangle$ can only emit one (gain), and the
limit $\Delta t\to0$ reproduces the Lindblad equation at any bath
temperature. We have verified this at $L=3$ and $\Gamma_-/\Gamma_+=3$,
where the deviation is halved when $\Delta t$ is halved, whereas the
opposite assignment of the two bath states does not converge. We first
simulated the circuit at the level of individual trajectories for $L=3$,
with two bath qubits and the control qubit, using full state vectors,
exact layer unitaries, Born-rule measurements of the bath qubits, and
the expectation value of the control qubit for each trajectory. At
$\Gamma=0$ the trajectories are deterministic and reproduce the
closed-system OTOC to machine precision. The deterministic collision
channel, in which the bath qubit is traced out after each layer,
deviates from the Lindblad solution by $\approx0.024\,\Delta t$, with
first-order convergence, and the trajectory Monte Carlo with a
\emph{thermal reset} after each measurement agrees with it within the
statistical error. If, instead, the bath qubit is only measured and
enters the next layer in its post-measurement state, the result agrees
up to $t\approx1.5/J$ and then develops a systematic deviation of $-0.1$
to $-0.15$. The reason is that the measurement removes the memory of the
bath but does not rethermalize it, so that the bath populations drift
with the absorbed excitations and become correlated with the system
trajectory. The thermal reset costs little on hardware: it consists of
the same measurement followed by a classically conditioned $X$ gate,
supplemented at finite temperature by a random bit flip. On hardware
each run yields a single $\pm1$ outcome of variance $1-F^2$ rather than
an expectation value per trajectory, which increases the required
statistics by a factor of about $2$--$3$.

The thermal reset has an important algorithmic consequence, which
reduces the simulation to the sampling of Kraus operators: each bath
qubit takes part in exactly one collision and never needs to be stored.
Note that the thermal reset is far simpler than, and should not be
confused with, thermal-state preparation. It only requires each bath
qubit to be initialized in its single-qubit thermal state, a classical
mixture of $\ket{\downarrow}$ and $\ket{\uparrow}$ realized by a randomly
applied bit flip, whereas preparing the Gibbs state of an interacting
many-body Hamiltonian is in general a hard
problem~\cite{Alhambra23a,Chen25a}. The whole sequence of attaching,
coupling, measuring and resetting the bath qubit of bond $l$ is
equivalent to applying one of four $4\times4$ Kraus operators
$K_{m|b}=\langle m|e^{-i\Delta tH_{{\rm int},l}}|b\rangle$ to the two
system qubits of the bond, where the incoming thermal bit $b$ is sampled
with probability $p(b)$ and the outcome $m$ with the Born probability
$\|K_{m|b}\psi\|^2$. Only the $2^{L+1}$-dimensional state of the control
qubit and the system is propagated. The scan over
the probe site requires almost no additional cost. For a
$\sigma^z$-product state the initial controlled-$W$ gate acts as the
sign $\langle\psi|\sigma^z_{l_W}|\psi\rangle$ on the $|1\rangle_c$
branch, and the final controlled-$W^\dagger$ gate amounts to a readout
of $\sigma^z_{l_W}$, so that a single batch of forward, butterfly and
backward evolutions yields $F$ for \emph{all} probe sites at once. The
leak factor is obtained from the same forward evolution followed by a
second backward evolution without the butterfly. If the two backward
evolutions are driven by the \emph{same} random numbers, the difference
taken trajectory by trajectory is an unbiased estimator of $\dF$ whose
variance is far smaller than that of either term, because the two
evolutions share their collision outcomes until the perturbation by the
butterfly has propagated to the bond being measured. The method extends
to unequal rates with thermal statistics of the bath bits, and it agrees
with the two-propagator sector solver at $L=3$ to $10^{-5}$. Results for
$L=11$ are presented in Sec.~\ref{sec:spin:LightCones}
(Fig.~\ref{fig:spin_mc11}).

\subsection{Light Cones of Local Spins}
\label{sec:spin:LightCones}

\begin{figure*}
\centering
\includegraphics[width=\textwidth]{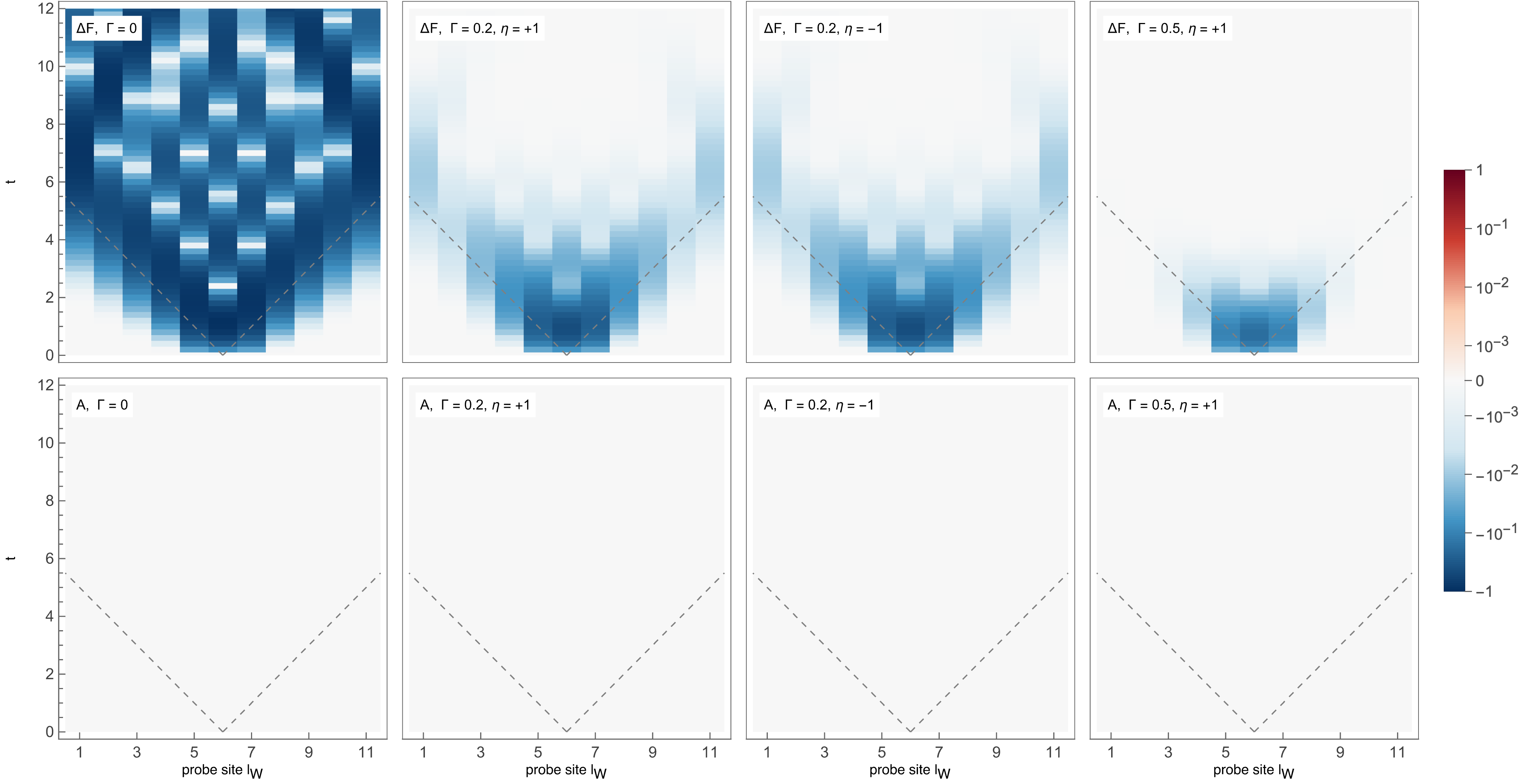}
\caption{Light cones of the local spin model at infinite temperature,
$\rho_0=\one/2^L$, with the butterfly at the center ($l_V=6$) of the
chain ($L=11$) and the probe scanned. Panels show the butterfly-induced
deviation $\dF$ (top row) and its antisymmetric part $A$ (bottom row) on
a signed-logarithmic color scale (linear below $10^{-3}$), for the
isolated chain, for $\Gamma=0.2$ with $\eta=\pm1$, and for $\Gamma=0.5$
with $\eta=+1$ (columns, left to right). Dashed lines mark the cone
$|x|=Jt$.}
\label{fig:spin_inft}
\end{figure*}

\begin{figure*}
\centering
\includegraphics[width=\textwidth]{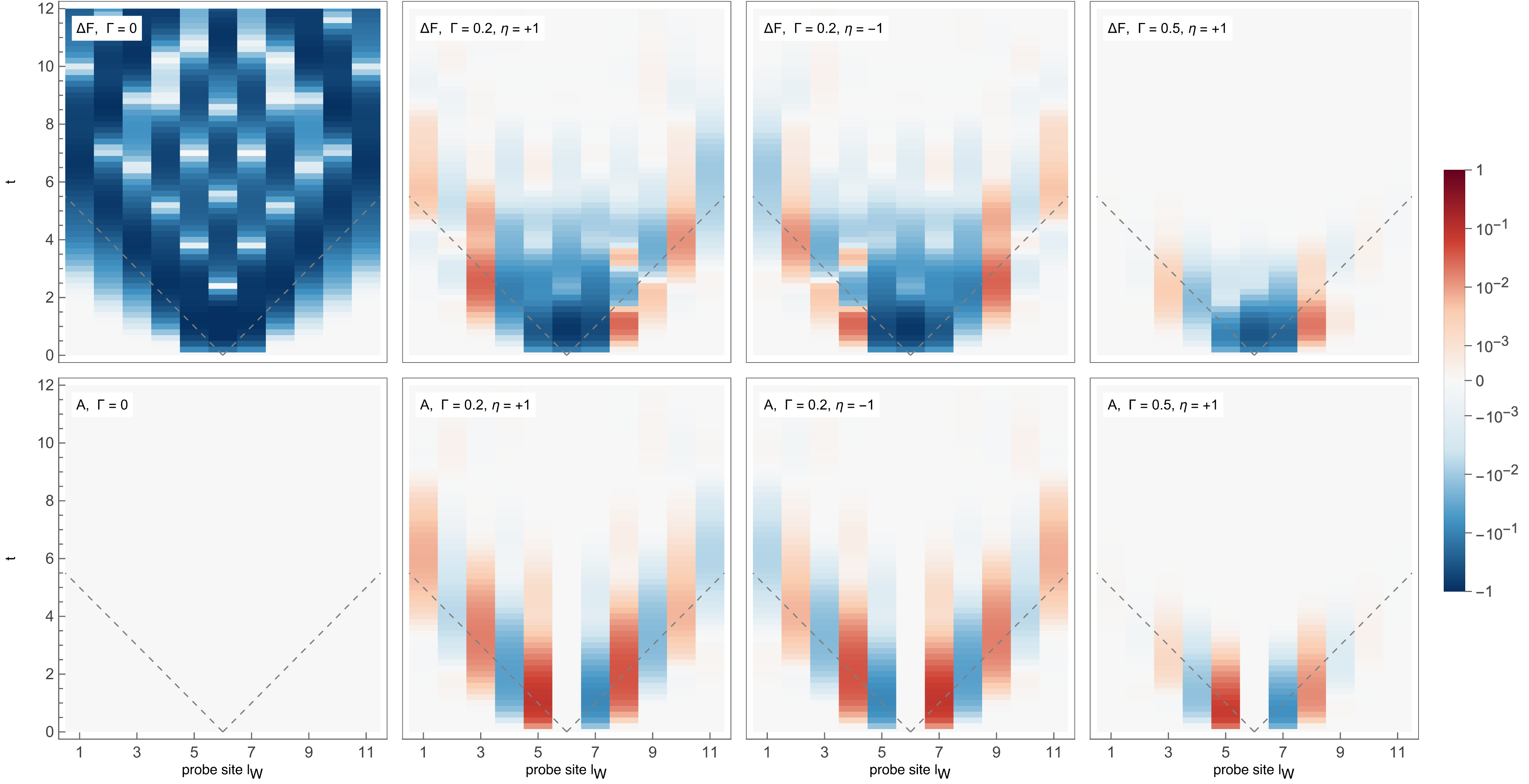}
\caption{Light cones of the local spin model for the N\'eel reference
state, with the butterfly at $l_V=6$ and the probe scanned. Rows and
columns as in Fig.~\ref{fig:spin_inft}.}
\label{fig:spin_neel}
\end{figure*}

\begin{figure*}
\centering
\includegraphics[width=\textwidth]{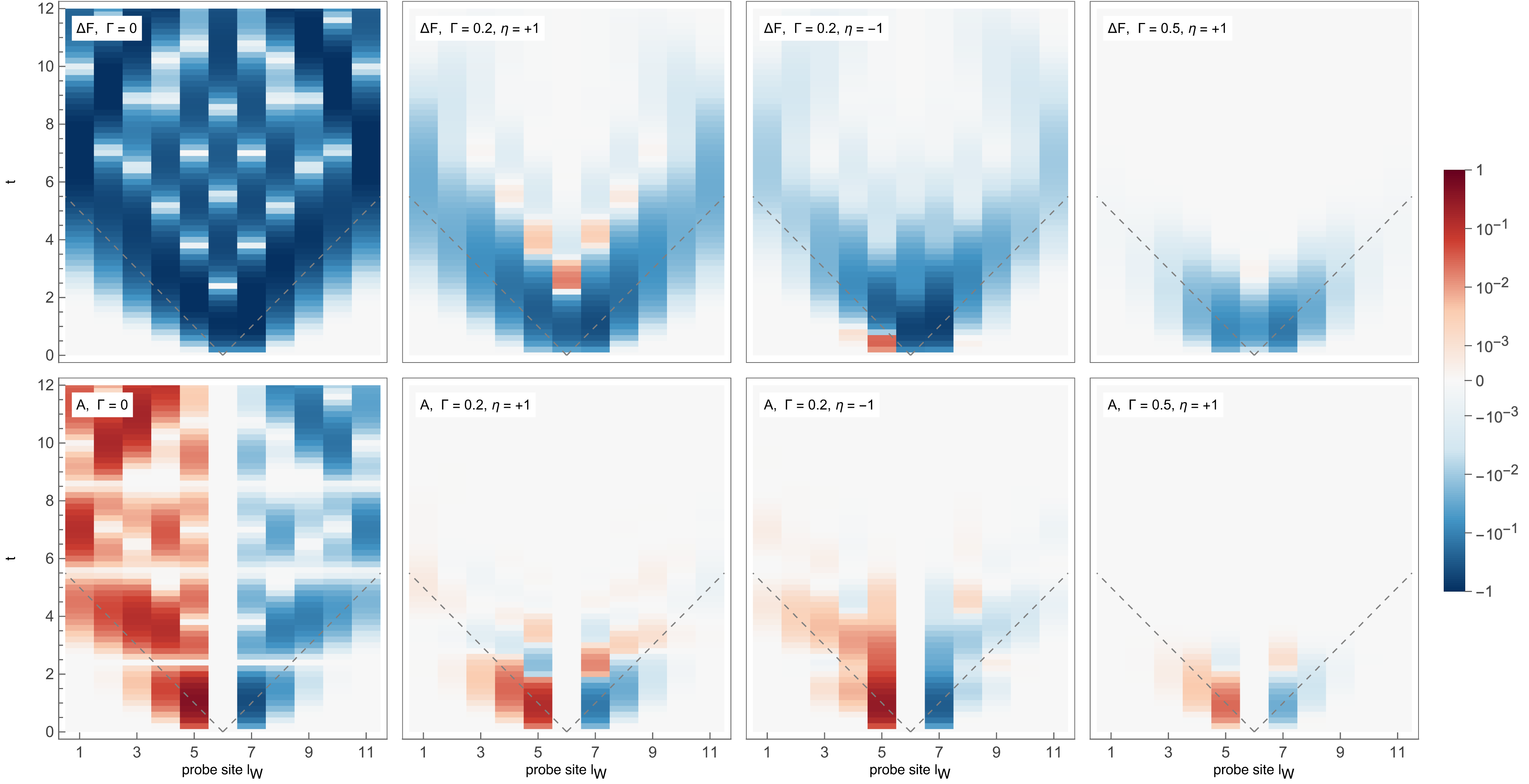}
\caption{Light cones of the local spin model for the single-wall
reference state $|\!\downarrow^{\otimes6}\uparrow^{\otimes5}\rangle$
(wall on bond 6--7), with the butterfly at $l_V=6$ and the probe scanned.
Rows and columns as in Fig.~\ref{fig:spin_inft}.}
\label{fig:spin_dw}
\end{figure*}

\begin{figure*}
\centering
\includegraphics[width=\textwidth]{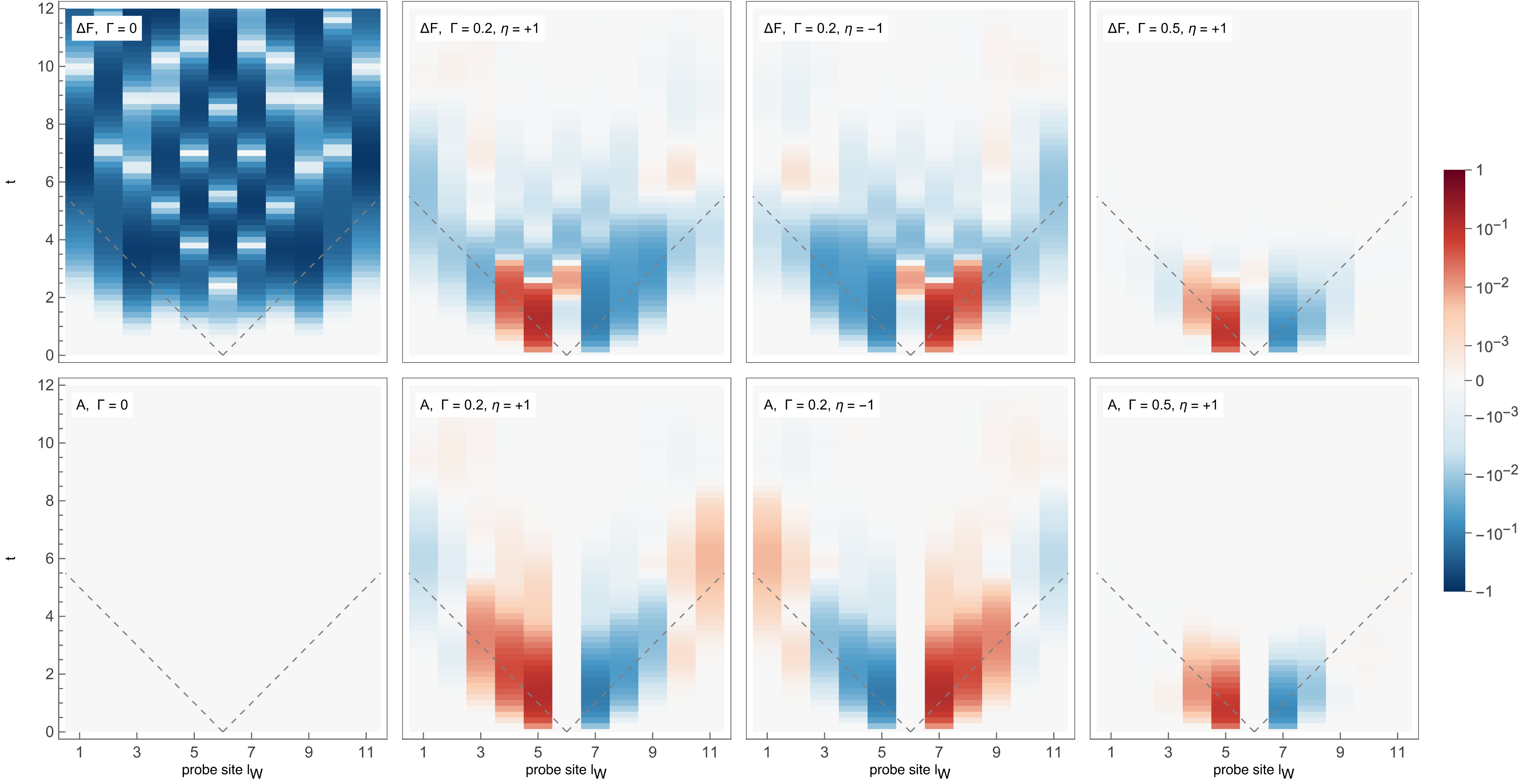}
\caption{Light cones of the local spin model for the symmetric two-wall
reference state
$|\!\downarrow\downarrow\downarrow\uparrow\uparrow\uparrow\uparrow\uparrow
\downarrow\downarrow\downarrow\rangle$, with the butterfly at $l_V=6$ and
the probe scanned. Rows and columns as in Fig.~\ref{fig:spin_inft}.}
\label{fig:spin_w2}
\end{figure*}

We compute the dissipative OTOC exactly for $L=11$, with the butterfly
$V=\sigma^z_6$ fixed at the center and the probe $W=\sigma^z_{l_W}$
scanned, for the infinite-temperature ensemble and for several pure
product states, at $J=1$, $\Gamma\in\{0,0.2,0.5\}$ and
$\eta\in\{0,\pm1\}$. The achiral bath, $\eta=0$, serves as a control;
for every statement below it plays the same role as the isolated chain.
The main exact result is a selection rule that decides whether the
chirality of the dissipation is visible in the OTOC and, if it is not,
in which observable it appears instead.

At infinite temperature the OTOC is exactly insensitive to the
chirality. Figure~\ref{fig:spin_inft} shows the maps of $\dF$ and $A$
for $\rho_0=\one/2^L$. The isolated chain displays the expected
symmetric ballistic cone with the butterfly velocity $v_B\simeq J$.
With dissipation, $F$
itself is suppressed almost uniformly by the leak
[$F_I(12/J,l_W{=}6)\sim10^{-4}$ at $\Gamma=0.2$ and $10^{-8}$ at
$\Gamma=0.5$], and the cone is no longer visible in a map of $F$ beyond
$t\sim2$--$3/J$, as found in Ref.~\cite{Zhang19a}. The difference $\dF$
removes the leak additively: it vanishes outside the cone up to
Lieb--Robinson tails and is negative inside by Eq.~\eqref{eq:dFinf}, and
on the logarithmic scale the cone and its oscillatory interior remain
visible down to $|\dF|\sim10^{-3}$, throughout the time range in which
the leak dominates $F$. The central observation is a null result with an
exact origin: the maps for $\eta=\pm1$ coincide to $4\times10^{-14}$,
each is reflection symmetric to $2\times10^{-14}$, and $A$ vanishes
identically. The chiral spectrum \eqref{eq:spectrum_spin} thus leaves
\emph{no} trace in the OTOC. The reason is the anti-chiral symmetry of
Sec.~\ref{sec:Symmetries}. The global spin flip
$X_\Pi=\prod_l\sigma^x_l$ satisfies
$X_\Pi A^-_lX_\Pi=A^+_l|_{\eta\to-\eta}$, so that at equal rates it maps
the set of jump operators onto itself with $\eta\to-\eta$, and the four
sign changes it produces in the two-$G$ form of $F$ cancel. Since $\one$
is invariant under the flip, $F(\eta)=F(-\eta)$ identically;
transposition provides a second, independent proof. The remaining
dependence on $\eta$ is even: the leak is slightly slower at $|\eta|=1$,
because one of the two chiral branches is undamped, but the OTOC does
not reveal which one. At $\eta=0$ the bond operators reduce to
$A^-_l\propto\sigma^-_l$ and leave site $L$ undamped, so that a probe at
that site leaks much more slowly and the scan of $F$ at $\eta=0$ is not
reflection symmetric. This boundary effect of the bond dissipators is
contained in $F_I$ and cancels in $\dF$. Since the maximally mixed state
is the steady state, the result also applies to the steady state: the
OTOC of the NESS is exactly even in $\eta$, and every signature of
chirality in this model at infinite temperature is transient.
Consistently, the late-time leak rate at $|\eta|=1$, $0.62$, is twice
the Liouvillian gap.

For the N\'eel state the chirality appears as a mirror-covariant
asymmetry. For a pure state the forward slot of Eq.~\eqref{eq:ZHC}
carries $\rho_0$ rather than a second copy of $G$, and the N\'eel state
$|\!\uparrow\downarrow\uparrow\cdots\uparrow\rangle$ is real and diagonal
but \emph{not} invariant under the spin flip, so neither anti-chiral
argument applies. Indeed (Fig.~\ref{fig:spin_neel}),
$\max|F_{+1}-F_{-1}|=0.17$, attained at $t=1/J$ one site away from the
butterfly, while the mirror covariance \eqref{eq:mirror} holds to
$3\times10^{-14}$: the maps for the two chiralities are mirror images of
each other, and each is asymmetric; accordingly $A$ is nonzero, with
opposite signs for $\eta=\pm1$. The asymmetry $\mathrm{asym}_d[\dF]$
is $-0.173$ at $d=1$ and $t=1/J$, $+0.083$ at $d=2$ and $t=2/J$, and
$-0.035$ at $d=3$ and $t=3.2/J$, whereas the $\eta=0$ control stays
below $10^{-3}$. The scrambling front thus reaches the probes on the two
sides with different amplitudes, and the sign of the difference
alternates with the distance. The magnetization of $\rho_\fwd(t)$ shows
the same chirality in a scalar quantity: the staggered order relaxes
symmetrically at $\eta=0$ but asymmetrically at $\eta=+1$, with an
accumulation reminiscent of the skin effect~\cite{Song19a}. Two
practical points should be noted. First, the leak decays \emph{faster}
than at infinite temperature, because
$F_I=s_{l_W}\Tr\{\Vf[\sigma^z_{l_W}]\,\rho_\fwd\}$ involves two decaying
factors. Second, $F_I$ is no longer positive: for some probe sites it
changes sign (first at $t=4.6/J$ and $l_W=9$ for $\Gamma=0.2$), beyond
which a correction based on dividing by $F_I$ fails. The difference
$\dF$ is not affected by this problem.

For a single domain wall the chirality appears in scalar observables as
well. The state $|\!\downarrow^{\otimes6}\uparrow^{\otimes5}\rangle$
breaks parity in addition to the spin flip (Fig.~\ref{fig:spin_dw}), so
that the maps of $\dF$ for $\eta=\pm1$ are neither equal nor mirror
images of each other; they differ by up to $0.30$. We find
$\max|F_{+1}-F_{-1}|=0.25$, and the leak factor at the central
probe differs between the two chiralities by the ratio
$F_I^{(-1)}/F_I^{(+1)}=2.3$ at $t=12/J$; that is, the correlator with
$V=\one$ already detects the sign of $\eta$. The mechanism is most
easily understood in the fermion language, with $\downarrow$ standing
for an occupied site. The filled left block melts by emitting particles
with momenta centered on $q=+\pi/2$, which for $\eta=-1$ is the dark mode
of the loss channel. Moreover, the filled block blocks the gain channel
by the exclusion principle, so that the melting is governed by the
chiral rate $\kappa_l(q)$ rather than by $\kappa_l+\kappa_g$. For
$\eta=-1$ the front is therefore protected from dissipation, whereas
for $\eta=+1$ the same modes are maximally damped and the profile is
flattened. The domain wall thus acts as a \emph{polarizer} for the
chirality. Two limitations should be kept in mind. The chirality of the
leak depends on the probe site (the central probe leaks faster at
$\eta=+1$, the edge probes at $\eta=-1$), so a probe in the bulk should
be used. Moreover, for a state that breaks parity, $A$ is \emph{not} an
observable of the chirality, since the $\eta=0$ control is as large as
the signal. Reversing the orientation of the wall exchanges
$\eta\leftrightarrow-\eta$ exactly.

The symmetric double wall isolates the chirality exactly. This state,
with walls on bonds 3--4 and 8--9 and the butterfly on the central site
(Fig.~\ref{fig:spin_w2}), removes the ambiguity of the single wall: the
state, the Hamiltonian and the geometry are even under parity, while
parity maps $\eta\to-\eta$. Three exact consequences follow, all
confirmed to machine precision: the map for the isolated chain is
reflection symmetric; $F_{+1}(t,l_W)=F_{-1}(t,L{+}1{-}l_W)$; and the same
relation holds for the leak factor of each probe, so that at the central
probe $F_I$ is exactly even in $\eta$. The reflection-odd part of a
single scan therefore equals the $\eta$-odd part at the same probe (an
identity verified to $10^{-14}$). It is an observable of the chirality
alone, which requires neither an $\eta=0$ control nor a correction by
$1/F_I$. The use of $\dF$ rather than $F$ is essential here. The
decomposition
$\mathrm{asym}_d[F]=\mathrm{asym}_d[F_I]+\mathrm{asym}_d[\dF]$ at
$\eta=+1$ gives $\mathrm{asym}[F]=-0.210=+0.043-0.253$ at $d=1$ and
$t=1.2/J$, where the chirality resides in the commutator term, but
$\mathrm{asym}[F]=+0.231=+0.236-0.004$ at $d=3$ and $t=1.4/J$, where it
resides in the leak \emph{only}. The sites $l_V\pm3$ are the wall sites,
and what a map of $F$ registers there is the chiral melting of the
walls, which is the blocking mechanism described above and is present
already for $V=\one$; it is not scrambling. In $\dF$ the chirality has a
definite sign at every distance: $-0.253$ at $d=1$ ($t=1.2/J$), $-0.099$
at $d=2$ and $-0.032$ at $d=3$, against an $\eta=0$ control of
$1.7\times10^{-4}$. A further feature is that $\dF$ is \emph{positive}
on one side of the butterfly ($+0.134$ at $t=1.2/J$ and $l_W=5$, where
$F=0.50$ exceeds $F_I=0.37$), with the mirror value $-0.120$ at $l_W=7$.
On the side where the loss channel is dark, inserting the butterfly thus
makes the echo of the probe \emph{larger} than without it. This is
impossible at infinite temperature by Eq.~\eqref{eq:dFinf}, and it does
not occur in the isolated chain. By Eq.~\eqref{eq:dFcomm} it is an
overlap between the evolved probe and the coherences at site $l_V$
created by the melting walls, with a sign set by the chirality. At
$\Gamma=0.2$ about $90\%$ of the map is reflection odd at $d=1$; for
this state the chirality constitutes almost the entire signal. This is
the configuration that we propose for an experiment.

The selection rule is confirmed by one further case. The $x$-polarized
two-wall state $|\!-\!-\!-\!+\!+\!+\!+\!+\!-\!-\!-\rangle$, with
$\langle H_0\rangle=-1.5J$, is invariant under the spin flip despite its
spatial structure, and the OTOC is again exactly even in $\eta$ (to
$3\times10^{-14}$). The chirality appears instead in the channel that is
odd under $X_\Pi$, namely the magnetization, which is \emph{generated}
from zero by the chiral dissipation with an exactly antisymmetric
profile, $m_l=-m_{L+1-l}$ and $m^{(+1)}_l=-m^{(-1)}_l$. Taken together,
these cases establish the rule of Sec.~\ref{sec:Symmetries}
(Table~\ref{tab:sym}). If $\rho_0$ is invariant under an anti-chiral
symmetry (the global spin flip, or transposition for the two-$G$
structure at infinite temperature), the $\sigma^z$--$\sigma^z$
dissipative OTOC is exactly even in $\eta$; the chirality is then
invisible in it and survives only in observables that are odd under the
symmetry. If $\rho_0$ breaks all anti-chiral symmetries, the chirality
appears as a mirror-covariant asymmetry of the probe scan when parity is
intact, and in addition in scalar quantities such as the leak factor at
the central probe when parity is broken as well.

\begin{table}[tbp]
\centering
\small
\setlength{\tabcolsep}{4pt}
\begin{tabular}{lccp{3.1cm}}
\toprule
reference state & $X_\Pi$ & parity & chirality visible in\\
\midrule
$\one/2^L$ & yes & yes & nothing ($F$ exactly $\eta$-even)\\
N\'eel & no & yes & light-cone asymmetry (mirror covariant)\\
single $z$ wall & no & no & light cone \emph{and} scalars (central-probe $F_I$ ratio $2.3$)\\
symmetric $z$ walls & no & yes & $A=\eta$-odd part of $\dF$ ($10^{-14}$; $>10\sigma$ by Monte Carlo)\\
$x$-basis walls & yes & yes & only $\eta$-odd observables (generated $\langle\sigma^z\rangle$)\\
\bottomrule
\end{tabular}
\caption{Visibility of the dissipative chirality for five reference
states (columns 2--3: invariance under the global spin flip and under
parity), butterfly fixed at the center, probe scanned,
infinite-temperature bath. All exact statements verified at the
$10^{-14}$--$10^{-16}$ level.}
\label{tab:sym}
\end{table}

\begin{figure}[tbp]
\centering
\includegraphics[width=\columnwidth,height=0.72\textheight,keepaspectratio]{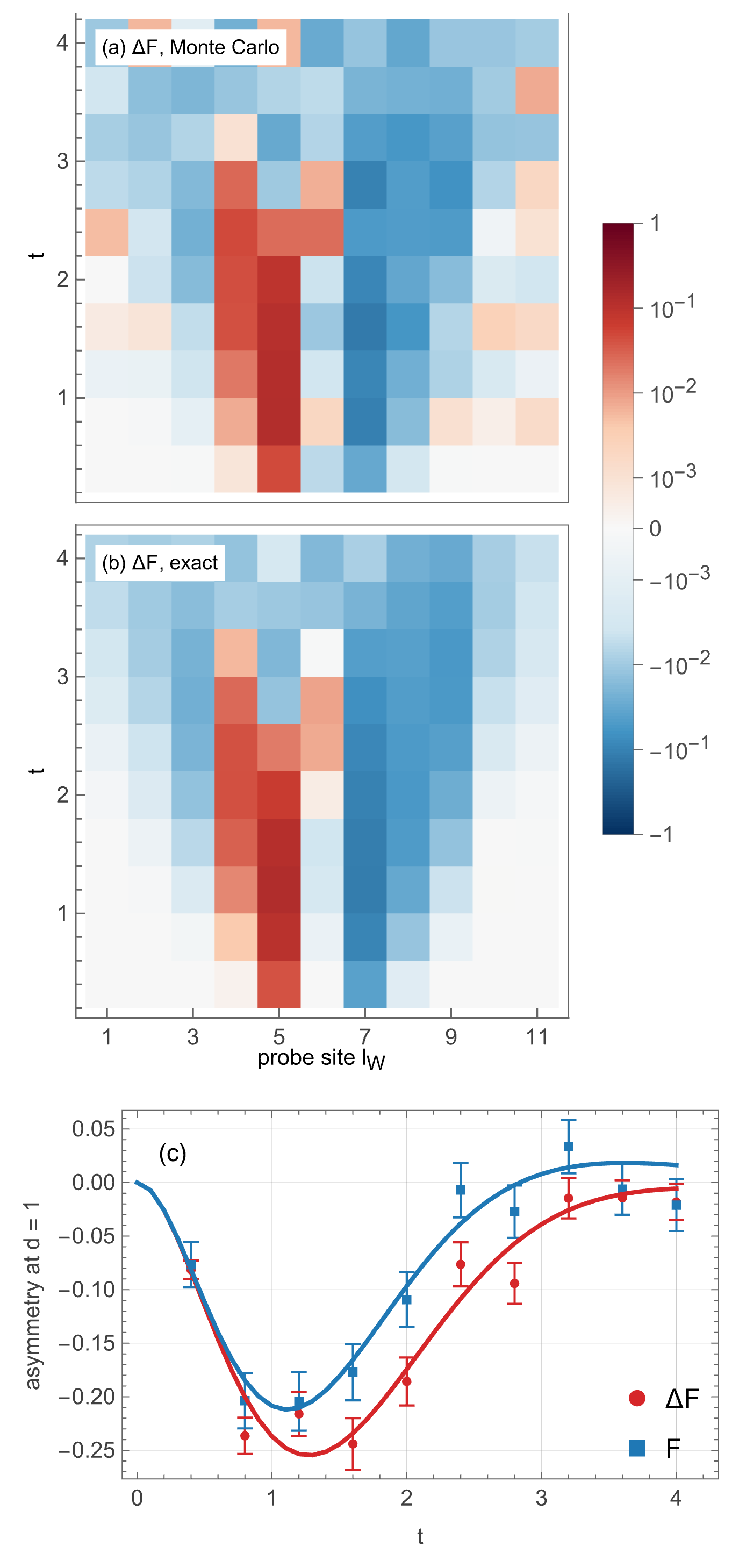}
\caption{Circuit Monte Carlo with thermal reset for the symmetric
two-wall reference state ($L=11$, $\Gamma=0.2$, $\eta=+1$,
$\Delta t=0.1/J$, butterfly at $l_V=6$, $1024$ trajectories per time
slice). (a) Monte Carlo map of $\dF$. (b) Exact Lindblad result on the
same grid. (c) Asymmetry at distance $d=1$ about the butterfly from the
same trajectories, for $\dF$ and for $F$, compared with the exact
curves.}
\label{fig:spin_mc11}
\end{figure}

Finally, the hardware-efficient circuit reproduces these maps.
Figure~\ref{fig:spin_mc11} shows the circuit Monte Carlo of
Sec.~\ref{sec:spin:Methods} for the symmetric two-wall state at
$\Gamma=0.2$ and $\eta=+1$, with $10$ time slices $t=0.4$--$4/J$, $11$
probe sites each, and $1024$ trajectories per slice. The Monte Carlo
data agree with the exact results for $F$, $F_I$ and $\dF$ within the
statistical errors, with $\chi^2/\mathrm{dof}=1.07$, $0.95$ and $1.19$
over $110$ points. The estimator with common random numbers reduces the
error of $\dF$ to $\sigma\approx0.008$, below the $\sigma\approx0.02$ of
$F$ itself; at this accuracy the $O(\Delta t)$ systematic error of the
collision model and the Trotter splitting is no longer negligible, which
may account for the slightly larger $\chi^2$ of $\dF$. Most importantly,
the chiral asymmetry is resolved directly by the trajectory data
[Fig.~\ref{fig:spin_mc11}(c)]: $\dF(l_V{+}1)-\dF(l_V{-}1)=-0.216\pm0.021$
at $t=1.2/J$, where the exact asymmetry is largest ($10.4\sigma$; the
exact value is $-0.253$), $-0.186\pm0.022$ at $t=2/J$ ($8.3\sigma$) and
$-0.076\pm0.021$ at $t=2.4/J$ ($3.7\sigma$), and it is consistent with
zero from $t\approx3.2/J$ on. It exceeds $10\sigma$ in all four time
slices $t=0.8$--$1.6/J$, with the largest significance, $14.0\sigma$, at
$t=0.8/J$ ($-0.237\pm0.017$; the exact value is $-0.200$). The same
trajectories give at most $7.9\sigma$ for the asymmetry of $F$. Detecting the dissipative chirality on hardware
by this protocol thus requires only a few thousand shots per time slice,
with $F$ and $F_I$ obtained in the same run.

\subsection{Liouvillian Skin Effects of Local Spins}
\label{sec:spin:LongTime}

At infinite bath temperature the steady state is $\one/2^L$ and its OTOC
is exactly even in $\eta$; the Liouvillian skin effect of this model can
therefore be observed only at finite bath temperature. We now take
$\Gamma_->\Gamma_+$, keeping $\Gamma=0.2$, so that the imaginary DM term,
and with it the Hatano--Nelson asymmetry and the loss chirality, is
unchanged, while the imaginary Zeeman term $\delta$ of
Eq.~\eqref{eq:heff_spin} becomes nonzero. The reference case is
$\Gamma_-/\Gamma_+=3$ ($\beta\omega=\ln3$); the colder bath
$\Gamma_-/\Gamma_+=19$ is used to check the trend. Two structural changes
occur. The identity \eqref{eq:identity} no longer holds, so both
propagators must be evolved. Moreover, the maximally mixed state is no
longer
stationary; the natural reference state, which requires no preparation, is
the thermal product state $\rho_\beta$ of Eq.~\eqref{eq:rhobeta}.

\begin{figure}[tbp]
\centering
\includegraphics[width=\columnwidth,height=0.72\textheight,keepaspectratio]{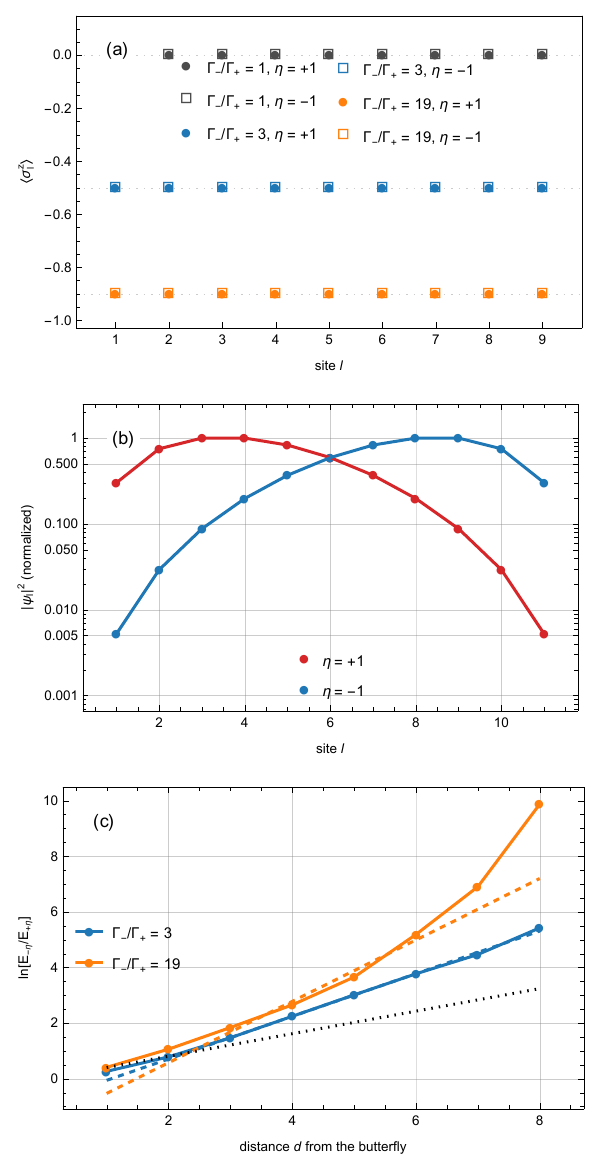}
\caption{Local spin model at finite bath temperature ($\Gamma=0.2$,
$J=1$). (a) Steady-state magnetization profile at $L=9$ for
$\Gamma_-/\Gamma_+=1,3,19$ and $\eta=\pm1$, obtained by relaxing
$\one/2^L$ under $\Lf$; dotted lines mark $-\tanh(\beta\omega/2)$.
(b) Profile of the slowest right eigenmode of $\Heff$ in the
single-fermion sector (one $\downarrow$ spin) at $L=11$ and
$\Gamma_-/\Gamma_+=3$, normalized to its maximum and compared with
$e^{\mp2\kappa_{\rm NH}(l-1)}\sin^2[\pi l/(L{+}1)]$. (c) Logarithm of
the ratio of the arm envelopes, $\ln[E_{-\eta}(d)/E_{+\eta}(d)]$ of
Eq.~\eqref{eq:kcone}, versus the probe distance $d$ for the steady-state
OTOC with the butterfly at the edge ($l_V=1$, $L=9$), at
$\Gamma_-/\Gamma_+=3$ and $19$; the dashed lines are linear fits over
$d=3$--$6$ and the dotted line is $2\kappa_{\rm NH}d$.}
\label{fig:spin_finiteT}
\end{figure}

We begin with the postselected length. The chirality of $\Heff$ belongs
to the no-jump evolution. In the sector of a single Jordan--Wigner
fermion, that is, one $\downarrow$ spin in the $\uparrow$-polarized
background, its slowest right eigenmode is the gauge-transformed
standing wave
$|\psi_l|^2\propto e^{\mp2\kappa_{\rm NH}(l-1)}\sin^2[\pi l/(L{+}1)]$
with $\kappa_{\rm NH}=1/\lNH$ given by Eq.~\eqref{eq:kappa} with
$\gamma\to\Gamma$, that is, $\lNH=4.9$ sites at $\Gamma=0.2$ and
$\eta=1$, independent of $\delta$ [Fig.~\ref{fig:spin_finiteT}(b)]. This
is the first of three lengths: it is well defined and temperature
independent, but unobservable without postselection.

At finite temperature the OTOC of the steady state is chiral. We apply
the echo protocol to $\rho_\beta$ itself. The transposition argument no
longer applies, because it exchanges the two propagators, which are now
different, and the spin flip maps $\Gamma_+\leftrightarrow\Gamma_-$;
only parity remains. The scans for $\pm\eta$ are still exact mirror
images of each other (to $2\times10^{-15}$ at $L=9$) but are no longer
equal: $\max|F_{+1}-F_{-1}|=0.16$ at $\Gamma_-/\Gamma_+=3$, with
$\mathrm{asym}_1[\dF]=0.07$ already at $t=1/J$. The insensitivity found
at infinite temperature is therefore a property of the
infinite-temperature \emph{bath}, not of steady states in general: at
any finite bath temperature the echo protocol applied to the achiral
thermal steady state, without any state preparation, resolves the
nonreciprocity of the environment. The mechanism is given by
Eq.~\eqref{eq:sumrule_spin}. The partial polarization suppresses the
gain channel by the factor $e^{\beta\omega}$ relative to the loss
channel, so that the chiral parts of $\kappa_l$ and $\kappa_g$ no longer
cancel and the two arms of the cone leak at different rates. Such a
chirality at the level of the rates cannot arise in the fermionic model
with equal rates (Sec.~\ref{sec:Anatomy}). In the $L=11$ map with the
butterfly at the center, $\dF$ is negative throughout, and the chirality
appears as an asymmetry in the \emph{amplitude} of the two arms
[$\dF(l_V{-}1)-\dF(l_V{+}1)=-0.068$ at $t=1/J$ for $\eta=+1$]. In
addition, the arm that propagates against the skin direction is
\emph{amplified at the edge}: $\dF(l_W{=}1)$ grows from $-0.008$ at
$t=4/J$ to $-0.042$ at $t=6/J$, when the bulk of the cone has decayed to
$3\times10^{-4}$, while the opposite edge remains at $-0.0014$. The
perturbation created at the center thus accumulates at the edge where
the slowest Liouvillian modes are localized.

The transient cone defines a second length. With the butterfly at the
edge, $l_V=1$, the single arm of the cone propagates \emph{along} the
dark direction for one sign of $\eta$ and \emph{against} it for the
other, and the envelopes
$E_{\pm\eta}(d)=\max_t|\dF_{\pm\eta}(t,l_V{+}d)|$ of the arm obey
[Fig.~\ref{fig:spin_finiteT}(c)]
\begin{equation}
  \ln\frac{E_{-\eta}(d)}{E_{+\eta}(d)}\simeq2\kappa_{\rm cone}\,d,
  \qquad\lcone=1/\kappa_{\rm cone},
  \label{eq:kcone}
\end{equation}
with $2\kappa_{\rm cone}=0.77$ at $\Gamma_-/\Gamma_+=3$ (increments per
site $0.52,0.69,0.77,0.77,0.76,0.69$ for $d=2$--$7$ at $L=9$), that is,
$\lcone\approx2.6$ sites. At equal rates the ratio is identically one,
and for the colder bath the increments are larger
($2\kappa_{\rm cone}\approx1.1$, not converged within $L=9$). Thus
$\lcone$ is a second, \emph{temperature-dependent} length: it is
infinite at $\beta=0$, equals $2.6$ sites at $\beta\omega=\ln3$, and
decreases further at lower temperature. It differs from $\lNH$, which is
temperature independent and about twice as long. A simple estimate from
the differential damping of the two arms, $4\delta\teta/J=0.40$ per
site, underestimates the slope by a factor of two. A theory of
$\kappa_{\rm cone}$, in which the forward and backward evolutions are
chiral under different generators, remains to be developed.

\begin{figure}[tbp]
\centering
\includegraphics[width=\columnwidth,height=0.72\textheight,keepaspectratio]{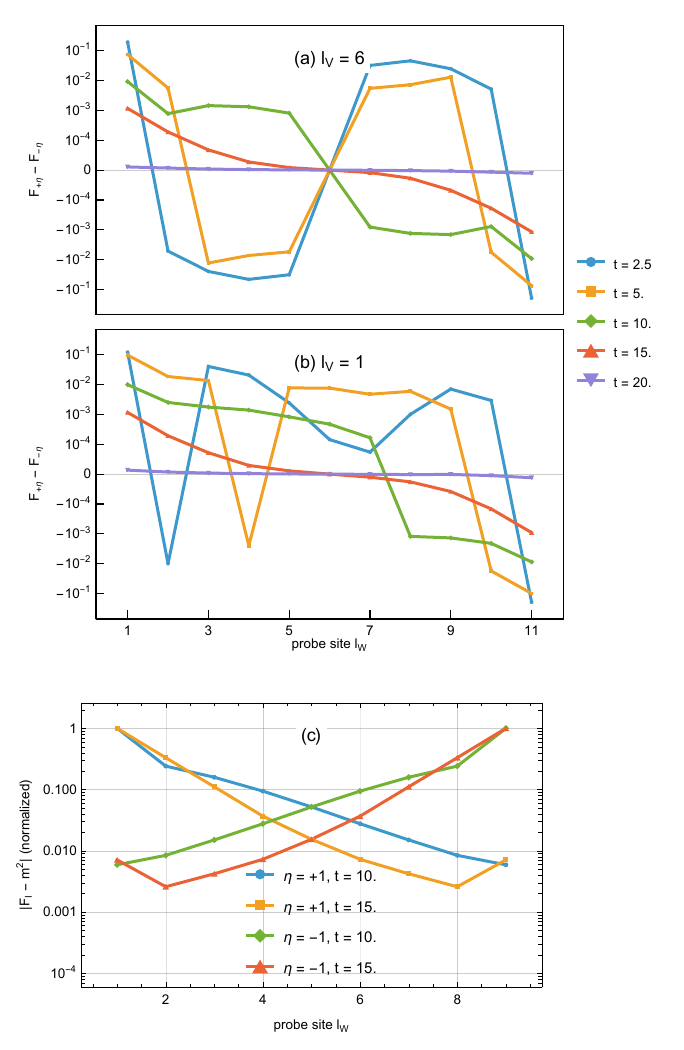}
\caption{Late-time behavior of the steady-state OTOC of the local spin
model ($\Gamma=0.2$, $\Gamma_-/\Gamma_+=3$, $|\eta|=1$). (a),~(b) The
$\eta$-odd part $F_{+\eta}-F_{-\eta}$ versus the probe site at several
times, for $L=11$ with the butterfly at the center (a) and at the edge
(b). (c) Late-time leak factor $|F_I-m^2|$ versus the probe site for
$L=9$ and $\eta=\pm1$ at $t=10/J$ and $15/J$, normalized at the edge.}
\label{fig:spin_finiteTlate}
\end{figure}

At late times the Liouvillian skin effect appears in the leak factor,
which defines the third length. The transient cone is confined to
$\Gamma t\lesssim2$. Afterwards both $F$ and $F_I$ relax to
$m^2=\tanh^2(\beta\omega/2)=0.25$, and the $\eta$-odd part of $F$
is dominated by that of the leak factor, $F_I(+\eta)-F_I(-\eta)$, which
does not depend on the butterfly (Fig.~\ref{fig:spin_finiteTlate}). The
butterfly does not drop out, however. Relative to the $\eta$-odd part
of the leak factor, the maximum over the probe sites of the $\eta$-odd
part of $\dF$ amounts to $18\%$, $0.4\%$ and $21\%$ at $\Gamma t=2$,
$3$ and $4$ for the butterfly at the center, and to $13\%$, $23\%$ and
$20\%$ for the butterfly at the edge. The contribution of the butterfly
thus decays at the same rate as that of the leak factor, with a relative
weight that oscillates in time and is typically about one fifth. At
earlier times, $\Gamma t\lesssim1$, the odd
part is dominated by the contribution of the butterfly, superposed on
peaks of $\pm0.19$ at the two edges (at $\Gamma t=0.5$) that belong to
the leak factor. At
late times the remaining profile is localized at the edges, positive at
the left edge and negative at the right edge for $+\eta$. It is
mainly that of the leak of the probe itself, which we therefore analyze
directly through $F_I$. The leak
profile is exponential and localized at the edge: the normalized values
are $1,0.33,0.11,0.037,0.015$ on sites $1$--$5$ at $L=9$ and $t=15/J$,
corresponding to $\kappa\approx1.0$ at the edge and $\kappa\approx0.55$
further inside ($\lleak\approx1$--$1.8$ sites). Its amplitude decays at
the rate $0.54$--$0.60$, close to twice the Liouvillian gap
($2\times0.3227=0.645$), as expected for a product of two decaying
factors. The mechanism is the Liouvillian skin
effect~\cite{Haga21a,Song19a}. The slowest relaxation modes of $\Lf$
form a complex pair $\lambda_1=-0.3227\pm0.2639i$, whose right
eigenvector is localized at the right edge for $+\eta$ and whose
\emph{left} eigenvector is localized at the left edge ($\kappa=0.28$);
for $\Lb^\dagger$ the corresponding left eigenvector has $\kappa=0.76$.
The late-time leak factor,
\begin{equation}
  F_I(t,l_W)-m^2\propto e^{2\Re\lambda_1t}\,
  \langle l^\fwd_1|\rho_\beta\sigma^z_{l_W}\rangle\,
  \langle l^\bwd_1|\sigma^z_{l_W}\rangle,
  \label{eq:lateleak}
\end{equation}
is the product of the profiles of the two left eigenvectors. It
reproduces sites $1$--$4$ ($1,0.334,0.131,0.037$ against
$1,0.33,0.11,0.037$) and has a node at site $5$, which is a consequence
of the complex pair. The echo protocol thus detects the Liouvillian skin
effect \emph{without postselection}, through the leak factor of each
probe, which is the quantity isolated by scanning the probe and by
separating $F$ into $F_I$ and $\dF$. At $\Gamma=0.2$, $|\eta|=1$ and
$\Gamma_-/\Gamma_+=3$ the three lengths are ordered as
\begin{equation}
  \underbrace{\lNH=4.9}_{\text{postselected}}
  \;>\;\underbrace{\lcone\approx2.6}_{\text{transient cone}}
  \;>\;\underbrace{\lleak\approx1\text{--}1.8}_{\text{late-time skin}}.
  \label{eq:hierarchy}
\end{equation}
They characterize three distinct objects: the first is independent of
$\beta$, the second is defined for $\Gamma t\lesssim2$, and the third
for $\Gamma t\gtrsim2$. At equal rates, where the OTOC is even in
$\eta$, the last two disappear: $\lcone$ is infinite and the late-time
skin profile is absent, while $\lNH$ remains but is unobservable without
postselection. Whether
$\lleak$ is related to $\lNH/2$ once the quantum jumps are taken into
account, and how the three lengths depend on $\beta$, are open
questions.

\section{Summary}
\label{sec:Summary}

\begin{table*}[tbp]
\centering
\small
\begin{tabular}{p{3.4cm}p{6.2cm}@{\qquad}p{6.2cm}}
\toprule
 & Hatano--Nelson Lindbladian (fermions) & local XX spin model\\
\midrule
jump operators & $A_l\propto c_l+i\eta c_{l+1}$, $B_l\propto\cdag_l+i\eta\cdag_{l+1}$ (same chirality) & $A^-_l\propto S^-_l+i\eta S^-_{l+1}$, $A^+_l$ (opposite chirality)\\
effective Hamiltonian $\Heff$ & HN chain, $J_{L/R}=\tfrac12(J\mp\gamma\teta)$ & same HN chain $+$ imaginary Zeeman at $\delta\neq0$\\
$\lNH$ & Eq.~\eqref{eq:kappa}, self-dual, postselected & same; $\beta$-independent\\
damping matrix & exactly diagonal, $\eta$-independent in the bulk & $\kappa_l+\kappa_g=2\Gamma-2\delta\teta\sin q$\\
steady state & chiral current, skin profile, $\lNESS=1/[2\,\mathrm{arcsinh}(\gamma/J)]$ & uniform $e^{-\beta\omega S^z_{\rm tot}}/Z$, achiral for all $\eta$\\
$\Vb^\dagger=\Vf$? & no & yes at $\Gamma_+=\Gamma_-$\\
chirality in $\dF$ & sign reversal between arms; $A=\dF_{\rm src}$ universal up to $s_l$ & selection rule: $\eta$-even for anti-chiral-invariant $\rho_0$; mirror-covariant asymmetry otherwise; amplitude asymmetry at $\delta\neq0$\\
reset strategy / leak & $D=4s\delta$ in NESS $\Rightarrow\lOTOC=\lNESS$ & late-time $\eta$-odd leak $\Rightarrow\lleak$ (Liouvillian skin mode)\\
exact method & Gaussian solution, Stinespring dilation & $\binom{2L}{L}$ sector evolution\\
trajectory Monte Carlo & shared record $+$ reweight (Gaussian) & thermal-reset Kraus sampling\\
\bottomrule
\end{tabular}
\caption{Comparison of the two models (HN: Hatano--Nelson).}
\label{tab:compare}
\end{table*}

We have investigated how the Liouvillian skin effect appears in the
dissipative OTOC and how it can be measured with quantum circuits, for
two models that share the same effective non-Hermitian Hamiltonian, the
Hatano--Nelson chain, and hence the same non-Hermitian skin depth
$\lNH$. Table~\ref{tab:compare} compares the two models. Their full
Lindblad dynamics differ qualitatively. In the fermionic model the gain
channel has the same chirality as the loss channel, and the steady state
carries a chiral current and a skin profile. In the spin model the gain
operators are the adjoints of the loss operators, and the steady state is
a uniform thermal product state. The dissipative OTOC distinguishes the
two cases, and three conclusions hold for both models.
\begin{enumerate}[leftmargin=*]
\item \emph{The postselected and the full Lindblad dynamics are governed
by different lengths.} The non-Hermitian skin depth $\lNH$ belongs to the
no-jump evolution and is accessible only at an exponential cost. The
observables of the full Lindblad dynamics---the steady-state profile,
the difference between the two reset strategies, and the late-time leak
factor---are governed by other lengths, which the Liouvillian sets
through its damping and its slowest relaxation modes. For the fermionic
model the skin depth obtained from the OTOC with the two reset
strategies equals the Liouvillian skin depth,
$\lOTOC=\lNESS=1/[2\,\mathrm{arcsinh}(\gamma/J)]$, which is independent
of $\eta$ and differs from the self-dual $\lNH$. The same distinction
separates the dilation in which the two interferometer branches share
their bath registers, which gives the dissipative OTOC, from the one
with separate registers for each branch, which gives the postselected
quantity.
\item \emph{The leak factor carries physical information.} Subtracting
$F_I$ from the OTOC, rather than dividing by it, separates the chirality
of the scrambling, contained in $\dF$, from the chirality of the
relaxation of the probe, contained in $F_I$. In the fermionic model the
difference between the two reset strategies resides entirely in $F_I$
and measures the Liouvillian skin profile. In the spin model at finite
temperature the $\eta$-odd part of the late-time leak factor has the
profile of the skin-localized relaxation modes. In both cases the
steady-state skin effect is obtained from the leak factor without
postselection.
\item \emph{Symmetry determines the visibility of the chirality.} Mirror
covariance makes the reflection-odd part of a single probe scan an
observable of the chirality alone, whereas anti-chiral symmetries of the
reference state can remove the chirality from the OTOC altogether. In
the fermionic model the jump set is not self-adjoint, no anti-chiral
symmetry exists even at infinite temperature, and the chiral component
is independent of the reference state up to the sign factor $s_l$. In
the spin model the chirality is visible only for reference states that
break the global spin flip; it is isolated exactly by the reflection-odd
part of $\dF$ for the symmetric two-wall state, which we propose for an
experiment.
\end{enumerate}

Concerning the methods, the fermionic model is solved exactly, at a cost
polynomial in $L$, by the Gaussian solution of the Lindblad equation. The
solution is verified by the Stinespring dilation and sampled by a
quantum-trajectory estimator within the fermionic Gaussian formalism, in
which the two interferometer branches share one measurement record of
the bath and are reweighted. The spin model is solved exactly within its
$m=0$ sector and sampled by a Monte Carlo simulation of the circuit with
thermal reset, which resolves the chirality at more than $10\sigma$ with
about
$10^3$ trajectories per time slice. The measure-and-reset circuits are
at the same time the protocols proposed for quantum hardware; they
require mid-circuit measurement and reset of $O(L)$ bath qubits and no
postselection.

Our work is related to several lines of research. OTOCs have been
studied for non-Hermitian Hamiltonians~\cite{Zhai20a,Zhao22a}, which
corresponds to the postselected quantity. The space-time duality of
Ref.~\cite{Zhou22a} maps the OTOC of a unitary circuit to a non-Hermitian
circuit with a boundary perturbation, whereas we consider the
Liouvillian skin effect and the dissipative OTOC of one and the same
open system. The Liouvillian skin effect and its relaxation time scales
have been investigated in
Refs.~\cite{Haga21a,Yang22a,Lee23a,Zhou22b,Kawabata23a,Feng24a}, and
chiral damping~\cite{Song19a} is the analogue, for a two-point function,
of the asymmetry found here for a four-point function. The robustness of
scrambling measurements against decoherence was analyzed in
Ref.~\cite{Swingle18a}.

Several questions remain open. A theory of the transient cone length
$\lcone$ of the spin model, and of the temperature dependence of its
three lengths, is lacking. The chiral component $A/s_l$ of the fermionic
model, being independent of the reference state, may allow the
parameters $\teta$ and $\gamma$ to be extracted from OTOC data without
knowledge of the reference state. Finally, in the presence of
interactions the Gaussian methods no longer apply, but the sector
solver, the circuit Monte Carlo and the symmetry arguments remain valid.

\begin{acknowledgments}
This work was supported by the National Research Foundation of Korea
(Grants No.~RS-2023-NR068116, No.~2023R1A2C1005588,
No.~RS-2024-00432563, and No.~RS-2024-00404854) and by the Korea
Institute for Advanced Study (KIAS) through its Visiting Scholar
program. We thank Yinan Fang for useful discussions.
\end{acknowledgments}

\appendix

\section{System--Bath Model of the Hatano--Nelson Lindbladian}
\label{app:HNbath}

\subsection{Born--Markov Derivation in the Wideband Limit}

We follow the standard weak-coupling procedure~\cite{Breuer02a,Daley14a}, but
with one departure from the textbook route: the final step is a
\emph{wideband (flat spectral density) limit}, \emph{not} a secular
approximation. Iterating the von Neumann equation once in the
interaction picture and tracing over the bath,
\begin{equation}
  \dot{\tilde\rho}_{\sys}(t)
  =-\int_0^t\!ds\,\Tr_{\bath}\big[\tilde V(t),[\tilde V(s),\tilde\rho_{\rm tot}(s)]\big].
  \label{eq:double_commutator}
\end{equation}
The Born factorization $\tilde\rho_{\rm tot}(s)\approx\tilde\rho_{\sys}(s)
\otimes\rho_{\bath}$ and the Markov replacement $\tilde\rho_{\sys}(s)\to
\tilde\rho_{\sys}(t)$ are standard. Collecting the bath modes of each
channel into $R_{a,l}=\sum_kg^{(a)}_ka_{l,k}$ and $S_{b,l}=\sum_k
g^{(b)}_kb^\dagger_{l,k}$, the Gaussian baths contribute exactly two
nonvanishing correlation functions,
\begin{align}
  \big\langle\tilde R_{a,l}(t)\tilde R^\dagger_{a,l'}(s)\big\rangle
  &=\delta_{ll'}C_a(t-s),
  \label{eq:Ca_def}\\
  \big\langle\tilde S_{b,l}(t)\tilde S^\dagger_{b,l'}(s)\big\rangle
  &=\delta_{ll'}C_b(t-s),
  \label{eq:Cb_def}
\end{align}
with
\begin{align}
  C_a(\tau)&=\int\!d\omega\,J_a(\omega)\big[1-f_a(\omega)\big]e^{-i\omega\tau},
  \label{eq:Ca}\\
  C_b(\tau)&=\int\!d\omega\,J_b(\omega)f_b(\omega)e^{+i\omega\tau},
  \label{eq:Cb}
\end{align}
and $J_\alpha(\omega)=\sum_k|g^{(\alpha)}_k|^2\delta(\omega-\omega^{(\alpha)}_k)$.
(Each reservoir conserves its own particle number, so anomalous averages
vanish; cross-bond and cross-channel averages vanish because the
reservoirs are independent; the conjugate orderings are proportional to
$f_a=0$ and $1-f_b=0$. Although $V$ is built from fermionic operators,
each term is a product of one odd bath and one odd system operator, so
no extra Jordan--Wigner signs arise.) Bond-diagonality of the final
dissipator is therefore \emph{exact}: no ``spatial secular'' argument is
needed. The only delicate structure is in frequency. The second-order
equation reads
\begin{widetext}
\begin{align}
  \dot{\tilde\rho}_{\sys}(t)=-\sum_l\int_0^t\!ds\,\Big\{
  &C_a(t-s)\big[\tilde A^\dagger_l(t)\tilde A_l(s)\tilde\rho_{\sys}(s)
    -\tilde A_l(s)\tilde\rho_{\sys}(s)\tilde A^\dagger_l(t)\big]\nonumber\\
  +\,&C_b(t-s)\big[\tilde B^\dagger_l(t)\tilde B_l(s)\tilde\rho_{\sys}(s)
    -\tilde B_l(s)\tilde\rho_{\sys}(s)\tilde B^\dagger_l(t)\big]
  +\mathrm{h.c.}\Big\}.
  \label{eq:redfield}
\end{align}
\end{widetext}
Every system operator appears at one of \emph{two} times, tied by the
memory kernels; all routes to a Markovian equation differ only in how
they treat the two-time products $\tilde A^\dagger_l(t)\tilde A_l(s)$.

In the wideband limit---flat spectral densities over a bandwidth
$\Lambda\gg J,\gamma_\alpha$,
\begin{equation}
  J_\alpha(\omega)=\frac{\gamma_\alpha}{2\pi},\qquad|\omega|\lesssim\Lambda,
  \label{eq:flatband}
\end{equation}
together with the flat occupations---the kernels collapse,
\begin{equation}
  C_a(\tau)\to\frac{\gamma_a}{2\pi}\int_{-\Lambda}^{\Lambda}\!d\omega\,e^{-i\omega\tau}
  =\gamma_a\frac{\sin\Lambda\tau}{\pi\tau}\xrightarrow[\Lambda\to\infty]{}\gamma_a\delta(\tau),
  \label{eq:Ca_collapse}
\end{equation}
and similarly $C_b\to\gamma_b\delta(\tau)$. The time $s$ in
\eqref{eq:redfield} is then pinned to $t$: every two-time product
collapses to an equal-time one, $\tilde A_l(s)\to\tilde A_l(t)$, and with
$\int_0^tds\,\delta(t-s)F(s)=\tfrac12F(t)$ (the $\delta$-function sits at
the endpoint; each Hermitian-conjugate piece contributes half the rate)
one obtains, back in the Schr\"odinger picture, exactly
Eq.~\eqref{eq:LindbladHN} with the \emph{local} jump operators $A_l$, $B_l$
and $\gamma_a=\gamma_b=:\gamma$. At no point was an eigenoperator
decomposition needed: the $\delta$-kernel treats all Bohr frequencies
identically, and complete positivity is automatic---the usual role of the
secular approximation in repairing the Redfield equation is not needed
because there is nothing to repair.

\subsection{Time Scales, Validity, and the Singular-Coupling Limit}

The wideband limit is an instance of the singular-coupling limit of
Refs.~\cite{Gorini76a,Palmer77a,Breuer02a}, which considers the family
\begin{equation}
  H(\epsilon)=H_{\sys}+\epsilon^{-1}V+\epsilon^{-2}H_{\bath},\qquad\epsilon\to0 .
  \label{eq:SCL_family}
\end{equation}
The rescaling stretches every bath frequency and amplifies every
coupling, which on the spectral densities combine into
$J^{(\epsilon)}_\alpha(\omega)=J_\alpha(\epsilon^2\omega)$: the plateau
value is invariant while the support stretches,
$\Lambda\to\Lambda/\epsilon^2$. The correlation functions become
$C^{(\epsilon)}_\alpha(\tau)=\epsilon^{-2}C_\alpha(\tau/\epsilon^2)$, a
nascent $\delta$-function with fixed weight $\gamma_\alpha$ and shrinking
width $1/\Lambda$; for the box spectrum the family is the Dirichlet
kernel of Eq.~\eqref{eq:Ca_collapse}. The memory integral in
\eqref{eq:redfield} is one-sided, and since the kernel is real and even
exactly half of its weight lies at $\tau>0$, $\int_0^\infty d\tau\,
C^{(\epsilon)}_\alpha=\gamma_\alpha/2$, the Lamb shift vanishing at band
center by symmetry---the origin of the endpoint rule used above. The
leading correction is controlled by the first half-line moment of the
kernel, $\int_0^\infty d\tau\,\tau\,C^{(\epsilon)}_\alpha=\gamma_\alpha/\pi\Lambda$,
i.e., by relative errors $O((J+\gamma)/\Lambda)$. In this limit the reduced
dynamics converges rigorously, uniformly on bounded time intervals, to
the GKSL semigroup whose jump operators are the \emph{bare} coupling
operators $A_l$, $B_l$, with no secular step~\cite{Gorini76a,Palmer77a}.

Four time scales appear: the bath memory $\tau_B\simeq1/\Omega$, where
$\Omega$ is the width of the window over which $J_\alpha$ \emph{and} $f_\alpha$ are
simultaneously flat (in a Fermi-sea realization the smaller of $\Lambda$
and the detunings of the two Fermi levels from the band edges); the
system time $1/J$; the relaxation time $1/\gamma$; and the secular time
$\tau_{\rm sec}\gtrsim L/2\pi J$ (as large as $L^2/J$ near the band
edges), the inverse of the minimal Bohr-frequency spacing, which plays a
role only in the Davies route. The derivation requires precisely the
hierarchy \eqref{eq:validity}, $\Omega\gg\max(J,\gamma)$: the bath is
white on the system's internal time scale [corrections $O(J/\Omega)$]
and the Born--Markov truncation holds [corrections $O(\gamma/\Omega)$].
The leading finite-$\Omega$ corrections are the Lamb shift
$\Delta_\alpha(\omega)=\frac{\gamma_\alpha}{2\pi}\ln\frac{\Lambda+\omega}
{\Lambda-\omega}\simeq\gamma_\alpha\omega/\pi\Lambda$, an $O(\gamma J/\Lambda)$
renormalization of the dispersion, and non-Markovian corrections from the
$1/\Lambda\tau$ tails of the sinc kernel. The per-step dilations of
Sec.~\ref{sec:Methods} realize the same hierarchy with $\tau_B=\Delta t$:
a fresh, uncorrelated bath register per step gives a bath memory of
exactly one step, hence an effective flat spectrum of width
$\sim2\pi/\Delta t$, and the coupling rescaling $\sqrt{\gamma/\Delta t}$ is
the discrete counterpart of the $\delta$-correlated noise
$C(\tau)=\gamma\delta(\tau)$.

\subsection{The Secular (Davies) Alternative}
\label{app:secular}

It is instructive to see what the textbook secular approximation
\emph{would} produce, both to justify avoiding it and because the
exercise exposes the chiral-damping structure in momentum space. We work
on a ring with $c_l=L^{-1/2}\sum_ke^{ikl}c_k$, $\varepsilon(k)=-J\cos k$,
and group velocity $v_k=J\sin k$. The Davies construction~\cite{Davies74a,Breuer02a}
decomposes each coupling operator into eigenoperators of $H_{\sys}$ and
discards all frequency off-diagonal terms. Since $c_k$ is itself an
eigenoperator,
\begin{equation}
  A_l=\frac{1}{\sqrt L}\sum_ke^{ikl}u_kc_k,\qquad
  u_k=\frac{1+i\eta e^{ik}}{\sqrt{1+\eta^2}},
  \label{eq:eigenops}
\end{equation}
so each local jump operator is a coherent superposition over the entire
band of Bohr frequencies; similarly $B_l$ with
$\tilde u_k=(1+i\eta e^{-ik})/\sqrt{1+\eta^2}$. The weights carry the
chirality, $|u_k|^2=1-\teta\sin k$ and $|\tilde u_k|^2=1+\teta\sin k$. The secular
approximation is legitimate only if $\gamma\ll\min_{\omega\neq\omega'}
|\omega-\omega'|\sim(2\pi J/L)|\sin k|$, which for a gapless chain fails in
the thermodynamic limit at \emph{any} fixed $\gamma$---the well-known
breakdown of the global master equation for transport in extended
systems~\cite{Cattaneo19a}. More importantly, even where formally valid
it is \emph{unwanted}: it replaces the local jump operators by the
energy-filtered, delocalized $A_l(\omega)$, and the very structure the
derivation is meant to produce lives in the cross-frequency coherences
it throws away. Locality survives if and only if the half-Fourier
transform $\int_0^\infty d\tau\,e^{i\omega\tau}C_\alpha(\tau)$ is
$\omega$-independent across the system band---exactly the wideband
condition \eqref{eq:flatband}.

Carrying the construction through anyway (with flat rates), the cross
terms within a degenerate shell $\{k,-k\}$ carry $\sum_le^{2ikl}=0$, so the
result is diagonal in momentum, with the chiral rates
\begin{align}
  \gamma^{\rm loss}_k&=\gamma\Big(1-\teta\frac{v_k}{J}\Big),
  \label{eq:chiral_rates}\\
  \gamma^{\rm gain}_k&=\gamma\Big(1+\teta\frac{v_k}{J}\Big).
  \label{eq:chiral_rates_gain}
\end{align}
For $\eta>0$ right-movers are weakly lost and strongly fed, left-movers
the reverse: the momentum-space fingerprint of chiral
damping~\cite{Song19a}, the fermionic counterpart of
Eq.~\eqref{eq:spectrum_spin}. The occupations obey independent rate
equations with $\gamma^{\rm gain}_k+\gamma^{\rm loss}_k=2\gamma$ and
\begin{equation}
  n^{\rm ss}_k=\frac{1+\teta\sin k}{2},
  \label{eq:nk_ss}
\end{equation}
and the steady state carries a net chiral current $\teta J/4$ per
site---a check that the $i\eta$ phase biases transport. The
anti-Hermitian part of $\Heff$ is the same with or without the secular
approximation on the ring, since $\sum_l(A_l^\dagger A_l+B_l^\dagger B_l)$ is
translation invariant and hence already frequency diagonal; the two
master equations differ only in the recycling terms, local
$A_l\rho A_l^\dagger$ versus momentum-diagonal $c_k\rho\cdag_k$. Being
translation covariant and diagonal in $k$, the secular equation cannot
produce any boundary accumulation---the Liouvillian skin effect lives in
the frequency coherences that the secular step discards; every mode
relaxes at the same rate $2\gamma$, so it misses the skin-induced
separation between Liouvillian gap and relaxation time~\cite{Haga21a}.
The Hatano--Nelson Lindbladian is a \emph{local} (wideband/singular-coupling)
Lindblad equation, not a \emph{global} (secular) one, and only the local
one exhibits the skin phenomenology that motivates the model.

\section{Derivation of the NESS Skin Depth}
\label{app:kernel}

This appendix derives the NESS skin depth of Eq.~\eqref{eq:lness},
$1/\lNESS=2\,\mathrm{arcsinh}(\gamma/J)$, from the Lyapunov equation
\eqref{eq:Lyapunov}. The argument proceeds in four steps. The equation
for the deviation $\delta C=C-\tfrac12\one$ is set up, which shows that
all $\eta$ dependence sits in a current-like source while the resolvent
that fixes the decay is $\eta$-free. The source is then split into a bulk
part, which drives a uniform current but no density, and an emergent,
edge-localized residual. The decay rate is read off from the evanescent
zeros of the bulk resolvent symbol. Finally, a Bessel-kernel
representation supplies the closed form and the $x^{-1/2}$ prefactor
that biases finite-window fits.

\subsection{Setup}
We use the matrices of Sec.~\ref{sec:GaussianSolution}: the hopping
matrix $h$ of Eq.~\eqref{eq:Hsys}, real and symmetric, the dyad matrix
$M=\alpha^\dagger\alpha$ of Eq.~\eqref{eq:Mdef} and the drift $D$ of
Eq.~\eqref{eq:Ddef}. For the Hatano--Nelson jump vectors the off-diagonal
entries of $M$ are purely imaginary, $M_{l,l+1}=i\eta/(1+\eta^2)$, so
$\Re M$ is \emph{exactly diagonal},
\begin{equation}
  \Re M=\mathrm{diag}\Big(\frac{1}{1+\eta^2},1,\dots,1,\frac{\eta^2}{1+\eta^2}\Big):
  \label{eq:ReM}
\end{equation}
the Hatano--Nelson cross terms of loss and gain cancel bond by bond
(Sec.~\ref{sec:GaussianSolution}), leaving the $\eta$-independent bulk
damping rate $2\gamma$ of $\delta C$, the only exceptions being the two
terminated edge bonds, whose damping deficits are asymmetric and swap
under $\eta\leftrightarrow1/\eta$---the microscopic origin of the
height's broken self-duality. The chirality resides entirely in $\Im M$,
which is purely off-diagonal and current-like, with entries
$\pm\eta/(1+\eta^2)$ on the bonds. Substituting $C=\tfrac12\one+\delta C$
into Eq.~\eqref{eq:Lyapunov} and using $D+D^\dagger=-2\gamma\Re M$ gives
\begin{equation}
  D\,\delta C+\delta C\,D^\dagger=-i\gamma\,\Im M,
  \label{eq:lyapS}
\end{equation}
so that \emph{all} $\eta$ dependence resides in the source, of amplitude
$\gamma\eta/(1+\eta^2)$, while the resolvent that fixes the decay is
$\eta$-free---the analytic origin of the $\eta$-independent depth. The
solution is
\begin{equation}
  \begin{split}
  \delta C&=i\gamma\int_0^\infty\!dt\;e^{Dt}\,\Im M\,e^{D^\dagger t}\\
  &\xrightarrow[\text{bulk}]{\Re M\to\one}
  i\gamma\int_0^\infty\!dt\;e^{-2\gamma t}e^{iht}\,\Im M\,e^{-iht}.
  \end{split}
  \label{eq:intrep}
\end{equation}

\subsection{Bulk--Edge Decomposition and the Exact Residual}
The matrix $\Im M$ is bulk-uniform, not edge-localized; the boundary source is
emergent. In the bulk $h$ and $\Im M$ are both diagonal in momentum,
$h(k)=-J\cos k$ and $(\Im M)(k)=2i\eta\sin k/(1+\eta^2)$, so
$[h,\Im M]=0$ and the infinite-chain solution is the uniform current
pattern $\delta C_{\rm bulk}=\tfrac{i}{2}\Im M$, whose diagonal vanishes:
the uniform source maintains a steady current but no density. On the
open chain the remainder $\delta C_{\rm e}=\delta C-\delta C_{\rm bulk}$
obeys Eq.~\eqref{eq:lyapS} with the residual source
\begin{equation}
  R=\frac12\big[h,\Im M\big]+\frac{i\gamma}{2}\big\{\Re M-\one,\Im M\big\}
  \label{eq:Rres}
\end{equation}
in place of $-i\gamma\Im M$, verified numerically to machine precision.
Both terms are edge-localized: the commutator reduces to the diagonal
corner entries $([h,\Im M])_{11}=-([h,\Im M])_{LL}=J\eta/(1+\eta^2)$, a
real on-site source $\pm J\eta/2(1+\eta^2)$ at the two corner sites, and
the edge-damping correction lives on the two terminated bonds.
$\delta C_{\rm e}$ carries all of the density.

\subsection{Exact Rate from the Resolvent Symbol}
Away from the edges
$\delta C_{\rm e}=-(2\gamma-i\,\mathrm{ad}_h)^{-1}R$, and the spatial decay
is set by the evanescent zeros of the bulk symbol
$\Sigma(k,k')=2\gamma-i[\varepsilon(k)-\varepsilon(k')]$. Decay along the
diagonal at rate $\kappa$ corresponds to the complex pair
$k=q+i\kappa/2$, $k'=q-i\kappa/2$, for which
$\varepsilon(k)-\varepsilon(k')=2iJ\sin q\sinh(\kappa/2)$, so
$\Sigma=0\iff\gamma=J\sin q\sinh(\kappa/2)$. The slowest evanescent pair
minimizes $\kappa$ at $\sin q=1$ (band center, maximal group velocity),
giving
\begin{equation}
  \frac{1}{\lNESS}=\kappa_{\min}=2\,\mathrm{arcsinh}\frac{\gamma}{J},
  \label{eq:arcsinh}
\end{equation}
Eq.~\eqref{eq:lness} with its exact lattice completion; the leading term
is the ballistic estimate $v_{\max}/2\gamma$. The derivation is exact and
boundary independent: the open boundary determines the source and the
prefactor, never the exponent. The deviation is carried by an evanescent
particle--hole pair, each member a damped single-particle resolvent
$[(\gamma-ih)^{-1}]_{x,0}\sim e^{-x\,\mathrm{arcsinh}(\gamma/J)}$.

\subsection{Kernel Representation and Prefactor}
On the infinite chain
the Jacobi--Anger expansion gives $(e^{iht})_{x,y}=(-i)^{x-y}J_{x-y}(Jt)$,
with $J_n$ the Bessel function (the semi-infinite Dirichlet propagator
adds an image term of the same exponential order that modifies only the
prefactor). The remainder density at depth $x$ is a sum of products of
two Bessel functions of nearly equal index with the representative kernel
\begin{equation}
  K(x)=\int_0^\infty\!dt\;e^{-2\gamma t}J_x(Jt)^2
  =\frac{1}{\pi J}\,Q_{x-1/2}\!\Big(1+\frac{2\gamma^2}{J^2}\Big),
  \label{eq:closed}
\end{equation}
by the classical Laplace transform of a Bessel product, $Q_\mu$ the
Legendre function of the second kind. With
$Q_\nu(\cosh\alpha)\simeq\sqrt{\pi/2\nu\sinh\alpha}\;e^{-(\nu+1/2)\alpha}
[1+O(\nu^{-1})]$ and $\cosh\alpha=1+2\gamma^2/J^2$, i.e.,
$\alpha=2\,\mathrm{arcsinh}(\gamma/J)$, one finds $K(x)\propto
x^{-1/2}e^{-\alpha x}$, recovering \eqref{eq:arcsinh}. The $x^{-1/2}$
prefactor is the finite-window fit bias: a log-linear fit over a window
centered at $\bar x$ reads the rate $\alpha+1/2\bar x$. Numerically, the
local rates $\ln[Q_{x-1/2}/Q_{x+1/2}]$ at $\gamma/J=0.7$ are
$1.3517,1.3216,1.3108$ at $x=10,30,90$ against $\alpha=1.3053$---excesses
$0.046,0.016,0.006$ matching $1/2x=0.050,0.017,0.006$.

\section{System--Bath Model of the Local Spin Lindbladian}
\label{app:spinbath}

Each bath spin $T_l$ is a spin-$\tfrac12$ with splitting $\omega$ which,
through its coupling to the common reservoir, flips down and up at the
rates $\kappa_-$ and $\kappa_+$ (the sign labeling the flip operator
$T^\mp_l$, as for $A^\mp_l$ and $\Gamma_\mp$),
\begin{equation}
  \dot\rho_T=-i[\omega T^z_l,\rho_T]
  +\kappa_-\mathcal D[T^-_l]\rho_T+\kappa_+\mathcal D[T^+_l]\rho_T,
  \label{eq:bathspin}
\end{equation}
obeying detailed balance at the reservoir temperature,
$\kappa_+/\kappa_-=e^{-\beta\omega}$, with total rate
$\kappa=\kappa_-+\kappa_+$; the stationary state is the thermal one, with
occupations $p_\downarrow=\kappa_-/\kappa$, $p_\uparrow=\kappa_+/\kappa$.
The reservoir is common in that it fixes one $\beta$ for every bond; we
assume that it does not correlate different bath spins at the
Born--Markov level, so cross-bond correlators vanish and the
bond-diagonality of the dissipator is exact. Writing $R_l :=  gT^-_l$,
the coupling \eqref{eq:Hint_spin} takes the form
$H_{\rm int}=\sum_l\big(R^\dagger_lA^-_l+A^+_lR_l\big)$ (bath and system
operators commute), and by the quantum regression theorem the stationary
two-time correlators of the damped spin are
\begin{align}
  C_\downarrow(\tau)& := \big\langle\tilde R_l(t)\tilde R^\dagger_l(s)\big\rangle
  =g^2\,p_\downarrow\,e^{-i\omega\tau-\kappa|\tau|/2},
  \label{eq:Cdown}\\
  C_\uparrow(\tau)& := \big\langle\tilde R^\dagger_l(t)\tilde R_l(s)\big\rangle
  =g^2\,p_\uparrow\,e^{+i\omega\tau-\kappa|\tau|/2},
  \label{eq:Cup}
\end{align}
$\tau=t-s$, using $T^-T^+=|\!\downarrow\rangle\langle\downarrow\!|$,
$T^+T^-=|\!\uparrow\rangle\langle\uparrow\!|$ and the decay of the
coherence $T^\pm$ at $\kappa/2$. In Appendix~\ref{app:HNbath} the two
orderings belonged to two different reservoirs and one of them vanished
for each; here both are nonzero for the \emph{same} bath spin, weighted
by its thermal occupations, and they obey the detailed-balance
[Kubo--Martin--Schwinger (KMS)] relation
\begin{equation}
  C_\uparrow(\tau)=e^{-\beta\omega}\,C_\downarrow(-\tau),
  \label{eq:KMS_spin}
\end{equation}
which is the origin of Eq.~\eqref{eq:beta_spin}.

Inserting $H_{\rm int}$ into the double commutator
\eqref{eq:double_commutator} and applying the Born factorization, the bath
trace produces
\begin{widetext}
\begin{align}
  \dot{\tilde\rho}(t)=-\sum_l\int_0^t\!ds\,\Big\{
  &C_\downarrow(t-s)\big[\tilde A^+_l(t)\tilde A^-_l(s)\tilde\rho(s)
    -\tilde A^-_l(s)\tilde\rho(s)\tilde A^+_l(t)\big]\nonumber\\
  +\,&C_\uparrow(t-s)\big[\tilde A^-_l(t)\tilde A^+_l(s)\tilde\rho(s)
    -\tilde A^+_l(s)\tilde\rho(s)\tilde A^-_l(t)\big]
  +\mathrm{h.c.}\Big\},
  \label{eq:redfield_spin}
\end{align}
\end{widetext}
the exact counterpart of Eq.~\eqref{eq:redfield} with $A_l\to A^-_l$,
$B_l\to A^+_l$, $C_a\to C_\downarrow$, $C_b\to C_\uparrow$. The memory
kernels are one-sided exponentials of width $\tau_B=2/\kappa$, and their
half-Fourier transforms
\begin{align}
  \int_0^\infty\!d\tau\,e^{i\nu\tau}C_{\downarrow}(\tau)
  &=\frac{g^2p_\downarrow}{\kappa/2-i(\nu-\omega)},
  \label{eq:lorentz_down}\\
  \int_0^\infty\!d\tau\,e^{i\nu\tau}C_{\uparrow}(\tau)
  &=\frac{g^2p_\uparrow}{\kappa/2-i(\nu+\omega)},
  \label{eq:lorentz_up}
\end{align}
are Lorentzians of width $\kappa$ centered at $\nu=\pm\omega$. For
$\kappa\gg J,\omega$ they are flat across the system band $|\nu|\le J$---the
bath is white on the system's time scale---so, exactly as in
Eq.~\eqref{eq:Ca_collapse} and the steps following it, the kernels
collapse to $\Gamma_{-,+}\delta(\tau)$ with the endpoint rule, every
two-time product becomes an equal-time one, the local operators $A^\mp_l$
survive undecomposed, and Eq.~\eqref{eq:redfield_spin} reduces to the GKSL
form \eqref{eq:lindblad_spin} with
$\Gamma_{-,+}=2\,\Re\!\int_0^\infty d\tau\,C_{\downarrow,\uparrow}(\tau)$,
i.e., Eq.~\eqref{eq:rates_spin}. The imaginary parts of the same
integrals, $\mp\Gamma_{-,+}\omega/\kappa$, give the Lamb shift
$H_{\rm LS}=\frac{\omega}{\kappa}\sum_l\big(\Gamma_+ A^-_lA^+_l
-\Gamma_- A^+_lA^-_l\big)$, by Eq.~\eqref{eq:AAspin} a Zeeman shift of
order $\Gamma\omega/\kappa$ plus a hopping correction of order
$\delta\omega/\kappa$, both of relative order $\omega/\kappa$ and dropped.
The validity hierarchy is that of Eq.~\eqref{eq:validity} with
$\tau_B=2/\kappa$: $\kappa\gg J$ keeps the jump operators local (the
argument of Appendix~\ref{app:secular} applies verbatim), $\kappa\gg\Gamma$
(i.e., $\kappa\gg g$) is the Born--Markov condition, and $\kappa\gg\omega$
makes the Lorentzians flat across the band and the Lamb shift
negligible---hence $\kappa\gg J,g,\omega$. No condition involves $L$. The
bath spin's own damping is what supplies the memory cutoff: $T_l$ is
refreshed by the reservoir at rate $\kappa$, so it plays the role of the
fresh bath qubit of each collision. Because both channels are fed by the same
thermal spin, $\Gamma_-/\Gamma_+=p_\downarrow/p_\uparrow=e^{\beta\omega}$ for
every bond---the KMS relation that the two engineered reservoirs of
Appendix~\ref{app:HNbath} deliberately violate. Since $A^-_l$ lowers
$S^z_{\rm tot}$ by exactly one, this is quantum detailed balance with
respect to $e^{-\beta\omega S^z_{\rm tot}}$, and it is why the steady state
is the uniform thermal product state \eqref{eq:rhobeta} for any $\phi$.

\bibliography{References}

\end{document}